\def\comments{0}
\def\llncs{0}
\documentclass[11pt,letterpaper]{article}
\usepackage{graphicx} 
\usepackage{fullpage}
\usepackage{amsmath}

\usepackage{amsthm}
\usepackage{amssymb}
\usepackage{physics}
\usepackage{aliascnt}
\usepackage{hyperref}
\usepackage{cleveref}
\usepackage{dsfont}
\usepackage{xcolor}
\usepackage{mathtools}
\usepackage{breakcites}
\usepackage{subcaption}
\usepackage[shortlabels]{enumitem}

\newcommand{\newsharedtheorem}[2]{%
    \newaliascnt{#1}{theorem}%
    \newtheorem{#1}[#1]{#2}%
    \aliascntresetthe{#1}%
}

\ifnum\llncs=0
\newtheorem{theorem}{Theorem}[section]
\newsharedtheorem{proper}{Property}
\newsharedtheorem{corollary}{Corollary}
\newsharedtheorem{definition}{Definition}
\newsharedtheorem{lemma}{Lemma} 
\newtheorem{claim}{Claim}      
\newsharedtheorem{proposition}{Proposition}
\newsharedtheorem{conjecture}{Conjecture}
\newsharedtheorem{remark}{Remark}

\theoremstyle{definition}

\newsharedtheorem{note}{Note}

\newtheorem{construction}{Construction}
\fi

\newcommand{\resultrefnames}[5]{%
    \crefname{#1}{#2}{#3}%
    \Crefname{#1}{#4}{#5}%
    \expandafter\def\csname #1autorefname\endcsname{#4}%
}
\resultrefnames{theorem}{theorem}{theorems}{Theorem}{Theorems}
\resultrefnames{proper}{property}{properties}{Property}{Properties}
\resultrefnames{corollary}{corollary}{corollaries}{Corollary}{Corollaries}
\resultrefnames{definition}{definition}{definitions}{Definition}{Definitions}
\resultrefnames{lemma}{lemma}{lemmas}{Lemma}{Lemmas}
\resultrefnames{claim}{claim}{claims}{Claim}{Claims}
\resultrefnames{proposition}{proposition}{propositions}{Proposition}{Propositions}
\resultrefnames{conjecture}{conjecture}{conjectures}{Conjecture}{Conjectures}
\resultrefnames{remark}{remark}{remarks}{Remark}{Remarks}
\resultrefnames{note}{note}{notes}{Note}{Notes}
\resultrefnames{notation}{notation}{notations}{Notation}{Notations}
\resultrefnames{observation}{observation}{observations}{Observation}{Observations}
\resultrefnames{invariant}{invariant}{invariants}{Invariant}{Invariants}
\resultrefnames{construction}{construction}{constructions}{Construction}{Constructions}

\newcommand{\A}{\mathcal{A}}
\newcommand{\B}{\mathcal{B}}
\newcommand{\D}{\mathcal{D}}
\renewcommand{\O}{\mathcal{O}}
\newcommand{\N}{\mathbb{N}}
\newcommand{\Z}{\mathbb{Z}}

\newcommand{\C}{\mathbb{C}}
\newcommand{\E}{\mathop{\mathbb{E}}}

\newcommand{\SAME}{\mathsf{SAME}}
\newcommand{\DIFF}{\mathsf{DIFF}}

\newcommand{\poly}{\mathsf{poly}}
\newcommand{\negl}{\mathsf{negl}}

\newcommand{\wt}[1]{\widetilde{#1}}
\newcommand{\ol}[1]{\overline{#1}}
\newcommand{\floor}[1]{\left\lfloor #1 \right\rfloor}
\newcommand{\ceil}[1]{\left\lceil #1 \right\rceil}

\DeclareMathOperator{\ctrl}{ctrl}

\DeclareMathOperator{\Ima}{\mathsf{Im}}
\DeclareMathOperator{\Supp}{\mathsf{Supp}}
\DeclareMathOperator{\Sim}{\mathsf{Sim}}

\DeclareMathOperator{\TD}{\mathsf{TD}}

\DeclareMathOperator{\Span}{Span}
\DeclareMathOperator{\Sw}{Sw}

\newcommand{\xor}{\oplus}

\newcommand{\Haar}{\mathsf{Haar}}

\ifnum\comments=1
\newcommand{\todo}[1]{{\color{blue} (TODO: #1) }}
\newcommand{\eli}[1]{{\color{purple} (Eli: #1) }}
\else
\newcommand{\todo}[1]{}
\newcommand{\eli}[1]{}
\fi

\newcommand{\Reg}[1]{\mathbf{#1}}

\newcommand{\Hyb}{\mathsf{Hyb}}

\title{On Random Purifications of Isometries and Channels}
\author{Eli Goldin\\Princeton University}

\begin{document}

\maketitle

\begin{abstract}
    Acorn tricks have emerged recently as a powerful tool used to turn weak quantum resources into stronger ``randomized'' ones. The most successful recent example of this is the random purification channel, which turns $t$ copies of a mixed state into $t$ copies of a random purification. It is an open question whether similar techniques apply for quantum channels. In detail, can queries to an arbitrary isometry channel be transformed into queries to a random unitary completion? Similarly, can queries to an arbitrary quantum channel be transformed into queries to a random isometry dilation?

    We answer both of these questions. In particular, for an isometry from dimension $N$ to dimension $M$, we give an efficient, stateful simulator emulating $T$ sequential, forward queries to a fixed Haar random unitary completion. The simulator makes at most one forward isometry query per simulated call and achieves total error $O\left(T\sqrt{\frac{N}{M-T}}\right)$ in trace distance.

    The existence of such a simulator implies that all isometry query lower bounds extend to algorithms with the ability to query an arbitrary unitary completion. Viewing state preparation isometries in this lens, we also can translate arbitrary copy complexity lower bounds to the setting of forward access to state preparation unitaries. For both lifting theorems, the loss is controlled by the relevant dimensions and can be made arbitrarily small for most applications.

    Conversely, we give a pair of distributions on quantum channels which can be distinguished exactly using two queries to an arbitrary dilation but which require exponentially many channel queries to distinguish with nonnegligible advantage. This conclusively rules out efficient simulation of random dilations of a channel.

    As a result of our lower bound technique, we also give a lower bound on approximately implementing the quantum switch. We show that simulating the quantum switch to any constant error below $1/2$ requires exponentially many channel queries, even when the channels are identical. This gives an exponential query advantage for indefinite causal order compared to standard quantum circuits, improving on the restricted separation of (Bavaresco et al., 2025).

    Of independent interest, we show how to ``decontrol'' controlled forward query access to an isometry oracle. We provide a transformation that converts an algorithm making controlled queries to $\phi V$ for a random phase $\phi$ into an algorithm making the same number of uncontrolled forward queries to $V$. This generalizes the analogous result for unitaries.
\end{abstract}

\newpage
\tableofcontents
\newpage
\section{Introduction}

\paragraph*{The random purification channel}
One of the most powerful recent techniques in quantum information is a trick that has come to be known as the ``random purification channel''~\cite{TWZ26,AG26}. The random purification channel is a map which turns $t$ copies of any mixed state $\rho$ into $t$ copies of a random purification of $\rho$. Thus, any algorithm which uses copies of the purification of a mixed state (and is agnostic to the choice of purification) can be converted into an algorithm using copies of the original mixed state itself. This has enabled applications across many fields, including quantum algorithms~\cite{UNWT25}, quantum cryptography~\cite{AG26}, quantum learning theory~\cite{PSTW25,PSTW26,FGSS26,CWZ25}, and quantum information theory~\cite{GML26,Zhou26}.

{\emergencystretch=5em The random purification channel is an example of something known as an ``acorn trick''~\cite{TWZ26}, a trick where a weak quantum resource can be transformed into a randomized variant of a stronger quantum resource. This trick first appeared in its modern form in~\cite{C:GolZha25}, although a number of prior works used similar ideas for more targeted applications~\cite{Kretschmer21,ITCS:Zhandry24b}. Alongside the random purification channel, acorn tricks have been used to give lifting theorems for a number of other resources. These include simulating controlled query access to a unitary using only forward query access~\cite{TW26}, simulating state preparation unitaries using copies of the state~\cite{C:GolZha25,TWZ26}, and generalizations to bosonic Gaussian states~\cite{WW25,MGFL25,CFGL+26}.\par}

\paragraph*{Generalization to isometries and channels}
Quantum maps generally come in three flavors, unitaries, isometries, and general quantum channels. Unitary maps are arbitrary reversible quantum maps sending $n$ qubits to $n$ qubits. Operationally, unitary maps can be understood as arbitrary reversible maps acting on $n$ qubits. Isometries are the generalization of unitaries to maps with different input and output spaces. An isometry is thus a map which takes in any $n$ qubit pure state (of unit norm) and outputs an $m > n$ qubit pure state (of unit norm). Quantum channels are the most general kind of quantum map. These represent general computation, with the ability to apply arbitrary unitaries/isometries as well as perform arbitrary measurements (in any order).

Just as mixed states can be ``purified'' to pure states, isometries can be ``completed'' to unitaries and channels can themselves be ``dilated'' to isometries (and thus again completed to unitaries). Formally, for every isometry $V$, there is a unitary $D^V$ such that
$$V\ket{\phi} = D^V(\ket{\phi}\otimes \ket{0})$$
For every quantum channel $\Phi$, there is an isometry $V^\Phi$ such that 
$$\Phi(\rho) = \Tr_{\Reg{B}}(V^\Phi\rho V^{\Phi,\dagger})$$

Thus, one may ask whether acorn tricks can also be used to give lifting theorems for query access to isometries or general quantum channels. Can isometry queries be lifted to unitary queries? Can channel queries be lifted to isometry queries? Tang, Wright, and Zhandry formulate these two conjectures as follows.

\begin{conjecture}[Completing isometries, Conjecture~1.10 of~\cite{TWZ26}]
    \label{conj:isometry-to-unitary}
    Let $V:\Reg A\to\Reg A\otimes\Reg E$ be an isometry. Every
    $q$-query algorithm $\A$ for unitaries on $\Reg A\otimes\Reg E$
    can be transformed into an algorithm $\Sim_\A$ making $\poly(q)$
    queries to $V$ such that
    \[
        \Sim_\A^V= \E_{D\gets\mathcal D}[\A^D],
    \]
    for some distribution $\mathcal D$ over unitaries $D$ satisfying
    \[
        D(\ket{\psi}_{\Reg A}\otimes\ket0_{\Reg E})=V\ket{\psi}
        \qquad\text{for every }\ket{\psi}\in\Reg A.
    \]
\end{conjecture} 

\begin{conjecture}[Dilating channels, Conjecture~1.9 of~\cite{TWZ26}]
    \label{conj:channel-to-isometry}
    Let $\Phi$ be a channel on $\Reg A$, and let $\Reg E$ be a sufficiently
    large purification register. Every $q$-query algorithm $\A$ for
    isometries $\Reg A\to\Reg A\otimes\Reg E$ can be transformed into
    an algorithm $\Sim_\A$ making $\poly(q)$\footnote{Note that the original paper required the simulator for channels to make only $q$ queries. As we refute this conjecture, weakening it by strengthening the simulator only strengthens our result.} queries to $\Phi$ such that
    \[
        \Sim_\A^\Phi= \E_{W\gets\mathcal D}[\A^W],
    \]
    for some distribution $\mathcal D$ over isometries $W$ satisfying
    \[
        \Tr_{\Reg E}(W\rho W^\dagger)=\Phi(\rho)
        \qquad\text{for every state }\rho.
    \]
\end{conjecture}

Prior to this work, both of these conjectures have been proven false in the exact setting~\cite{MSM19conjugation,YM26conjugation,YNUTM25,GdTQB26}. As such, we consider the approximate versions of these conjectures, where the simulator is allowed to differ from its target state by some error parameter.

We prove an approximate version of the first conjecture. That is, the simulator will approximately implement the channel up to error depending on the completion dimension. In particular, this error is inverse exponential in the number of ancilla qubits needed to implement the completion. 

We also refute the second conjecture, including its approximate versions. That is, we show that simulating a random dilation of a quantum channel up to constant error requires an exponential number of queries.

Our results and techniques immediately give three important related results. 

\begin{enumerate}
    \item(Controlled queries to isometries do not help) It is known that controlled queries to a unitary can be simulated using only forward queries up to a random global phase. As a key step in our argument, we show that an analogous result holds for isometries.
    \item(State preparation lower bounds) An algorithm making $t$ forward queries to a random state preparation unitary in dimension $M$ can be efficiently simulated using $t$ copies of the corresponding state, with trace-distance error $O(t/\sqrt{M-t})$. Thus, a copy-complexity lower bound $S$ gives a forward-query lower bound $\Omega(\min\{S,\sqrt M\})$; see~\Cref{sec:applications}.
    \item(Lower bounds for the quantum switch) The techniques used for ruling out simulating random dilations of channels also immediately give a lower bound on implementing a superchannel known as the quantum switch. In particular, we show that implementing the quantum switch up to error $\frac{1}{2}-\epsilon$ requires an exponential number of queries to the underlying channels.
\end{enumerate}

\subsection{Our Results}

\paragraph{Barriers for simulating isometry queries}

A number of results in the literature implicitly disprove~\Cref{conj:isometry-to-unitary} in the exact setting. To give one example, complex conjugation of an unknown $M$-dimensional unitary can be implemented exactly using $M-1$ forward queries~\cite{MSM19conjugation}. If an exact simulator as in~\Cref{conj:isometry-to-unitary} existed, then one would also be able to conjugate an unknown isometry from dimension $N$ to $M>N$ using $\poly(M)$ queries. However, exact complex conjugation of an unknown isometry is impossible with any finite number of queries~\cite{YM26conjugation}. 

One may hope to bypass this impossibility by allowing the simulator to emulate a random completion of $V$ up to some error. We show that this is indeed possible, and furthermore the corresponding error is quite small.

\paragraph{Approximately simulating isometry queries}

We establish an explicit simulator proving an approximate version of~\Cref{conj:isometry-to-unitary}.

\begin{theorem}[Informal version of \Cref{thm:exact-main}]
    Let $T\in \N$ and let $\Reg{A},\Reg{E}$ be arbitrary Hilbert spaces, where $\Reg{E}$ is of dimension $D\geq 2$. Then there is a stateful simulator $\Sim^{(\cdot)}$ with the following property. Let $V:\Reg{A}\to \Reg{A}\otimes \Reg{E}$ be an isometry and let $\A^{(\cdot)}$ be a $T$ query algorithm. Then
    $$\TD\left(\E_{D^V}[\A^{D^V}], \A^{\Sim^V}\right) = O\left(\frac{T}{\sqrt{D}}\right)$$
    where the expectation is taken as the Haar distribution over all completions $D^V$ of $V$.
    
    Furthermore, $\Sim^V$ can be efficiently implemented in time $\poly(T,\log (\dim(\Reg{AE})))$ and makes exactly one query to $V$ per call.
\end{theorem}

For any oracle algorithm $\A^{(\cdot)}$ making queries to a random unitary completion of $V$, we can use the above theorem to approximate the algorithm's average output using the same number of forward queries to $V$, with error decreasing as the completion dimension grows.

In particular, for an isometry mapping $n$ qubits to $n+m$ qubits, the error scales with $2^{-m/2}$. As such, for sufficiently expanding isometries, the error is effectively inverse exponential.

We also observe that for any isometry $V$, the isometry $V\otimes \ket{0^m}$ is also an isometry for which query access is similarly useful. Under our theorem, simulating random completions of $V\otimes \ket{0^m}$ will always have inverse exponential error in $m$. As such, error depending on the completion dimension can be made sufficient for many purposes, even when $V$ itself is not very expanding.

\paragraph{Simulating controlled queries to isometry oracles}

As an important lemma for the proof of~\Cref{thm:exact-main}, we show that controlled queries to an isometry can be exactly simulated using forward queries. This is an extension of~\cite{TW26}, which showed the same result holds for unitaries.

\begin{theorem}[Simulating controlled isometry queries, informal]\label{thm:intro-decontrol-isometry}
    Let $T\in \N$ and let $\Reg{A},\Reg{E}$ be arbitrary Hilbert spaces. Then there is a stateful simulator $\Sim_{ctrl}^{(\cdot)}$ with the following property. Let $V:\Reg{A}\to \Reg{A}\otimes \Reg{E}$ be an isometry and let $\A^{(\cdot)}$ be an algorithm making at most $T$ controlled queries to its oracle. Then
    $$\E_{\theta \gets [0,2\pi)}[\A^{\ctrl[e^{i\theta}V]}] = \A^{\Sim_{ctrl}^V}$$
    
    {\emergencystretch=2em Furthermore, $\Sim_{ctrl}^V$ can be efficiently implemented up to error $\epsilon$ in time $\poly(\log(1/\epsilon),\allowbreak T,\allowbreak\log (\dim(\Reg{AE})))$ and makes exactly one query to $V$ per call.\par}
\end{theorem}

\paragraph{Applications of simulating random completions}

As our main theorem is very general, it immediately translates any algorithm which works for all completions of an isometry into an algorithm making direct queries to the isometry. The implications of this are most clear in the setting of lower bounds. As examples, we introduce new lower bounds for the tasks of isometry tomography and isometry inversion when query access is given to an arbitrary completion of the isometry. The lower bound for isometry inversion is tight for sufficiently expanding isometries.

We also give a specialization of our main theorem to the setting of state preparation unitaries. For a (potentially mixed) state $\rho$, a state preparation unitary for $\rho$ is any unitary $U_\rho$ such that
$$\rho_{\Reg{A}} = \Tr_{\Reg{B}}(U_\rho \ketbra{0}_{\Reg{AB}} U_\rho^\dagger)$$

We show that copy complexity lower bounds translate directly to state preparation lower bounds, with error loss scaling inversely with the dimension of the corresponding preparation unitary. 

\begin{theorem}[Informal version of~\Cref{thm:perfectlb} and~\Cref{cor:state-lifting-lb}]
    Let $D\geq N$ and consider a decision task involving an unknown $N$ dimensional state $\rho$. Suppose that solving the task with success probability $2/3$ requires $\Omega(L)$ copies of $\rho$. Then solving the same task with success probability at least $2/3$ using forward queries to a state preparation unitary for $\rho$, with a purification register of dimension $D$, requires
    $$\Omega(\min(L,\sqrt{N\cdot D}))$$
    queries for algorithms required to succeed for every compatible preparation unitary. For pure state problems, the bound holds for every $D\geq 1$.
\end{theorem}

This generalizes the pure-state lifting theorem of Isogai, Yoshida, and Murao (Theorem~3.3 of~\cite{IYM26PAC}), and gives an efficient simulation without their promise on a known small subspace orthogonal to $\ket0$. For more details, see~\Cref{sec:applications}.

Note that extending our result to give exactly $\Omega(L)$ without some term depending on $D$ is impossible. There are some situations (such as quantum state estimation) where forward access to a state preparation oracle can achieve more efficient query complexity~\cite{Chen25inversefree}.

We use our lifting theorem to recover recent results on PAC learning and state distinguishing as well as producing new lower bounds on state tomography and mixedness testing. 

See~\Cref{tab:intro-application-lower-bounds} for a table summarizing our lower bounds.

\begin{table}[htbp]
    \centering
    \small
    \setlength{\tabcolsep}{4pt}
    \renewcommand{\arraystretch}{1.3}
    \begin{tabular}{@{}p{0.18\linewidth}p{0.28\linewidth}p{0.34\linewidth}p{0.14\linewidth}@{}}
        \hline
        \raggedright Task & \raggedright Known copy/isometry\newline complexity & \raggedright Forward-query\newline lower bound & \raggedright References \tabularnewline
        \hline
        \raggedright Isometry tomography
        & \raggedright $\Theta(NM/\epsilon^2)$\newline Upper:~\cite{MB25tomography}\newline Lower:~\cite{CZY26tomography}
        & \raggedright $\Omega(\sqrt{M/N})$
        & \raggedright \Cref{cor:isometry-tomography-completion-lb} \tabularnewline
        \raggedright Isometry inversion
        & \raggedright $\Theta(N^2)$\newline Upper:~\cite{CMLZW24inversion,YSM25adjointation}\newline Lower:~\cite{CYZ26inversion}
        & \raggedright $\Omega(\min\{N^2,\sqrt{M/N}\})$
        & \raggedright \Cref{cor:isometry-inversion-completion-lb} \tabularnewline
        \raggedright Realizable PAC learning
        & \raggedright $\Theta\!\left(\frac{v+\log(1/\delta)}{\epsilon}\right)$\newline\cite{AdW18PAC}
        & \raggedright $\Omega\!\left(\min\left\{\frac{v+\log(1/\delta)}{\epsilon},\delta\sqrt M\right\}\right)$
        & \raggedright \Cref{cor:pac-preparation-lb};\newline\cite{IYM26PAC} \tabularnewline
        \raggedright Agnostic PAC learning
        & \raggedright $\Theta\!\left(\frac{v+\log(1/\delta)}{\epsilon^2}\right)$\newline\cite{AdW18PAC}
        & \raggedright $\Omega\!\left(\min\left\{\frac{v+\log(1/\delta)}{\epsilon^2},\delta\sqrt M\right\}\right)$
        & \raggedright \Cref{cor:pac-preparation-lb};\newline\cite{IYM26PAC} \tabularnewline
        \raggedright Pure-state distinguishing
        & \raggedright $\Theta(\epsilon^{-2})$\newline\cite{Watrous18TQI}
        & \raggedright $\Omega(\min\{\epsilon^{-2},\sqrt M\})$
        & \raggedright \Cref{cor:state-distinguishing-preparation-lb};\newline\cite{WCC26distinguishing} \tabularnewline
        \raggedright Rank-$r$ state tomography
        & \raggedright $\Theta(Nr/\epsilon^2)$\newline Upper:~\cite{PSTW25}\newline Lower:~\cite{SSW25tomography,CZY26tomography}
        & \raggedright $\Omega(\min\{Nr/\epsilon^2,\sqrt{ND}\})$
        & \raggedright \Cref{cor:state-tomography-preparation-lb} \tabularnewline
        \raggedright Mixedness testing
        & \raggedright $\Theta(N/\epsilon^2)$\newline\cite{OW21spectrum}
        & \raggedright $\Omega(\min\{N/\epsilon^2,\sqrt{ND}\})$
        & \raggedright \Cref{cor:mixedness-preparation-lb} \tabularnewline
        \hline
    \end{tabular}
    \caption{Known copy/isometry complexities and lower bounds for uncontrolled forward queries. The second column counts direct forward isometry queries for the first two rows and state copies for the remaining rows; each $\Theta$ entry includes matching upper and lower bounds. Isometry tomography uses diamond-norm error $\epsilon$ and assumes $M\geq2N$; inversion uses fixed constant diamond-norm error. The third column concerns unitary completions of isometries from dimension $N$ to $M$, or state preparation unitaries, respectively. For PAC learning and pure-state distinguishing, $M$ is the total preparation dimension, $v$ is the VC dimension, and $\delta$ is the learner's failure probability. For state tomography and mixedness testing, $N$ is the state dimension and $D\geq N$ is the purification dimension; for pure-state tomography, $r=1$ and any $D\geq1$ is allowed. State tomography uses trace-distance error $\epsilon$; distinguishing and mixedness testing use trace-distance separation $\epsilon$. Success probability is at least $2/3$, except for PAC learning where it is at least $1-\delta$. The last column also cites prior forward-query results recovered by our corollaries:~\cite{IYM26PAC} takes the worst case over ambient dimensions, and~\cite{WCC26distinguishing} assumes $M=\Omega(\epsilon^{-8})$. The referenced corollaries give the full parameter conditions.}
    \label{tab:intro-application-lower-bounds}
\end{table}

\paragraph{Impossibility of simulating dilations}

We give a concrete problem for which access to an arbitrary dilation of a quantum channel provides an exponential query advantage, even in the presence of error.

Several related simulation results and lower bounds were previously known. The random Stinespring superchannel constructions of~\cite{GMZFL25,YNUTM25} exactly simulate parallel queries to a Haar-random isometry dilation using the same number of parallel channel queries. There are also a number of partial lower bounds for sequential queries. In particular, efficient simulation of random dilations has been ruled out in the following settings
\begin{enumerate}
    \item Exact simulation of polynomial queries to a random dilation using a subexponential number of queries to the channel~\cite{YNUTM25} (using the quantum switch lower bound of~\cite{BKM+25}).
    \item {\emergencystretch=2em Constant-error approximate simulation of two queries to a random dilation using two channel queries~\cite{GdTQB26}.\par}
    \item {\emergencystretch=2em Exact simulation of three queries to a random dilation using finitely many channel queries~\cite{GdTQB26}.\par}
    \item Constant-error approximate simulation of a random dilation \textit{and its inverse} using a subexponential number of queries~\cite{AYTM26}.
\end{enumerate}

Note that all of these lower bounds are restricted among some practical dimension. Note that our resolution of~\Cref{conj:isometry-to-unitary} bypasses exact implementation lower bounds, so one may suspect that an approximate simulator for random dilations of a channel may likewise exist. Thus, even after all the above negative results, the primary open question remains unresolved. 
\begin{center}
    \textit{
    Can one approximately simulate a random dilation using only a polynomial number of queries to the channel? }
\end{center}

We answer this question in the negative. In particular, we give a pair of distributions over channels which can be perfectly distinguished with arbitrary dilation access, but which are indistinguishable given subexponential channel access.

\begin{theorem}[Informal version of~\Cref{thm:channel-dilation-separation}]
    For every $D\geq3$, there are two ensembles of channels on a $D$-dimensional space such that two adaptive forward queries to any fixed isometry dilation identify the ensemble perfectly, whereas every algorithm making at most $q$ ordinary forward channel queries has distinguishing advantage
    \[
        O\left(\frac{q^{3/2}}{\sqrt D}\right).
    \]
    Thus, constant distinguishing advantage requires $\Omega(D^{1/3})$ channel queries, even with arbitrary computation and reference registers between queries.
\end{theorem}

As a corollary, this rules out simulating two queries to a random dilation up to constant error using subexponential channel access. In addition, by applying a search-to-decision style argument, it rules out simulating polynomially many queries to a random dilation up to error arbitrarily close to $1$. See~\Cref{cor:no-channel-dilation-simulator,cor:channel-dilation-near-unit-error} for the formal statements.

\paragraph{The quantum switch}

The lower bound of~\cite{YNUTM25} works by reducing to implementing a superchannel known as the quantum switch~\cite{CDPV13,BKM+25}. For two channels $\Phi,\Psi$, the quantum switch $\Sw[\Phi,\Psi]$ is the quantum analogue of the program
\[
    \begin{array}{ll}
        \textbf{if } C=0: & \text{run }\Phi\text{ then }\Psi,\\
        \textbf{if } C=1: & \text{run }\Psi\text{ then }\Phi.
    \end{array}
\]
Here $\Reg C$ is a qubit whose state coherently controls the order of the two channels.

The quantum switch is important as an example of ``indefinite causal order''. That is, it cannot be implemented by a circuit making one query to the underlying channels $\Phi$ and $\Psi$. One of the fundamental questions of indefinite causal order is
\begin{center}
    \textit{Can transformations with indefinite causal order be efficiently simulated by algorithms making polynomially many queries to the underlying channels?}
\end{center}
A recent work of Bavaresco et al.~\cite{BKM+25} has given partial progress on this question in the case of the quantum switch. They show that any algorithm making one query to $\Phi$ and a subexponential number of queries to $\Psi$ cannot implement the quantum switch exactly. Two queries to any fixed isometry dilation of $\Psi$, together with one ordinary query to $\Phi$, suffice to implement the switch (Figure~8 of~\cite{YNUTM25}), yielding the lower bound for exactly simulating random dilations of~\cite{YNUTM25}.

The overall question of the complexity advantage of indefinite causal order thus has a potential resolution if a stronger lower bound on implementing the quantum switch is discovered. This is stated precisely in~\cite{BKM+25} as follows.

\begin{conjecture}[Quantum switch simulation, Conjecture~1 of~\cite{BKM+25}]
    Any quantum circuit with fixed or classically controlled causal order that exactly and deterministically simulates the quantum switch for all pairs of $n$-qubit channels, using $k_\Phi$ queries to one channel and $k_\Psi$ to the other, requires
    \[
        \max\{k_\Phi,k_\Psi\}=\Omega(2^n).
    \]
\end{conjecture}

As techniques for ruling out simulating random dilations seem closely tied to lower bounds of the quantum switch, one might hope that our lower bound also translates. In fact, it does. We establish the conjecture's exponential lower-bound conclusion with a cubic loss in dimension. Furthermore, our lower bound extends beyond exact implementations to algorithms with constant simulation error below $1/2$.

\begin{theorem}[Informal version of~\Cref{cor:quantum-switch-query-lower-bound}]
    For every fixed $\epsilon<1/2$, any quantum circuit with fixed or classically controlled causal order that simulates one quantum switch on arbitrary $n$-qubit channels, with trace-distance error at most $\epsilon$ on every input, requires
    \[
        \Omega(2^{n/3})
    \]
    total forward channel queries in the worst case. This holds even when the two channels supplied to the switch are promised to be identical, and with arbitrary computation and reference registers between queries.
\end{theorem}

The error threshold $\frac{1}{2}$ is tight. The quantum switch can be simulated with error $\frac{1}{2}$ with one query to each channel. This is done by measuring the control register and then classically controlling the order the channels are queried. For full details, see~\Cref{app:switch-half-error}.

\section{Technical Overview}

\subsection{Simulating random completions of isometries}

To get a sense of what a random completion of an isometry looks like, it is helpful to write down an operational description. In particular, let $V$ be an arbitrary isometry mapping $\Reg{A}\to \Reg{A}\otimes \Reg{B}$. Let $D^V$ be an arbitrary completion of $V$. Then $D^V$ is a unitary on $\Reg{A}\otimes \Reg{B}$ such that
$$D^V \ket{x}_{\Reg{A}}\ket{0}_{\Reg{B}} = V_{\Reg{A}\to \Reg{AB}}\ket{x}_{\Reg{A}}$$
A random completion is then a unitary which acts as $D^V$ on states of the form $\ket{x}\ket{0}$ and acts ``randomly'' on the orthogonal subspace. Define $\Pi_0$ to be the projector onto the space spanned by states $\ket{x}\ket{0}$. Define $\Pi_V = VV^\dagger$ to be the projector onto the image of $V$. Define $\Pi_0^\perp = I - \Pi_0$ and $\Pi_V^\perp = I - \Pi_V$ to be the projectors onto their orthogonal complements. We reserve overlines for complex conjugation. Seen in this light, a random completion should randomly map $\Ima(\Pi_0^\perp)$ to $\Ima(\Pi_V^\perp)$.

Let $\wt{U}$ be a Haar random unitary mapping $\Ima(\Pi_0^\perp)$ to $\Ima(\Pi_V^\perp)$. Then a random completion of $V$ is the distribution over unitaries
$$D^V_{\wt{U}} \coloneqq D^V \Pi_0 + \wt{U} \Pi_0^\perp$$

\paragraph{Path recording} There is a generic technique for statefully simulating queries to a Haar random unitary: the path recording method~\cite{STOC:MaHua25,GY25,FLMNW26}. The path recording method gives a stateful simulator $PR_{[N]}$ such that query access to $PR_{[N]}$ is indistinguishable from applying a Haar random unitary $U$ acting on the Hilbert space spanned by $\ket{x}$ for $x\in [N]$. 

One may then hope to give the very simple exact simulator for a random completion $D_V^{\wt{U}}$ of $V$: replace $\wt{U}$ with a path recording simulator $\wt{PR}$ mapping $\Ima(\Pi_0^\perp)$ to $\Ima(\Pi_V^\perp)$. This leads to the corresponding simulator
$$D^V\Pi_0 + \wt{PR} \Pi_0^\perp$$
In fact, given the ability to implement $\wt{PR}$, this map is efficiently implementable using one controlled query to $V$. We will define an idealized, exact (but not implementable) simulator $\wt{\Sim}^V$ as follows:
\begin{enumerate}
    \item Coherently measure $\{\Pi_0,\Pi_0^\perp\}$, XORing the result onto the ancilla $\Reg{C}$.
    \item Controlled on $\Reg{C}$ containing $1$, apply $V$ to $\Reg{A}$ and replace $\Reg{A}\Reg{B}$ with the output.
    \item Controlled on $\Reg{C}$ containing $0$, apply $\wt{PR}$ to $\Reg{A}\Reg{B}$.
    \item Coherently measure $\{I - \wt{PR} \wt{PR}^\dagger, \wt{PR} \wt{PR}^\dagger\}$, XORing the result onto the ancilla $\Reg{C}$.
    \item Trace out register $\Reg{C}$ and output $\Reg{AB}$.
\end{enumerate}
Here, $\wt{PR}\wt{PR}^\dagger$ is the projector checking if the state in $\Reg{AB}$ was previously produced as an output of $\wt{PR}$. Since the image of $\wt{PR}$ is orthogonal to the image of $V$, applying the measurement in step 4 perfectly resets $\Reg{C}$ to $\ket{0}$. As such, this circuit exactly simulates query access to $D_V^{\wt{U}}$ for a random $\wt{U}$. 

We remark that the action of $\wt{PR}\wt{PR}^\dagger$ \textit{does not} approximate $\wt{U}\wt{U}^\dagger$. While $\wt{PR}$ simulates $\wt{U}$ on forward query access, the same guarantee does not apply to $\wt{PR}^\dagger$. Nevertheless, applying the measurement in step 4 suffices to erase the ancilla $\Reg{C}$.

\paragraph{Making the simulator efficient} The only problem with this simulator is that since $\Ima(\Pi_V^\perp) = \Ima(\wt{PR})$ can be arbitrary, it is not possible to efficiently implement $\wt{PR}$ using only query access to $V$. Our solution for this is straightforward: replace $\wt{PR}$ with a path recording oracle $PR$ acting as a Haar random unitary on the whole space $\Reg{AB}\to\Reg{AB}$. Since $PR$ acts on the entire space, implementing it does not require any information about the image of the isometry and so can be done efficiently.

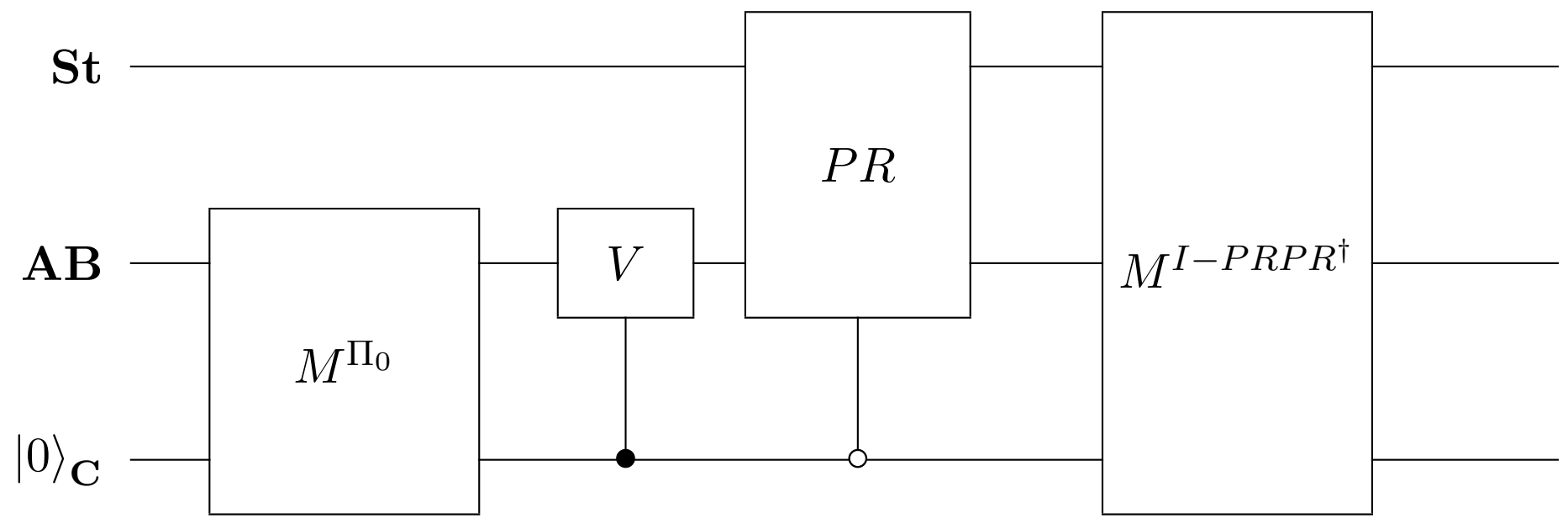
\begin{figure}[htbp]
    \centering
    \begingroup
    \setlength{\unitlength}{1pt}
    \begin{picture}(388,140)
        \put(50,116){\makebox(0,0)[r]{$\Reg{St}$}}
        \put(50,72){\makebox(0,0)[r]{$\Reg{AB}$}}
        \put(50,28){\makebox(0,0)[r]{$\ket{0}_{\Reg{C}}$}}
        \put(56,116){\line(1,0){138}}
        \put(244,116){\line(1,0){30}}
        \put(334,116){\line(1,0){42}}
        \put(56,72){\line(1,0){18}}
        \put(134,72){\line(1,0){18}}
        \put(182,72){\line(1,0){12}}
        \put(244,72){\line(1,0){30}}
        \put(334,72){\line(1,0){42}}
        \put(56,28){\line(1,0){18}}
        \put(134,28){\line(1,0){83}}
        \put(221,28){\line(1,0){53}}
        \put(334,28){\line(1,0){42}}
        \put(74,16){\framebox(60,68){$M^{\Pi_0}$}}
        \put(152,60){\framebox(30,24){$V$}}
        \put(167,28){\line(0,1){32}}
        \put(167,28){\makebox(0,0){$\bullet$}}
        \put(194,60){\framebox(50,68){$PR$}}
        \put(219,30){\line(0,1){30}}
        \put(219,28){\makebox(0,0){$\circ$}}
        \put(274,16){\framebox(60,112){$M^{I-PRPR^\dagger}$}}
    \end{picture}
    \endgroup
    \caption{The simulator using path recording on the entire space $\Reg{AB}$. The $PR$ gate acts jointly on $\Reg{AB}$ and the database register $\Reg{St}$, which is initially empty and retained between simulator invocations. The first block $M^{\Pi_0}$ coherently XORs membership in $\Ima(\Pi_0)$ into $\Reg{C}$, and the filled and open controls select $C=1$ and $C=0$, respectively. The $V$ gate replaces the input $\Reg{AB}$ with $V$ applied to $\Reg{A}$. The final block denotes the coherent measurement $\{I-PRPR^\dagger,PRPR^\dagger\}$ on $\Reg{AB}$ and $\Reg{St}$, XORing the first outcome's indicator into $\Reg{C}$. The ancilla $\Reg{C}$ is discarded after the invocation.}
    \label{fig:path-recording-dilation-circuit}
\end{figure}

As such, our final simulator $\Sim^V$ (using controlled access to $V$) will be as follows
\begin{enumerate}
    \item Coherently measure $\{\Pi_0,\Pi_0^\perp\}$, XORing the result onto the ancilla $\Reg{C}$.
    \item Controlled on $\Reg{C}$ containing $1$, apply $V$ to $\Reg{A}$ and replace $\Reg{A}\Reg{B}$ with the output.
    \item Controlled on $\Reg{C}$ containing $0$, apply $PR$ to $\Reg{A}\Reg{B}$.
    \item Coherently measure $\{I - PR PR^\dagger, PR PR^\dagger\}$, XORing the result onto the ancilla $\Reg{C}$.
    \item Trace out register $\Reg{C}$ and output $\Reg{AB}$.
\end{enumerate}
This is visualized in~\Cref{fig:path-recording-dilation-circuit}.

Note that our final simulator is not an exact simulator. However, as long as $\Reg{B}$ is sufficiently large, query access to $\Sim^V$ will be indistinguishable from query access to the exact simulator $\wt{\Sim}^V$. In particular, we will show that on some sufficient subspace, these two maps are close in operator norm.

This last step will be done by reasoning carefully about the specific implementation of $PR$ and $\wt{PR}$. In particular, we will show that on some invariant subspace of $\wt{\Sim}^V$, $\Sim^V$ and $\wt{\Sim}^V$ are close in operator norm. For the full details, see the main body of the paper. 

To get a sense of why this should be true, notice that in both games, $PR$ and $\wt{PR}$ will always be called on states in $\Ima(\Pi_0^\perp)$. On such states, $\wt{PR}$ acts as a Haar random unitary $\Ima(\Pi_0^\perp)\to \Ima(\Pi_V^\perp)$ and so will output a random state in $\Ima(\Pi_V^\perp)$. $PR$ will act as a Haar random unitary $\Reg{AB}\to \Reg{AB}$ and so will output a completely random state on registers $\Reg{AB}$. Say that $\Reg{A}$ consists of $n$ qubits and $\Reg{B}$ consists of $m$ qubits. Then the rank of $\Pi_V^\perp$ is the same as the rank of $\Pi_0^\perp$, which is $2^{m+n} - 2^n$. As such, a random state over $\Reg{AB}$ will have in expectation a
$$\frac{2^{m+n}-2^n}{2^{m+n}} = 1 - \frac{1}{2^m}$$ fraction of its weight on $\Ima(\Pi_V^\perp)$. Similarly, it will turn out that $PR$ and $\wt{PR}$ are approximately
$$\sqrt{\frac{1}{2^m}}$$
close in operator norm when restricted to the appropriate subspace (with some minor loss due to the number of queries).

\paragraph{Removing the control}

Note that $\Sim^V$ as defined uses controlled queries to $V$. We would like to construct a simulator which emulates a random completion of $V$ using only forward queries. Towards this, a different acorn trick seems like it would be useful. It is known that controlled queries to a unitary with a random global phase can be simulated by forward queries to the unitary.

\begin{theorem}[Informal version, Theorem~1.1 of~\cite{TW26}]
    Every algorithm making $T$ controlled forward queries to an unknown unitary $U$ can be simulated using $T$ ordinary forward queries to $U$, with a polynomial overhead in time and space. The simulated output is the original algorithm's output with $U$ replaced by $e^{i\theta}U$, averaged over a uniformly random $\theta\in[0,2\pi)$. The same phase $\theta$ is used for all queries.
\end{theorem}

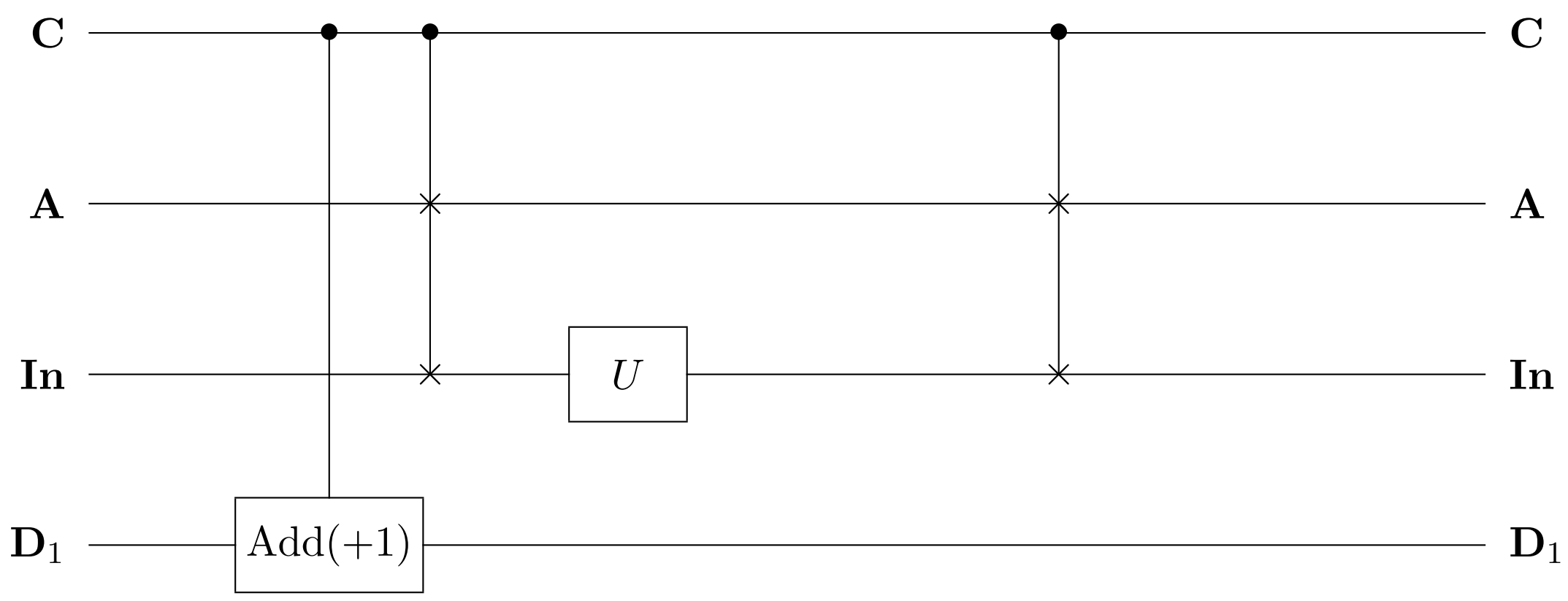
\begin{figure}[htbp]
    \centering
    \begingroup
    \setlength{\unitlength}{1pt}
    \begin{picture}(440,188)
        \put(46,160){\makebox(0,0)[r]{$\Reg{C}$}}
        \put(46,116){\makebox(0,0)[r]{$\Reg{A}$}}
        \put(46,72){\makebox(0,0)[r]{$\Reg{In}$}}
        \put(46,28){\makebox(0,0)[r]{$\Reg{D}_1$}}
        \put(52,160){\line(1,0){360}}
        \put(52,116){\line(1,0){360}}
        \put(52,72){\line(1,0){124}}
        \put(206,72){\line(1,0){206}}
        \put(52,28){\line(1,0){38}}
        \put(138,28){\line(1,0){274}}
        \put(90,16){\framebox(48,24){$\mathrm{Add}(+1)$}}
        \put(114,40){\line(0,1){120}}
        \put(114,160){\makebox(0,0){$\bullet$}}
        \put(140,72){\line(0,1){88}}
        \put(140,160){\makebox(0,0){$\bullet$}}
        \put(140,116){\makebox(0,0){$\times$}}
        \put(140,72){\makebox(0,0){$\times$}}
        \put(176,60){\framebox(30,24){$U$}}
        \put(302,72){\line(0,1){88}}
        \put(302,160){\makebox(0,0){$\bullet$}}
        \put(302,116){\makebox(0,0){$\times$}}
        \put(302,72){\makebox(0,0){$\times$}}
        \put(418,160){\makebox(0,0)[l]{$\Reg{C}$}}
        \put(418,116){\makebox(0,0)[l]{$\Reg{A}$}}
        \put(418,72){\makebox(0,0)[l]{$\Reg{In}$}}
        \put(418,28){\makebox(0,0)[l]{$\Reg{D}_1$}}
    \end{picture}
    \endgroup
    \caption{The forward-query simulator of~\cite{TW26}. Filled controls select $C=1$. The workspace $\Reg{In}$ has the same size as $\Reg{A}$ and is initially $\ket{0}$; the counter $\Reg{D}_1$ is initially zero. When $C=1$, the swaps route the input through the unconditional $U$ query and increment $\Reg{D}_1$. When $C=0$, $U$ instead acts on $\Reg{In}$. Both $\Reg{In}$ and $\Reg{D}_1$ are retained between invocations and traced out at the end of the algorithm.}
    \label{fig:decontrol-unitary}
\end{figure}

The corresponding simulator is described in~\Cref{fig:decontrol-unitary}. Unfortunately, the above theorem only applies to unitaries, and the simulator does not syntactically work for isometries. However, we nevertheless adapt the theorem to the isometry setting.

\begin{theorem}[Informal version,~\Cref{thm:simctrl}]\label{thm:techoverview-simctrl}
    Every algorithm making $T$ controlled forward queries to an unknown isometry $V$ can be simulated using $T$ ordinary forward queries to $V$, with a polynomial overhead in time and space. The simulated output is the original algorithm's output with $V$ replaced by $e^{i\theta}V$, averaged over a uniformly random $\theta\in[0,2\pi)$. The same phase $\theta$ is used for all queries.
\end{theorem}

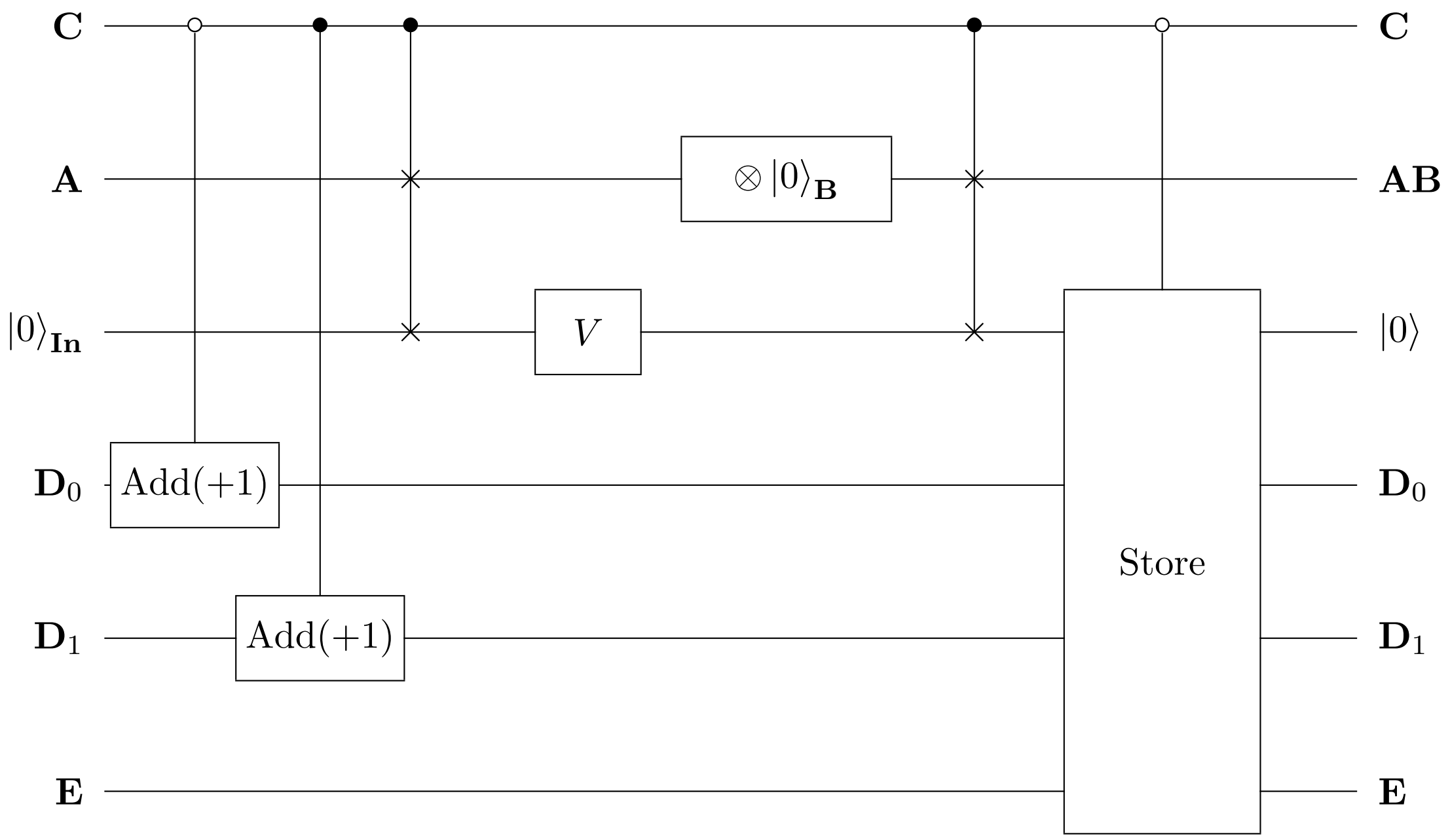
\begin{figure}[htbp]
    \centering
    \begingroup
    \setlength{\unitlength}{1pt}
    \begin{picture}(440,276)
        \put(46,248){\makebox(0,0)[r]{$\Reg{C}$}}
        \put(46,204){\makebox(0,0)[r]{$\Reg{A}$}}
        \put(46,160){\makebox(0,0)[r]{$\ket{0}_{\Reg{In}}$}}
        \put(46,116){\makebox(0,0)[r]{$\Reg{D}_0$}}
        \put(46,72){\makebox(0,0)[r]{$\Reg{D}_1$}}
        \put(46,28){\makebox(0,0)[r]{$\Reg{E}$}}
        \put(52,248){\line(1,0){24}}
        \put(80,248){\line(1,0){274}}
        \put(358,248){\line(1,0){54}}
        \put(52,204){\line(1,0){166}}
        \put(278,204){\line(1,0){134}}
        \put(52,160){\line(1,0){124}}
        \put(206,160){\line(1,0){122}}
        \put(384,160){\line(1,0){28}}
        \put(52,116){\line(1,0){2}}
        \put(102,116){\line(1,0){226}}
        \put(384,116){\line(1,0){28}}
        \put(52,72){\line(1,0){38}}
        \put(138,72){\line(1,0){190}}
        \put(384,72){\line(1,0){28}}
        \put(52,28){\line(1,0){276}}
        \put(384,28){\line(1,0){28}}
        \put(54,104){\framebox(48,24){$\mathrm{Add}(+1)$}}
        \put(78,128){\line(0,1){118}}
        \put(78,248){\makebox(0,0){$\circ$}}
        \put(90,60){\framebox(48,24){$\mathrm{Add}(+1)$}}
        \put(114,84){\line(0,1){164}}
        \put(114,248){\makebox(0,0){$\bullet$}}
        \put(140,160){\line(0,1){88}}
        \put(140,248){\makebox(0,0){$\bullet$}}
        \put(140,204){\makebox(0,0){$\times$}}
        \put(140,160){\makebox(0,0){$\times$}}
        \put(176,148){\framebox(30,24){$V$}}
        \put(218,192){\framebox(60,24){$\otimes\ket{0}_{\Reg{B}}$}}
        \put(302,160){\line(0,1){88}}
        \put(302,248){\makebox(0,0){$\bullet$}}
        \put(302,204){\makebox(0,0){$\times$}}
        \put(302,160){\makebox(0,0){$\times$}}
        \put(328,16){\framebox(56,156){$\mathrm{Store}$}}
        \put(356,172){\line(0,1){74}}
        \put(356,248){\makebox(0,0){$\circ$}}
        \put(418,248){\makebox(0,0)[l]{$\Reg{C}$}}
        \put(418,204){\makebox(0,0)[l]{$\Reg{AB}$}}
        \put(418,160){\makebox(0,0)[l]{$\ket{0}$}}
        \put(418,116){\makebox(0,0)[l]{$\Reg{D}_0$}}
        \put(418,72){\makebox(0,0)[l]{$\Reg{D}_1$}}
        \put(418,28){\makebox(0,0)[l]{$\Reg{E}$}}
    \end{picture}
    \endgroup
    \caption{Simulating controlled access to an isometry using one ordinary forward query. Filled controls select $C=1$ and open controls select $C=0$. The query wire changes from $\Reg{In}$ to $\Reg{Out}$ at the $V$ gate. Here $\Reg{E}=(\Reg{E}_1,\dots,\Reg{E}_T)$. The two controlled $\mathrm{Add}(+1)$ gates increment $\Reg{D}_0$ when $C=0$ and $\Reg{D}_1$ when $C=1$, respectively. The Store block swaps $\Reg{Out}$ with $\Reg{E}_i$ for $i=D_0$. The counters and stored dummy outputs are retained between invocations and traced out at the end of the algorithm.}
    \label{fig:decontrol-isometry}
\end{figure}

Our simulator for controlled isometry queries uses one ordinary forward query to $V$ per invocation. We retain two counters $\Reg{D}_0,\Reg{D}_1$, initially zero, and $T$ initially zero storage registers $\Reg{E}_1,\dots,\Reg{E}_T$, each the size of $\Reg{AB}$. Each invocation proceeds as follows:
\begin{enumerate}
    \item Coherently increment $\Reg{D}_0$ if $C=0$ and $\Reg{D}_1$ if $C=1$.
    \item Initialize a fresh register $\Reg{In}$ to $\ket{0}$, and swap it with $\Reg{A}$ if $C=1$.
    \item Apply $V$ unconditionally to $\Reg{In}$, producing a register $\Reg{Out}$ of the size of $\Reg{AB}$.
    \item Append $\ket{0}_{\Reg{B}}$ to $\Reg{A}$, and swap $\Reg{AB}$ with $\Reg{Out}$ if $C=1$.
    \item If $C=0$, swap $\Reg{Out}$ with the storage register $\Reg{E}_i$ indexed by the new value $i$ of $\Reg{D}_0$. 
    \item Output $\Reg{CAB}$ and retain the counters and storage registers for subsequent invocations.
\end{enumerate}
This simulator is illustrated in~\Cref{fig:decontrol-isometry}. When $C=1$, the query acts on the actual input and its output is returned on $\Reg{AB}$. When $C=0$, the input is preserved and the dummy output $V\ket{0}$ is stored in the next unused register $\Reg{E}_i$. In either case, the remaining workspace is reset to zero.

The only differences between our simulator and that of~\cite{TW26} are the addition of a second counter as well as the storage map in step 5 moving register $\Reg{Out}$ into register $\Reg{E}_i$. Intuitively, the simulator of~\cite{TW26} works since at the end of computation, the counter register will be enough to describe the other internal register $\Reg{In}$. In particular, the final internal state will be supported on states of the form
$$U^{T-t}\ket{0}_{\Reg{In}} \ket{t}_{\Reg{D}_1}$$
and so $\Reg{In}$ can be recovered from $\Reg{D}_1$ by applying $U^{T-t}$.

Similarly, our final internal state will be supported on states of the form
$$\left(\bigotimes_{i=1}^{t_0}(V\ket0)_{\Reg{E}_i}\right)\otimes\left(\bigotimes_{i=t_0+1}^{T}\ket0_{\Reg{E}_i}\right)\ket{t_0}_{\Reg{D}_0}\ket{t_1}_{\Reg{D}_1}$$
and so $\Reg{E}_1,\dots,\Reg{E}_T$ can be recovered from $\Reg{D}_0,\Reg{D}_1$ by preparing $V\ket0$ in the first $t_0$ storage registers. As such, similar proof techniques apply.

\paragraph{Putting everything together}
Since forward queries to a random completion of an isometry are invariant up to a global phase, we can think of the random phase of $\wt{U}$ as being applied to $V$. Thus, composing~\Cref{thm:techoverview-simctrl} with our simulator making controlled queries to $V$, we get a simulator for a random completion of an isometry making only forward queries to the isometry $V$.

\subsection{Impossibility of simulating random dilations of arbitrary channels}

Consider $\Phi$ any arbitrary quantum channel mapping a register $\Reg{A}\to \Reg{B}$. For simplicity, we will only consider the case $\Reg{B}=\Reg{A}$. Let $V^{\Phi}$ be an isometry dilation of $\Phi$. We want to find specific channels for which access to $V^{\Phi}$ is more powerful than direct access to $\Phi$, even when the exact structure of the dilation is unknown.

{\emergencystretch=2em \cite{GdTQB26} observes that query access to $V^{\Phi}$ allows one to detect correlations between purified environment registers. For intuition, suppose that $V^{\Phi}$ acts as follows for some states $\ket{\phi_0},\ket{\phi_1},\ket{e_0},\ket{e_1}$.\par}
$$V^\Phi \ket{0}_{\Reg{A}} = \ket{\phi_0}_{\Reg{A}}\ket{e_0}_{\Reg{E}}$$
$$V^\Phi \ket{\phi_0}_{\Reg{A}} = \ket{\phi_1}_{\Reg{A}}\ket{e_1}_{\Reg{E}}$$
Then applying $V^\Phi$ twice leads to the state 
$$\ket{\phi_1}_{\Reg{A}}\ket{e_0}_{\Reg{E}_0}\ket{e_1}_{\Reg{E}_1}$$
whereas applying $\Phi$ twice loses the environment registers. Note that any isometry dilation of $\Phi$ (of the same dimension) will be the same as $V^\Phi$ up to a unitary applied to the dilation registers. As such, applying any dilation of $\Phi$ to $\ket{0}$ twice will result in the state
$$\ket{\phi_1}_{\Reg{A}}\otimes (U\ket{e_0}_{\Reg{E}_0})\otimes (U\ket{e_1}_{\Reg{E}_1})$$
for some unitary $U$.

While it is impossible to tell from such a state any information about $\ket{e_0}$ and $\ket{e_1}$ specifically, it is possible to detect correlations between them. As an example, using the swap test, one can tell (with advantage $1/2$) whether $\ket{e_0}$ and $\ket{e_1}$ are the \textit{same} or \textit{orthogonal}.

In general, given an isometry dilation of a channel, one can exactly tell whether the purification states after two queries lie in the symmetric subspace or the antisymmetric subspace. Here, the symmetric subspace is the subspace of states preserved under permutations. The antisymmetric subspace is the subspace of states preserved under even permutations and whose sign is flipped under odd permutations.

\cite{GdTQB26} uses this observation to construct two distributions over quantum channels which can be distinguished more easily with two isometry dilation queries than with two direct channel queries. We give a different pair of distributions over channels which can be distinguished perfectly with two isometry dilation queries but are indistinguishable with two direct channel queries. We will then show how to modify our channels to achieve a lower bound even against exponentially many channel queries.

\paragraph{The two query lower bound}

Our initial goal will be to define two channels $\SAME_0$ and $\DIFF_0$ for which the purification of $\SAME_0$ applied twice will lead to the purification registers being in the symmetric subspace, and for $\DIFF_0$ the analogous process should lead to the antisymmetric subspace. Furthermore, $\SAME_0$ and $\DIFF_0$ should be indistinguishable to algorithms making only two channel queries.

Since these properties are invariant to the underlying dilation, we will define $\SAME_0$ and $\DIFF_0$ by defining their corresponding dilation isometries $V_{0,\SAME},V_{0,\DIFF}$.

The randomness of the channels will consist of two orthogonal Haar random states $\ket{\phi_0},\ket{\phi_1}$ over a high dimensional space. We will also set both states orthogonal to $\ket{0}$. Then, $V_{0,\SAME}$ will act as follows.
$$V_{0,\SAME} \ket{0}_{\Reg{A}} = \frac{1}{\sqrt{2}}\ket{\phi_0}_{\Reg{A}}\ket{0}_{\Reg{E}} + \frac{1}{\sqrt{2}}\ket{\phi_1}_{\Reg{A}}\ket{1}_{\Reg{E}}$$
$$V_{0,\SAME} \ket{\phi_b}_{\Reg{A}} = \ket{0}_{\Reg{A}}\ket{b\xor 1}_{\Reg{E}}$$
and for all orthogonal states $\ket{\psi}$,
$$V_{0,\SAME} \ket{\psi}_{\Reg{A}} = \ket{0}_{\Reg{A}}\ket{\psi,\bot}_{\Reg E}$$
where the state $\ket{\psi,\bot}$ is some state orthogonal to $\ket{0},\ket{1}$.\footnote{Technically, here we mean that the map $\ket{\psi}\mapsto \ket{\psi,\bot}_{\Reg{E}}$ is an isometry mapping to a subspace of $\Reg{E}$ orthogonal to $\ket{0},\ket{1}$.} 

$V_{0,\DIFF}$ will be the same as $V_{0,\SAME}$, except it will add a negative phase to $\ket{\phi_1}$. Formally,
$$V_{0,\DIFF} \ket{0}_{\Reg{A}} = \frac{1}{\sqrt{2}}\ket{\phi_0}_{\Reg{A}}\ket{0}_{\Reg{E}} + \frac{1}{\sqrt{2}}\ket{\phi_1}_{\Reg{A}}\ket{1}_{\Reg{E}}$$
$$V_{0,\DIFF} \ket{\phi_b}_{\Reg{A}} = (-1)^b\ket{0}_{\Reg{A}}\ket{b\xor 1}_{\Reg{E}}$$
and for all orthogonal states $\ket{\psi}$,
$$V_{0,\DIFF} \ket{\psi}_{\Reg{A}} = \ket{0}_{\Reg{A}}\ket{\psi,\bot}_{\Reg E}$$

Now, we can see that
\[
    \begin{aligned}
        V_{0,\SAME}V_{0,\SAME}\ket{0} &= \frac{1}{\sqrt{2}}\ket{0}_{\Reg{A}}\ket{0}_{\Reg{E}_1}\ket{1}_{\Reg{E_2}} + \frac{1}{\sqrt{2}}\ket{0}_{\Reg{A}}\ket{1}_{\Reg{E}_1}\ket{0}_{\Reg{E_2}}\\
        &= \ket{0}_{\Reg{A}} \otimes \left(\frac{1}{\sqrt{2}}\ket{0}_{\Reg{E}_1}\ket{1}_{\Reg{E}_2}+\frac{1}{\sqrt{2}}\ket{1}_{\Reg{E}_1}\ket{0}_{\Reg{E}_2}\right)
    \end{aligned}
\]
and so the state on $\Reg{E}_1\Reg{E}_2$ is a state in the symmetric subspace.

Similarly,$$V_{0,\DIFF}V_{0,\DIFF}\ket{0} = \ket{0}_{\Reg{A}} \otimes \left(\frac{1}{\sqrt{2}}\ket{0}_{\Reg{E}_1}\ket{1}_{\Reg{E}_2}-\frac{1}{\sqrt{2}}\ket{1}_{\Reg{E}_1}\ket{0}_{\Reg{E}_2}\right)$$
and so the state in $\Reg{E}_1\Reg{E}_2$ is a state in the antisymmetric subspace.

Thus, two queries to any isometry dilation of $\SAME_0,\DIFF_0$ can distinguish the two with probability $1$.

It then remains to show that these cannot be distinguished with two channel queries. However, since $\ket{\phi_0},\ket{\phi_1}$ are Haar random, any channel query algorithm cannot find these without first making a query. As such, the only useful first query an algorithm can make is querying $\SAME_0$ or $\DIFF_0$ on $\ket{0}_{\Reg{A}}$. Thus, distinguishing $\SAME_0$ and $\DIFF_0$ with two channel queries reduces to distinguishing $\SAME_0$ and $\DIFF_0$ with one channel query when given one of $\ket{\phi_0}$ or $\ket{\phi_1}$ at random. In this setting, it is impossible to tell if a relative phase has been added to $\ket{\phi_1}$ in the initial input, and so the average action of $\SAME_0$ and $\DIFF_0$ is (nearly) identical.

\paragraph{Extending to exponential queries}

Unfortunately, these channels can be distinguished using polynomially many channel queries. An adversary can query $\ket{0}$ multiple times to learn sufficient information about $\ket{\phi_0}$ and $\ket{\phi_1}$ to distinguish the channels.

To extend to an exponential lower bound, we simply increase the number of Haar random states. In particular, for each $x\in[L]$, independently sample an orthonormal Haar random pair $\ket{\phi_{x,0}},\ket{\phi_{x,1}}$. As before, we define $\SAME$ and $\DIFF$ by their dilation isometries. Define $V_{\SAME}$ by
\[
    \begin{aligned}
        V_{\SAME}\ket0
            &=\frac1{\sqrt{2L}}\sum_{x\in[L],\,b\in\{0,1\}}
                \ket{x,\phi_{x,b}}\ket{x,b}_{\Reg E},\\
        V_{\SAME}\ket{x,\phi_{x,b}}
            &=\ket0\ket{x,b\xor1}_{\Reg E},\\
        V_{\SAME}\ket\psi
            &=\ket0\ket{\psi,\bot}_{\Reg E}
    \end{aligned}
\]
for $\ket\psi$ in the remaining input subspace. Define $V_{\DIFF}$ by
\[
    \begin{aligned}
        V_{\DIFF}\ket0
            &=\frac1{\sqrt{2L}}\sum_{x\in[L],\,b\in\{0,1\}}
                \ket{x,\phi_{x,b}}\ket{x,b}_{\Reg E},\\
        V_{\DIFF}\ket{x,\phi_{x,b}}
            &=(-1)^b\ket0\ket{x,b\xor1}_{\Reg E},\\
        V_{\DIFF}\ket\psi
            &=\ket0\ket{\psi,\bot}_{\Reg E}.
    \end{aligned}
\]
Tracing out the environment defines $\SAME$ and $\DIFF$, respectively.

Applying each dilation twice to $\ket0$ gives
\[
    \begin{aligned}
        V_{\SAME}V_{\SAME}\ket0
            &=\frac{1}{\sqrt{2L}}\ket{0}\otimes \sum_{x\in[L]}
                \left(\ket{x,0}_{\Reg{E}_1}\ket{x,1}_{\Reg{E}_2}
                    +\ket{x,1}_{\Reg{E}_1}\ket{x,0}_{\Reg{E}_2}\right),\\
        V_{\DIFF}V_{\DIFF}\ket0
            &=\frac{1}{\sqrt{2L}}\ket{0}\otimes \sum_{x\in[L]}
                \left(\ket{x,0}_{\Reg{E}_1}\ket{x,1}_{\Reg{E}_2}
                    -\ket{x,1}_{\Reg{E}_1}\ket{x,0}_{\Reg{E}_2}\right).
    \end{aligned}
\]
Thus, the environment registers are again in the symmetric subspace for $\SAME$ and the antisymmetric subspace for $\DIFF$, so the same two-query dilation distinguisher works.

Now, channel queries to $\ket{0}$ will sample $x,b$ at random and return $\ket{x,\phi_{x,b}}$. As such, with high probability the distinguisher will never again make a query on $\ket{0}$ sampling the same $x$. Thus, for similar reasons as above the channels $\SAME$ and $\DIFF$ are indistinguishable.

Note that this is just an intuitive explanation for security. The formal argument will be slightly more careful and will thus save on parameters. However, the key insight indeed boils down to the fact that each query on $\ket{0}$ will with high probability return $\ket{x,\phi_{x,b}}$ for a new value $x$.

\paragraph{The quantum switch}

The key idea behind our lower bound is that query access to an isometry dilation allows one to test whether the purification registers are in the symmetric or antisymmetric subspace. It turns out that the ability to implement the quantum switch also provides this functionality. In particular, let $\Phi$ be any quantum channel such that the purification registers after two queries to a dilation of $\Phi$ on $\ket{0}$ lie in the symmetric subspace. Then
$$\Tr_{\Reg{A}}\left(\Sw[\Phi,\Phi](\ketbra{+}_{\Reg{C}}\otimes \ketbra{0}_{\Reg{A}})\right) = \ketbra{+}_{\Reg{C}}$$
Similarly, if the purification registers after two queries to a dilation of $\Phi$ on $\ket{0}$ lie in the antisymmetric subspace, then
$$\Tr_{\Reg{A}}\left(\Sw[\Phi,\Phi](\ketbra{+}_{\Reg{C}}\otimes \ketbra{0}_{\Reg{A}})\right) = \ketbra{-}_{\Reg{C}}$$

As such, one application of $\Sw[\SAME,\SAME]$ or $\Sw[\DIFF,\DIFF]$ suffices to perfectly distinguish $\SAME$ and $\DIFF$. As the same lower bound applies, we get that one application of the quantum switch can solve a task that requires exponentially many channel queries. Since the lower bound for channel queries limits the success probability by $\frac{1}{2} + \negl(n)$, this gives a lower bound on implementing the quantum switch up to error $\frac{1}{2} - \delta$ for any constant $\delta$.

\section{Preliminaries}\label{sec:preliminaries}
\subsection{Quantum Channels and Oracle Algorithms}

We use $[N]=\{0,\ldots,N-1\}$ and write $\TD(\rho,\sigma)=\frac12\|\rho-\sigma\|_1$ for trace distance.

\begin{definition}
    Let $\Reg{A},\Reg{B}$ be arbitrary quantum registers, and let $V:\Reg{A}\to \Reg{B}$ be an arbitrary linear map. Define the operator norm of $V$ to be
    $$\norm{V}_{op} = \sup_{\norm{\ket{\phi}}=1}\norm{V\ket{\phi}}$$
\end{definition}

\begin{definition}
    Let $\Reg{In},\Reg{Out},\Reg{St},\Reg{A}$ be arbitrary quantum registers.

    An oracle $\O$ mapping $\Reg{In}$ to $\Reg{Out}$ is a CPTP map
    $$\O:\mathcal{L}(\Reg{In}\otimes \Reg{St}) \to \mathcal{L}(\Reg{Out}\otimes \Reg{St})$$
     along with an initial state
    $\ket{\phi_0}_{\Reg{St}}$ in the register $\Reg{St}$.
    
    An oracle algorithm $\A^{(\cdot)}$ is a series of CPTP maps
     $\A^{(\cdot)} = (\A_1,\dots,\A_{t+1})$, where $\A_1,\dots,\A_t$ each have input $\Reg{Out},\Reg{A}$
     and output $\Reg{In},\Reg{A}$, and $\A_{t+1}$ maps $\Reg{Out},\Reg{A}$ to itself. Thus, the algorithm makes $t$ oracle queries.

    Let $\rho_{\Reg{Out},\Reg{A}}$ be an arbitrary quantum state. We write
    $$\A^{\O}(\rho) = \Tr_{\Reg{St}}\left(
        (\A_{t+1} \cdot \O \cdot \A_t\dots \O \cdot \A_1) 
        (\rho_{\Reg{Out},\Reg{A}}\otimes \ketbra{\phi_0}_{\Reg{St}}) \right)$$

    Without loss of generality, $\rho$ can be produced by $\A_1$. And so we will 
    write $\A^{\O} \coloneqq \A^{\O}(\ketbra{0}_{\Reg{Out},\Reg{A}})$.
\end{definition}

\begin{definition}[Important types of linear maps]
    Let $\Reg{A},\Reg{B}$ be arbitrary Hilbert spaces and let $V:\Reg{A}\to \Reg{B}$ be a linear map.
    
    We call $V$ a contraction if
    $$V^\dagger V \preceq I_{\Reg{A}},$$
    or, equivalently, $VV^\dagger\preceq I_{\Reg{B}}$.

    When $\Reg{B}=\Reg{A}$, we call $V$ a projector if $V=V^\dagger$ and $V^2 = V$.

    We call $V$ a partial isometry if $V^\dagger V$ is an orthogonal
    projector on $\Reg{A}$ (equivalently, $VV^\dagger$ is an orthogonal
    projector on $\Reg{B}$). For a partial isometry $V$, we define the image of $V$ as 
    $$\Ima(V)=\{V\ket{\phi}:\ket{\phi}\in \Reg{A}\} = \{\ket{\phi}\in \Reg{B} : VV^\dagger \ket{\phi} = \ket{\phi}\}$$
    and the support of $V$ as
    $$\Supp(V)=\{\ket{\phi}\in \Reg{A}:V^\dagger V \ket{\phi} = \ket{\phi}\}$$

    Note that if $V$ is a partial isometry or a projector, then it is a contraction.
\end{definition}

We will put special attention on the maps of the form $\rho \mapsto V\rho V^\dagger$ 
for $V$ a contraction. While such maps are generally not CPTP, they can be thought of as a CPTP map which outputs a special state $\ket{\bot} \notin \Ima(V)$ when the map fails. Formally, we define the following.

\begin{definition}
    Let $V:\Reg{A}\to \Reg{B}$ be a contraction, let
    $\Reg{B}_\bot=\Reg{B}\oplus\Span\{\ket{\bot}\}$, and regard $V$ as
    mapping into the natural copy of $\Reg{B}$ in $\Reg{B}_\bot$. We define
    the associated CPTP map
    $\Phi^V:\mathcal{L}(\Reg{A})\to\mathcal{L}(\Reg{B}_\bot)$ by
    $$\Phi^V(\rho)
      =V\rho V^\dagger
       +\Tr\!\left((I_{\Reg{A}}-V^\dagger V)\rho\right)\ketbra{\bot}.$$

    For $\A^{(\cdot)}$ an oracle algorithm, we will write $\A^V$ to denote $\A^{\Phi^V}$. We will typically assume without loss of generality that $\ket{\bot} \in \Reg{B}$ and write $\Reg{B}$ in place of $\Reg{B}_{\bot}$.
\end{definition}

The following standard observation shows that when two oracles are isometrically equivalent, they are also equivalent up to query access.

\begin{lemma}[Isometric equivalence of stateful oracles]
    \label{lem:oracle-isometric-equivalence}
    Let $W_1:\Reg{In}\otimes\Reg{St}_1\to
    \Reg{Out}\otimes\Reg{St}_1$, $W_2:\Reg{In}\otimes \Reg{St}_2\to \Reg{Out}\otimes \Reg{St}_2$ be oracles
    with private initial states $\ket{\phi_1},\ket{\phi_2}$.
    Suppose there is an isometry $J:\Reg{St}_1\to\Reg{St}_2$ such that
    \[
        J\ket{\phi_1}=\ket{\phi_2}
    \]
    and
    \[
        W_2(I_{\Reg{In}}\otimes J)
            =(I_{\Reg{Out}}\otimes J)W_1.
    \]
    Then, for every adaptive oracle algorithm $\A$,
    \[
        \A^{\Phi^{W_1}}=\A^{\Phi^{W_2}}.
    \]
\end{lemma}

\begin{proof}
    Note that $W_1 = (I_{\Reg{Out}}\otimes J^\dagger)W_2(I_{\Reg{In}}\otimes J)$. Telescoping then says that the state before tracing out the private register of $\A^{W_2}$ is $(I_{\Reg{Out}}\otimes J)$ applied to the same state from the computation of $\A^{W_1}$. Since $J$ is an isometry, tracing out the private register gives the same output state for both oracle algorithms.
\end{proof}

Given two contractions $V_1,V_2$, we will often want to bound the distance between $\Phi^{V_1}$ and $\Phi^{V_2}$ in terms of the operator norm of $V_1-V_2$. We will use the following lemma.

\begin{lemma}\label{lem:contraction-oracle-distance}
    Let $V_1,V_2$ be two contractions such that
    $$\norm{V_1-V_2}_{op} \leq \epsilon$$

    Then for any $t$ query oracle algorithm $\A$,
    $$\TD\left(\A^{\Phi^{V_1}},\A^{\Phi^{V_2}}\right) \leq 2t\epsilon$$
\end{lemma}

We will make use of the following useful fact about the operator norm

\begin{lemma}\label{lem:opprop}
    Let $V_1,V_2,W_1,W_2$ be partial isometries. Then the following relations hold
    $$\norm{V_1W_1 - V_2W_2}_{op} \leq \norm{V_1-V_2}_{op} + \norm{W_1-W_2}_{op}$$
    $$\norm{(V_1+V_2) - (W_1+W_2)}_{op} \leq \norm{V_1-W_1}_{op} + \norm{V_2-W_2}_{op}$$
    $$\norm{V_1}_{op} = \norm{V_1^\dagger}_{op}$$
\end{lemma}

For projectors, we will use a coherent implementation of the two-outcome
measurement $\{\Pi,I-\Pi\}$.

\begin{definition}
    Let $\Pi$ be a projector on register $\Reg{A}$ and let $\Reg{C}$ be a
    qubit. Define $M^\Pi$ to be conjugation by the unitary
    $$U_\Pi
      \coloneqq \Pi_{\Reg{A}}\otimes X_{\Reg{C}}
      +(I_{\Reg{A}}-\Pi_{\Reg{A}})\otimes I_{\Reg{C}}.$$
    Thus, when $\Reg{C}$ is initialized to $\ket{b}$, it coherently records
    outcome $b\xor 1$ on the image of $\Pi$ and outcome $b$ on its orthogonal
    complement.
\end{definition}

We explicitly define syntax for controlled queries to partial isometries.

\begin{definition}
    Let $\Reg{A},\Reg{A}_{pure}$ be registers. Define $\Reg{C} = \mathcal{H}(\{0,1\})$. Let $V:\Reg{A}\to \Reg{A}\otimes\Reg{A}_{pure}$ be a partial isometry. We define $\ctrl_b[V]:\Reg{C} \otimes \Reg{A} \to \Reg{C}\otimes \Reg{A}\otimes\Reg{A}_{pure}$ to be the map
    $$\ctrl_b[V] = \ketbra{b}_{\Reg{C}} \otimes V_{\Reg{A} \to \Reg{A}\Reg{A}_{pure}}  + \ketbra{b\xor 1}_{\Reg{C}} \otimes (I_{\Reg{A}} \otimes \ket{0}_{\Reg{A}_{pure}})$$

    For ease of notation, we write $\ctrl[V] \coloneqq \ctrl_1[V]$.
\end{definition}

Certain contractions composed of projectors and partial isometries have a simple implementation via an algorithm querying only the components of the contraction. As these will be useful for our analysis, we consider them explicitly. The following result is folklore (and has implicitly appeared in work of Ma and Huang~\cite{STOC:MaHua25}), but we include a full proof for completeness.

\begin{theorem}\label{thm:efficiently-implementable-partial-isometry}
    Let $\Pi_2$ be a projector on $\Reg{A}$, let $\Pi_1$ be a projector
    on $\Reg{A}\otimes\Reg{A}_{pure}$, and let
    $V_1,V_2:\Reg{A}\to\Reg{A}\otimes\Reg{A}_{pure}$ be partial isometries. Set
    $$W=\Pi_1V_1\Pi_2+(I_{\Reg{A}\Reg{A}_{pure}}-\Pi_1)V_2
      (I_{\Reg{A}}-\Pi_2).$$
    Then $\Phi^W$ has an efficient implementation making one ordinary query
    to each of $M^{\Pi_1},M^{\Pi_2}$ and one query to each of
    $\Phi^{\ctrl[V_1]}$ and
    $\Phi^{\ctrl[V_2]}$.
\end{theorem}

\begin{proof}
    To see that $W$ is a contraction, see that
    \begin{align*}
        W^\dagger W
        &=\Pi_2 V_1^\dagger \Pi_1 \Pi_1 V_1 \Pi_2 + \Pi_2 V_1^\dagger \Pi_1 (I_{\Reg{A}\Reg{A}_{pure}} - \Pi_1) V_2 (I_{\Reg{A}} - \Pi_2)\\
        &\quad+(I_{\Reg{A}}-\Pi_2)V_2^\dagger (I_{\Reg{A}\Reg{A}_{pure}} - \Pi_1) (I_{\Reg{A}\Reg{A}_{pure}} - \Pi_1) V_2 (I_{\Reg{A}}- \Pi_2)\\
        &\quad+(I_{\Reg{A}}-\Pi_2)V_2^\dagger (I_{\Reg{A}\Reg{A}_{pure}}-\Pi_1) \Pi_1 V_1 \Pi_2\\
        &=\Pi_2 V_1^\dagger \Pi_1 V_1 \Pi_2 + (I_{\Reg{A}}-\Pi_2) V_2^\dagger (I_{\Reg{A}\Reg{A}_{pure}}-\Pi_1) V_2 (I_{\Reg{A}}-\Pi_2)\\
        &\preceq \Pi_2+(I_{\Reg{A}}-\Pi_2)=I_{\Reg{A}}.
    \end{align*}
    Thus $W$ is a contraction and $\Phi^W$ is well defined.

    Consider the following implementation of $\Phi^W$.
    \begin{enumerate}
        \item Initialize a control qubit $\Reg{C}$ to $\ket{0}$ and apply
        $M^{\Pi_2}$ to the input register $\Reg{A}$, writing the result in $\Reg{C}$.
        \item Controlled on $\Reg{C}$ being $1$, apply $V_1$ to $\Reg{A}$. That is, apply $\Phi^{\ctrl_1[V_1]}$ to $\Reg{A}\Reg{C} \to \Reg{A}\Reg{A}_{pure}\Reg{C}$.
        \item Controlled on $\Reg{C}$ being $0$, apply $V_2$ to $\Reg{A}$. That is, apply $\Phi^{\ctrl_0[V_2]}$ to $\Reg{A}\Reg{C} \to \Reg{A}\Reg{A}_{pure}\Reg{C}$. This uses one query to $\Phi^{\ctrl[V_2]}$ by applying $X_{\Reg{C}}$ before and after the query.
        \item Apply $M^{\Pi_1}$ to the output register $\Reg{A}\otimes\Reg{A}_{pure}$, writing the result to $\Reg{C}$.
        \item Measure $\Reg{C}$ in the computational basis. Map outcome $1$ (as well as any failure in steps 2 and 3) to $\ket{\bot}$. On outcome $0$, discard
        $\Reg{C}$ and output the residual state.
    \end{enumerate}
    Here, we have written steps 2 and 3 as writing to the same register $\Reg{A}_{pure}$. This can be implemented coherently by initializing $\Reg{A}_{pure}$ to $\ket{0}$, having steps 2 and 3 put their outputs on fresh registers $\Reg{A}_{pure}^{(1)}$ and $\Reg{A}_{pure}^{(2)}$, and then doing a controlled swap with the same register $\Reg{A}_{pure}$.

    We verify the claim on a pure input $\ket{\psi}$; the general case follows
    by linearity, and the same calculation holds with an arbitrary reference
    register tensored throughout. After the first step, the joint state of the
    input and control qubit is
    $$\Pi_2\ket{\psi}\ket{1}+(I_{\Reg{A}}-\Pi_2)\ket{\psi}\ket{0}.$$
    After the two coherently controlled component queries, conditioned on both succeeding the resulting state is
    $$V_1\Pi_2\ket{\psi}\ket{1}
      +V_2(I_{\Reg{A}}-\Pi_2)\ket{\psi}\ket{0}.$$
    Applying $M^{\Pi_1}$ to the output and the same control qubit gives
    \begin{align*}
        &\Pi_1V_1\Pi_2\ket{\psi}\ket{0}
        +(I_{\Reg{A}\Reg{A}_{pure}}-\Pi_1)V_1\Pi_2\ket{\psi}\ket{1}\\
        &\quad+\Pi_1V_2(I_{\Reg{A}}-\Pi_2)\ket{\psi}\ket{1}
        +(I_{\Reg{A}\Reg{A}_{pure}}-\Pi_1)V_2(I_{\Reg{A}}-\Pi_2)\ket{\psi}\ket{0}\\
        &\qquad=W\ket{\psi}\ket{0}
        +\left((I_{\Reg{A}\Reg{A}_{pure}}-\Pi_1)V_1\Pi_2+\Pi_1V_2(I_{\Reg{A}}-\Pi_2)\right)\ket{\psi}\ket{1}.
    \end{align*}
    Consequently, conditioning on the final value on $\Reg{C}$ being $0$ produces exactly the subnormalized state
    $$W\ketbra{\psi}W^\dagger.$$
    All remaining branches (the outcome-$1$ branch above and the failure
    branches of $\Phi^{\operatorname{ctrl}_1[V_1]}$ and
    $\Phi^{\operatorname{ctrl}_0[V_2]}$) are mapped to $\ket{\bot}$. Since
    the entire procedure is trace preserving, their total probability is
    $$1-\Tr\left(W\ketbra{\psi}W^\dagger\right).$$
    The output is therefore
    $$W\ketbra{\psi}W^\dagger
      +\left(1-\Tr\left(W\ketbra{\psi}W^\dagger\right)\right)\ketbra{\bot}
      =\Phi^W(\ketbra{\psi})$$
\end{proof}

\begin{lemma}[Swap test, \cite{BCWdW01}]\label{lem:swap-test}
    Let $\ket{\psi}$ and $\ket{\phi}$ be states on isomorphic
    registers $\Reg A$ and $\Reg B$. The swap test is an efficient
    measurement that, on input $\ket{\psi}_{\Reg A}\ket{\phi}_{\Reg B}$,
    outputs a bit $b\in\{0,1\}$ satisfying
    \[
        \Pr[b=1]=\frac{1-\abs{\braket{\psi}{\phi}}^2}{2}.
    \]
    In particular, this probability is zero for identical states
    and $1/2$ for orthogonal states.
\end{lemma}

\subsection{The Quantum Switch}
\label{sec:quantum-switch}

For channels $\Phi$ and $\Psi$ on a common space, write
\[
    \Phi(\rho)=\sum_a K_a^\Phi\rho(K_a^\Phi)^\dagger,
    \qquad
    \Psi(\rho)=\sum_b K_b^\Psi\rho(K_b^\Psi)^\dagger.
\]
The quantum switch~\cite{CDPV13} is the channel
$\Sw[\Phi,\Psi]$ defined by
\[
    \Sw[\Phi,\Psi](\rho)
        =\sum_{a,b}S_{a,b}\rho S_{a,b}^\dagger,
\]
where
\[
    S_{a,b}=\ketbra{0}\otimes K_b^\Psi K_a^\Phi
        +\ketbra{1}\otimes K_a^\Phi K_b^\Psi.
\]
This definition is independent of the chosen operator-sum
representations. The first register is a control qubit that
coherently selects the order of the two channels: if $c=0$, apply
$\Phi$ then $\Psi$; otherwise, apply $\Psi$ then $\Phi$.

In this work, we consider only the case $\Phi=\Psi$, known as
self-switching~\cite{ESC18,CK19}. We refer to $\Sw[\Phi,\Phi]$
as the self-switch of $\Phi$~\cite{Kar26}. The self-switch also has a
nice operational description in terms of any purification of the
channel. The Stinespring representation of the switch
in Appendix C of~\cite{AWHMB20} leads to the following
implementation for identical channels.

\begin{construction}[Self-switch from an isometry dilation]
\label{con:self-switch-from-dilation}
    Let $\Phi$ be a channel on $\Reg{In}$ and let
    $V^\Phi:\Reg{In}\to\Reg{In}\otimes\Reg E$ be any fixed
    isometry dilation of $\Phi$. On input in the switch control
    $\Reg C$ and the target $\Reg{In}$, proceed as follows:
    \begin{enumerate}
        \item Apply $V^\Phi$ to $\Reg{In}$ and retain its
            environment in $\Reg E_1$.
        \item Apply $V^\Phi$ again to the resulting $\Reg{In}$
            and retain the second environment in $\Reg E_2$.
        \item Controlled on $\Reg C$ being $1$, swap $\Reg E_1$
            and $\Reg E_2$, and then discard both environments.
    \end{enumerate}
\end{construction}

\begin{lemma}\label{lem:self-switch-from-dilation}
    Let $\Phi$ and $V^\Phi$ be as in Construction~\ref{con:self-switch-from-dilation}. Then the construction implements $\Sw[\Phi,\Phi]$ exactly using two adaptive forward queries to $V^\Phi$.
\end{lemma}

\begin{proof}
    For some orthonormal basis $\{\ket{\phi_a}\}_a$ of $\Reg E$ we can write
    \[
        V^\Phi=\sum_a K_a^\Phi\otimes\ket{\phi_a}.
    \]
    We will also write
    $$\Phi(\rho)=\sum_a K_a^\Phi\rho(K_a^\Phi)^\dagger$$
    Before the controlled swap, the two queries act on a state $\ket{c}\ket{\psi}$ as follows
    \[
        \ket{c}\ket{\psi}\longmapsto
        \ket{c}\otimes \sum_{a,b}K_b^\Phi K_a^\Phi\ket{\psi}
            \ket{\phi_a}_{\Reg E_1}\ket{\phi_b}_{\Reg E_2}.
    \]
    After the controlled swap, we exactly get
    \[
        (\ctrl[SWAP]_{\Reg{C}\Reg{E}_1\Reg{E}_2}\cdot V^\Phi_{\Reg{In}\to \Reg{In}\Reg{E}_2} V^\Phi_{\Reg{In}\to \Reg{In}\Reg{E}_1})\ket{0}\ket{\psi}= \ket{0}\otimes \sum_{a,b}K^\Phi_b K^\Phi_a \ket{\psi}\ket{\phi_a}_{\Reg{E_1}} \ket{\phi_b}_{\Reg{E_2}}.
    \]
    and
    \[
        (\ctrl[SWAP]_{\Reg{C}\Reg{E}_1\Reg{E}_2}\cdot V^\Phi_{\Reg{In}\to \Reg{In}\Reg{E}_2} V^\Phi_{\Reg{In}\to \Reg{In}\Reg{E}_1})\ket{1}\ket{\psi}= \ket{1}\otimes \sum_{a,b}K^\Phi_a K^\Phi_b \ket{\psi}\ket{\phi_a}_{\Reg{E_1}} \ket{\phi_b}_{\Reg{E_2}}.
    \]
    
    As the following isometries agree on all pure states, they are equal, 
    \[
        \begin{aligned}
            &\ctrl[SWAP]_{\Reg{C}\Reg{E}_1\Reg{E}_2}\cdot
                V^\Phi_{\Reg{In}\to \Reg{In}\Reg{E}_2}
                V^\Phi_{\Reg{In}\to \Reg{In}\Reg{E}_1} \\
            &\quad=\sum_{a,b}\left(\ketbra{0}\otimes K_b^\Phi K_a^\Phi
                +\ketbra{1}\otimes K_a^\Phi K_b^\Phi\right)
                \otimes\ket{\phi_a}\ket{\phi_b} \\
            &\quad=\sum_{a,b}S_{a,b}\otimes\ket{\phi_a}\ket{\phi_b}.
        \end{aligned}
    \]
    Since $\{\ket{\phi_a}\}$ form an orthonormal basis, discarding the environments therefore gives the channel
    $$\rho\mapsto\sum_{a,b}S_{a,b}\rho S_{a,b}^\dagger,$$
    which is exactly $\Sw[\Phi,\Phi]$.
\end{proof}

\subsection{Young Diagrams and Schur--Weyl Duality}\label{sec:young-diagrams}

A Young diagram $\lambda$ consists of left-aligned rows of boxes with
lengths $\lambda_1\geq\cdots\geq\lambda_{\ell(\lambda)}>0$.
Write $\ell(\lambda)$ for its number of rows and
$|\lambda|=\sum_{i=1}^{\ell(\lambda)}\lambda_i$ for its number of boxes.
The box $(i,j)\in\lambda$ lies in row $i$ and column $j$, where
$1\leq j\leq\lambda_i$. Its hook length $h_\lambda(i,j)$ is one plus
the number of boxes strictly to its right in the same row and strictly
below it in the same column. The empty diagram has no rows or boxes.

Let $\mathcal H=\mathcal H([N])$, and let $S_t$ be the symmetric group
on $t$ elements. For $\pi\in S_t$, let $P_\pi$ permute the tensor
factors of $\mathcal H^{\otimes t}$. These permutations commute with
$U^{\otimes t}$ for every $U\in U(N)$. Schur--Weyl
duality (Theorem~3.11 of~\cite{FLMNW26}) gives the orthogonal decomposition
\[
    \mathcal H^{\otimes t}
        \cong\bigoplus_{\substack{|\lambda|=t\\\ell(\lambda)\leq N}}
            \mathcal Q_{\lambda,N}\otimes\mathcal P_\lambda,
\]
where $\lambda$ ranges over Young diagrams, and
$\mathcal Q_{\lambda,N}$ and $\mathcal P_\lambda$ are the corresponding
irreducible representation spaces of $U(N)$ and $S_t$. Write
\[
    d_{\lambda,N}=\dim\mathcal Q_{\lambda,N},
    \qquad f_\lambda=\dim\mathcal P_\lambda.
\]
Let $\Pi_{\lambda,t}$ be the orthogonal projector onto
$\mathcal Q_{\lambda,N}\otimes\mathcal P_\lambda$, with the ambient
dimension $N$ implicit.

For $|\lambda|=t$, the hook-length
formula~\cite[Lemmas~3.24--3.25]{FLMNW26} gives
\[
    f_\lambda=\frac{t!}{\prod_{(i,j)\in\lambda}h_\lambda(i,j)}.
\]
The Weyl dimension formula~\cite[Lemma~3.27]{FLMNW26}, in its
equivalent hook-length form, gives
\begin{equation}\label{eq:unitary-dimension}
    d_{\lambda,N}
        =\prod_{(i,j)\in\lambda}
            \frac{N+j-i}{h_\lambda(i,j)}.
\end{equation}
Empty products equal one, so $f_{\varnothing}=d_{\varnothing,N}=1$.

\subsection{Exact Path Recording}\label{sec:exact-path-recording}

We use the exact path recording oracle of Foxman, Lombardi, Ma, Nehoran,
and Wright~\cite[Section~5]{FLMNW26}, specialized to Haar-random unitaries.
Grinko and Yoshida~\cite{GY25} independently developed a concurrent path
recording construction in the updated version of their work.

Let $\mathcal H=\mathcal H([N])$. The ordered path register is
\[
    \mathcal R_t=\mathcal H^{\otimes t}\otimes
        \overline{\mathcal H}^{\otimes t},\qquad
    \ket{\vec x,\vec y}
        =\ket{x_1,\ldots,x_t}\otimes
            \overline{\ket{y_1,\ldots,y_t}}.
\]
Here $\mathcal R_0=\Span\{\ket{\varnothing}\}$; repeated entries are allowed.
Define the append map by
\[
    \operatorname{App}^{(t)}_{x,y}\ket{\vec x,\vec y}
        =\ket{(\vec x,x),(\vec y,y)}.
\]
The projector onto legal path states is
\[
    \Omega_t=\frac{1}{t!}\sum_{\pi\in S_t}
        P_\pi\otimes\overline{P_\pi},\qquad \Omega_0=I.
\]

Using the dimensions and projectors from \Cref{sec:young-diagrams},
define the reweightings acting only on the input slots by
\begin{align*}
    \Lambda_t&=\sum_{\substack{|\lambda|=t\\\ell(\lambda)\leq N}}
        \sqrt{\frac{f_\lambda}{d_{\lambda,N}}}
        \bigl(\Pi_{\lambda,t}\otimes I\bigr),\\
    \Lambda_t^+&=\sum_{\substack{|\lambda|=t\\\ell(\lambda)\leq N}}
        \sqrt{\frac{d_{\lambda,N}}{f_\lambda}}
        \bigl(\Pi_{\lambda,t}\otimes I\bigr).
\end{align*}
Here $I$ acts on the conjugate-output slots and the operator $\Lambda_t^+$
is inverse to $\Lambda_t$ on $\mathcal R_t$. Set $\Lambda_0=\Lambda_0^+=I$.

\begin{definition}[Exact path recording, Definition~5.1 and Remark~5.2 of~\cite{FLMNW26}]
    \label{def:exact-path-recording}
    For $N,T\in\N$, set
    $\Reg{In}=\mathcal H$ and
    $\Reg{St}=\bigoplus_{t=0}^T\mathcal R_t$.
    The exact path recording oracle $PR_{[N],T}$ is the
    partial isometry whose restriction to
    $\Reg{In}\otimes\mathcal R_t$, for $t<T$, is
    \[
        PR_{[N],T}\big|_{\Reg{In}\otimes\mathcal R_t}
        =\sum_{x,y\in[N]}\ketbra{y}{x}\otimes
            \Lambda_{t+1}\Omega_{t+1}
            \operatorname{App}^{(t)}_{x,y}\Lambda_t^+\Omega_t.
    \]
    It is zero on $\Reg{In}\otimes\mathcal R_T$.
    The recording register starts in $\ket{\varnothing}$.
\end{definition}

\begin{theorem}[Exact simulation, Theorem~5.3 and Section~2.1.1 of~\cite{FLMNW26}]
    \label{thm:exact-prsim}
    For every oracle algorithm $\A$ making $t\leq T$ controlled forward queries,
    \[
        \E_{U\gets\Haar([N])}\bigl[\A^{\ctrl[U]}\bigr]
            =\A^{\ctrl[PR_{[N],T}]}.
    \]
\end{theorem}

\begin{lemma}[Efficient implementation, Sections~4.2.3 and~5.3 of~\cite{FLMNW26}]
    \label{lem:exact-pr-efficient}
    Let $N,T\in\N$ and $0<\epsilon<1$. The channels associated
    with $PR_{[N],T}$, its adjoint, and their
    controlled versions can be implemented to within
    trace-distance error $\epsilon$, including an arbitrary
    reference register, using
    \[
        \poly(T,\log N,\log(1/\epsilon))
    \]
    gates. The underlying circuits also give coherent measurements
    of the initial-space and range projectors to the same accuracy
    with the same gate bound.
\end{lemma}

\subsection{Zhandry's Purified Oracle Method}

Compressed oracles, introduced by~\cite{C:Zhandry19}, are a technique for statefully and efficiently simulating queries to a random function $f:[N]\to [t]$. This is done in two steps, first Zhandry moves the randomness of the oracle to an external register (i.e. the oracle is purified) and views this register in the Fourier basis. Next, Zhandry shows how to efficiently represent the information contained in the purified register (i.e. the oracle is compressed). For this work, we only need the $N=2$ case; the ``compression'' step is unnecessary. As such, we will only present the so-called ``purified'' oracle, which simulates $f$ using $N\ceil{\log_2 t}$ qubits. We will focus on the phase-oracle formulation used by Ananth and Goldin~\cite{AG26}, i.e. maps of the form $\ket{x}\mapsto \omega_t^{f(x)}\ket{x}$, where $\omega_t$ is a $t$th root of unity. For simplicity, we consider only uncontrolled, forward queries.

\begin{definition}
    Let $t\in \N$, let $A$ be a set and let $f:A\to [t]$ be a function. Let $\omega_t = e^{\frac{2\pi i}{t}}$ be the $t$-th root of unity. We define $S^f$ to be the unitary on $\mathcal{H}(A)$ defined by
    $$S^f \ket{x} = \omega_t^{f(x)}\ket{x}$$
\end{definition}

\begin{theorem}[Generalized purified oracles, \cite{C:Zhandry19,AG26}]\label{thm:compext}
    Let $\ell,t\in \N$ and let $A =\{a_1,\dots,a_\ell\}$ be a set. Define $\Reg{St} = \mathcal{H}(\Z_t)^{\otimes \ell}$. Define the purified oracle $PO:\mathcal{H}(A) \otimes \Reg{St}$ by
    $$PO\ket{a_i}\ket{\{c_1,\dots,c_\ell\}} \mapsto \ket{a_i}\ket{\{c_1,\dots,c_{i-1},c_i+1,c_{i+1},\dots,c_\ell\}}$$
    where addition acts modulo $t$. This is the map which increases the $i$th state register by $1$, where $i$ is the index contained in the first register.
    We will set the default initial state to be $\ket{\{0,\dots,0\}}\in \Reg{St}$.

    Then query access to $PO$ is identical to query access to $S^f$ for a random function $f$. That is, for all oracle algorithms $\A^{(\cdot)}$,
    $$\E_{f:A\to [t]}[\A^{S^f}] = \A^{PO}$$
    Here, the initial state of the oracle's purification register is $\ket{\{0,\dots,0\}}$.
\end{theorem}

\begin{proof}
    For every function $f:A\to [t]$, define
    its Fourier-basis encoding by
    \[
        \ket{\widehat f}_{\Reg{St}}
        \coloneqq \frac{1}{t^{\ell/2}}
        \sum_{(c_1,\dots,c_\ell)\in \Z_t^\ell}
        \omega_t^{-\sum_{j=1}^{\ell} f(a_j)c_j}
        \ket{c_1,\dots,c_\ell}_{\Reg{St}}.
    \]
    The states
    $\{\ket{\widehat f}:f:A\to [t]\}$ form an orthonormal basis of
    $\Reg{St}$.  Moreover, for all $a_i\in A$, by setting $\wt{c}_i = c_i + 1$ and $\wt{c}_j = c_j$ for $j\neq i$, we have
    \begin{equation}\label{eq:po-fourier-action}
        \begin{split}
        PO\ket{a_i}\ket{\widehat f}\\
        = \frac{1}{t^{\ell/2}}
        \sum_{(c_1,\dots,c_\ell)\in \Z_t^\ell}
        \omega_t^{-\sum_{j=1}^{\ell} f(a_j)c_j}
        \ket{a_i}\ket{c_1,\dots,c_{i-1},c_i+1,c_{i+1},\dots,c_\ell}_{\Reg{St}}\\
        = \frac{1}{t^{\ell/2}}
        \sum_{(\wt{c}_1,\dots,\wt{c}_\ell)\in \Z_t^\ell} \omega_t^{f(a_i)}\omega_t^{-\sum_{j=1}^{\ell} f(a_j)\wt{c}_j}
        \ket{a_i}\ket{\wt{c}_1,\dots,\wt{c}_{i-1},\wt{c}_i,\wt{c}_{i+1},\dots,\wt{c}_\ell}_{\Reg{St}}\\
        = \omega_t^{f(a_i)}\ket{a_i}\ket{\widehat f}
        = (S^f\ket{a_i})\ket{\widehat f}.
        \end{split}
    \end{equation}
    Thus, in this basis, the purified oracle is block diagonal:
    \[
        PO = \sum_{f:A\to[t]} S^f\otimes
        \ketbra{\widehat f}_{\Reg{St}}.
    \]

    We can then write the initial state of the purification register as
    \[
        \ket{0,\dots,0}_{\Reg{St}}
        = \frac{1}{t^{\ell/2}}
        \sum_{f:A\to[t]}\ket{\widehat f}_{\Reg{St}}.
    \]

    Without loss of generality, we will assume that $\A^{(\cdot)}$ acts unitarily and that $\A^{S^f}$ is a pure state. We can do this by deferring all measurements to the end and purifying the final output. Let $\ketbra{\psi_f} = \A^{S^f}$. Since $\A$ does not act
    on $\Reg{St}$, equation~\eqref{eq:po-fourier-action} and induction over
    the oracle calls show before tracing out $\Reg{St}$, the final state of $\A^{PO}$ is
    \[
        \frac{1}{t^{\ell/2}}
        \sum_{f:A\to[t]} \ket{\psi_f}\ket{\widehat f}_{\Reg{St}}.
    \]
    Since the Fourier basis terms $\ket{\widehat f}$ are orthonormal, tracing out $\Reg{St}$ gives that
    \[
        \A^{PO}
        = \frac{1}{t^\ell}\sum_{f:A\to[t]}\ketbra{\psi_f}\\
        = \E_{f:A\to[t]}[\ketbra{\psi_f}]
        = \E_{f:A\to[t]}[\A^{S^f}],
    \]
    as claimed.
\end{proof}

In particular, we will only concern ourselves with the purified oracle for functions of the form $f:\{0,1\}\to [t]$. In this setting, the purification register is $\mathcal{H}(\Z_t) \otimes \mathcal{H}(\Z_t)$, and we have the map
$$PO\ket{0}\ket{t_0}\ket{t_1} \mapsto \ket{0}\ket{t_0+1}\ket{t_1}$$
and
$$PO\ket{1}\ket{t_0}\ket{t_1} \mapsto \ket{1}\ket{t_0}\ket{t_1+1}$$

\section{Simulating Controlled Queries to Isometries}

\begin{theorem}\label{thm:simctrl}
    Let $T,M,N\in \N$ such that $N|M$. There is a simulator $\Sim_{ctrl}^{(\cdot)}$ that makes at most one uncontrolled forward query to the oracle per invocation that has the following property. For every partial isometry $V$ sending $\mathcal{H}([N])$ to $\mathcal{H}([M])$ (with $\ket{0}$ in the support of $V$) and every oracle algorithm $\mathcal{A}^{(\cdot)}$ making at most $T$ controlled queries to its oracle, we have
    $$\E_{\theta \gets [0,2\pi)}[\A^{\ctrl[e^{i\theta}\cdot V]}] = \A^{\Sim_{ctrl}^{V}}$$
    Furthermore, $\Sim_{ctrl}^{(\cdot)}$ can be implemented to within error $\epsilon$ using $\poly(T, \log M, \log(1/\epsilon))$ gates. 
\end{theorem}

\begin{proof}
    Let $\Reg{A} = \mathcal{H}([N])$ and $\Reg{B} = \mathcal{H}([M/N])$, and let $\Reg{C} = \mathcal{H}(\{0,1\})$ be a control register. Set $t = T+1$.

    Define $\Reg{D}_0\otimes \Reg{D}_1 = \mathcal{H}([t])\otimes \mathcal{H}([t])$. For $i\in \{1,\dots,T\}$, define $\Reg{E}_i$ to be a copy of $\Reg{A}\otimes \Reg{B}$. Our simulator will use internal state $\Reg{D}_0\otimes \Reg{D}_1$ all initialized to $\ket{0}$. $\Reg{E}_1,\dots,\Reg{E}_T$ will all be initialized to $\ket{0}$.
    \begin{enumerate}
        \item First, controlled on the value in $\Reg{C}$ being $0$, add $1$ to $\Reg{D}_0$.
        \item Controlled on the value in $\Reg{C}$ being $1$, add $1$ to $\Reg{D}_1$.
        \item Generate a new register $\Reg{In}$ as a copy of $\Reg{A}$ initialized to $\ket{0}$.
        \item Controlled on the value in $\Reg{C}$ being $1$, swap registers $\Reg{A}$ and $\Reg{In}$.
        \item Apply $\Phi^V$ to $\Reg{In}$, producing a new state on a register $\Reg{Out}$.
        \item Initialize $\Reg{B}$ to $\ket{0}$.
        \item Controlled on the value in $\Reg{C}$ being $1$, swap registers $\Reg{A}\otimes \Reg{B}$ and $\Reg{Out}$. 
        \item For all $i\in \{1,\dots,T\}$, controlled on the value in $\Reg{C}$ being $0$ and the value in $\Reg{D_0}$ being $i$, swap $\Reg{E}_i$ and $\Reg{Out}$.
        \item Output $\Reg{CAB}$.
    \end{enumerate}

    For now, instead of considering $\theta$ sampled uniformly from $[0,2\pi)$, we use phases sampled from the discrete set $\{0,\frac{1}{t}2\pi,\frac{2}{t}2\pi,\dots,\frac{t-1}{t}2\pi\}$, and write $\omega_t=e^{2\pi i/t}$. We will also apply a separate random phase to the identity portion of $\ctrl[V]$. In particular, for $\theta_0,\theta_1\in [t]$, define $\ctrl[V]^{\theta_0,\theta_1}$ to be the map
    $$\ctrl[V]^{\theta_0,\theta_1}_{\Reg{C}\Reg{A}\mapsto \Reg{C}\Reg{A}\Reg{B}} = \omega_t^{\theta_0}\ketbra{0}_{\Reg{C}} \otimes I_{\Reg{A}} \otimes \ket{0}_{\Reg{B}} + \omega_t^{\theta_1}\ketbra{1}_{\Reg{C}}\otimes V_{\Reg{A}\to \Reg{A}\Reg{B}}$$
    We will show that for all $V$,
    $$\A^{\Sim_{ctrl}^V} = \E_{\theta_0,\theta_1\gets [t]}[\A^{\ctrl[V]^{\theta_0,\theta_1}}]$$
    As 
    \begin{equation}
        \begin{split}
            \E_{\theta\gets [0,2\pi)}[\A^{\ctrl[e^{i\theta}V]}]\\
            =\E_{\theta_0 \gets [t]}[\E_{\theta\gets [0,2\pi)}[\A^{\omega_t^{\theta_0}\cdot \ctrl[e^{i\theta}V]}]]\\
            =\E_{\theta_0,\theta_1\gets [t]}[\E_{\theta\gets [0,2\pi)}[\A^{\omega_t^{\theta_0}\cdot \ctrl[\omega_t^{-\theta_0}\omega_t^{\theta_1}e^{i\theta}V]}]]\\
            =\E_{\theta_0,\theta_1\gets [t]}[\E_{\theta\gets [0,2\pi)}[\A^{\ctrl[e^{i\theta}V]^{\theta_0,\theta_1}}]]\\
            =\E_{\theta \gets [0,2\pi)}\E_{\theta_0,\theta_1\gets [t]}[\A^{\ctrl[e^{i\theta}V]^{\theta_0,\theta_1}}]
        \end{split}
    \end{equation}
    we get that
    \begin{equation}
        \begin{split}
            \A^{\Sim_{ctrl}^V} = \E_{\theta\gets [0,2\pi)}[\A^{\Sim_{ctrl}^{e^{i\theta}V}}]\\
            = \E_{\theta_0,\theta_1\gets [t]}[\E_{\theta\gets [0,2\pi)}[\A^{\ctrl[e^{i\theta}V]^{\theta_0,\theta_1}}]]\\
            =\E_{\theta \gets [0,2\pi)}\E_{\theta_0,\theta_1\gets [t]}[\A^{\ctrl[e^{i\theta}V]^{\theta_0,\theta_1}}]\\
            =\E_{\theta\gets [0,2\pi)}[\A^{\ctrl[e^{ i\theta}V]}]
        \end{split}
    \end{equation}
    and so this will be enough to establish the theorem.

    For $f:\{0,1\}\to[t]$, recall that its phase oracle is
    $S^f\ket b=\omega_t^{f(b)}\ket b$.
    Define $f_{\theta_0,\theta_1}:\{0,1\}\to [t]$ by $f_{\theta_0,\theta_1}(b) = \theta_b$. Observe that $\ctrl[V]^{\theta_0,\theta_1} = \ctrl[V]\cdot (I_{\Reg{A}}\otimes S^{f_{\theta_0,\theta_1}}_{\Reg{C}})$. Thus, averaging over $\theta_0,\theta_1$ is equivalent to averaging over $f:\{0,1\}\to[t]$. That is,
    $$\E_{\theta_0,\theta_1\gets [t]}[\A^{\ctrl[V]^{\theta_0,\theta_1}}] = \E_{f:\{0,1\}\to [t]}[\A^{\ctrl[V]\cdot (I \otimes S^f)}]$$

    Let $PO$ be the purified oracle for functions $f:\{0,1\}\to[t]$ acting on $\Reg{C}\otimes \Reg{D}_0\otimes \Reg{D}_1$. Recall that $PO$ is defined by
    $$PO\ket b\ket{c_0,c_1} =\ket b\ket{c_0+(b\xor 1),c_1+b}$$
    Define an intermediate simulator $\Sim^V_{hyb}$ that acts on $\Reg{A}\otimes \Reg{C}\otimes \Reg{D}_0\otimes \Reg{D}_1 \to \Reg{A}\otimes \Reg{B}\otimes \Reg{C}\otimes \Reg{D}_0\otimes \Reg{D}_1$ by
    $$\Sim^V_{hyb} = (\ctrl[V]_{\Reg{C}\Reg{A}\to \Reg{C}\Reg{A}\Reg{B}} \otimes I_{\Reg{D}_0\Reg{D}_1})\cdot (I_{\Reg{A}} \otimes PO_{\Reg{C}\Reg{D}_0\Reg{D}_1})$$
    Then by~\Cref{thm:compext}, we have that
    $$\E_{f:\{0,1\}\to [t]}[\A^{\ctrl[V]\cdot (I\otimes S^f)}] = \A^{\Sim^V_{hyb}}$$
    And so it remains to show that $\A^{\Sim^V_{hyb}} = \A^{\Sim_{ctrl}^V}$.

    By observation, $\Sim_{ctrl}^V$ is identical to the following
    \begin{enumerate}
        \item Apply $\Phi^{\Sim^V_{hyb}}$. On failure, output $\ket{\bot}$ and terminate.
        \item For all $i\in \{1,\dots,T\}$, controlled on the value in $\Reg{C}$ being $0$ and the value in $\Reg{D}_0$ being $i$, store $V\ket{0}$ in $\Reg{E}_i$.
    \end{enumerate}

    In particular, consider the isometry $EXPAND$ defined by
    $$EXPAND\ket{x}_{\Reg{D}_0} \mapsto \ket{x}_{\Reg{D}_0}\otimes V\ket{0}_{\Reg{E}_1} \otimes \dots \otimes V\ket{0}_{\Reg{E}_x}\otimes \ket{0}_{\Reg{E}_{x+1}}\otimes \dots \otimes \ket{0}_{\Reg{E}_{T}}$$

    Then the purified internal state of $\A^{\Sim_{ctrl}^V}$ is equal to the purified internal state of $EXPAND$ applied to $\A^{\Sim^V_{hyb}}$. Since $EXPAND$ only acts on the internal state registers, we get
    $$\A^{\Sim_{ctrl}^V} = \A^{\Sim^V_{hyb}} = \E_{\theta_0,\theta_1}[\A^{\ctrl[V]^{\theta_0,\theta_1}}] = \E_{\theta\gets [0,2\pi)}[\A^{\ctrl[e^{i\theta}V]}]$$
\end{proof}

Combining this with~\Cref{thm:efficiently-implementable-partial-isometry} gives the following useful corollary, allowing us to simulate nicely formulated contractions by only making forward queries to important components.

\begin{corollary}\label{cor:randimp}
    Let $\Pi_1,\Pi_2$ be two projectors, and let $V_1,V_2$ be two partial isometries sending some register $\Reg{A}_{in}$ to $\Reg{A}_{out}$ (with $\ket{0}$ in the support of $V_1$).
    Consider $W_{\theta} = e^{i\theta}\Pi_1 V_1 \Pi_2 + (I - \Pi_1) V_2 (I-\Pi_2)$. Then there is a simulator $\Sim^{(\cdot)}_{t,contract,W}$ making one ordinary query to each of $M^{\Pi_1},M^{\Pi_2},\Phi^{V_1}$ and one query to the controlled oracle $\Phi^{\ctrl[V_2]}$ such that for all oracle algorithms $\A^{(\cdot)}$ making at most $t$ queries,
    $$\E_{\theta\gets [0,2\pi)}[\A^{\Phi^{W_{\theta}}}] = \A^{\Sim_{t,contract,W}^{M^{\Pi_1},M^{\Pi_2},\Phi^{V_1},\Phi^{\ctrl[V_2]}}}$$
\end{corollary}

\begin{proof}
    By~\Cref{thm:efficiently-implementable-partial-isometry}, there is an efficient algorithm implementing $W_{\theta}$ making one forward query to $M^{\Pi_1},M^{\Pi_2}$ and one query to each of the controlled oracles $\Phi^{\ctrl[e^{i\theta}V_1]},\Phi^{\ctrl[V_2]}$. Replacing the controlled query to $\Phi^{\ctrl[e^{i\theta}V_1]}$ with the corresponding simulator given by~\Cref{thm:simctrl} proves the corollary.
\end{proof}

\section{Simulating Unitary Completions of Isometries}\label{sec:positive-result}
\subsection{Construction}

\begin{definition}
    Let $M,N\in\N$ with $1\leq N<M$, and identify $[N]$ with its natural inclusion in $[M]$. Define
    \[
        M_0 \coloneqq [N]
        \qquad\text{and}\qquad
        M_1 \coloneqq [M]\setminus M_0.
    \]
    Thus $[M]=M_0\mathbin{\dot\cup}M_1$, $|M_0|=N$, and $|M_1|=M-N$.

    Define $\Pi_{M_0}$ to be the projector onto $\mathcal{H}(M_0)$ and define $\Pi_{M_1}=I-\Pi_{M_0}$, the projector onto $\mathcal{H}(M_1)$.

    We will also define basis-shifted versions of these projectors. In particular, for any unitary $U$ on $\mathcal{H}([M])$, define $\Pi_{M_0}^U=U\Pi_{M_0}U^\dagger$ and $\Pi_{M_1}^U=U\Pi_{M_1}U^\dagger$.
\end{definition}

\begin{definition}
    Let $M,N\in\N$ with $1\leq N<M$, and let $V$ be an isometry sending $\mathcal{H}([N])$ to $\mathcal{H}([M])$. Using the identification $\mathcal{H}([N])=\mathcal{H}(M_0)\subseteq\mathcal{H}([M])$, a unitary $D^V$ on $\mathcal{H}([M])$ is a completion of $V$ if, for all $x\in[N]$,
    $$D^V\ket{x} = V\ket{x}.$$

    Fix any completion $D^V$ of $V$. For $U_{M_1}\in U(\mathcal{H}(M_1))$, define
    $$D_{U_{M_1}}^V \coloneqq D^V (\Pi_{M_0} + (I_{\mathcal{H}(M_0)} \oplus U_{M_1}) \Pi_{M_1}).$$

    The random completion ensemble for $V$ is obtained by sampling $U_{M_1}\gets\Haar(\mathcal{H}(M_1))$ and outputting $D^V_{U_{M_1}}$.
\end{definition}

\begin{remark}
    Note that the random completion ensemble is independent of the choice of completion $D^V$ up to relabeling. In particular, if $D^V$ and $\wt{D}^V$ are two completions of $V$, then there exists a unitary $R$ on $\mathcal{H}(M_1)$ such that $D^V = \wt{D}^V(I_{\mathcal{H}(M_0)} \oplus R)$. For a unitary $W$ on $\mathcal{H}(M_1)$, write $\wt{D}_W^V\coloneqq\wt{D}^V(I_{\mathcal{H}(M_0)}\oplus W)$. Hence, for every unitary $U_{M_1}$ on $\mathcal{H}(M_1)$,
    \begin{align*}
        D_{U_{M_1}}^V
        &= D^V(I_{\mathcal{H}(M_0)} \oplus U_{M_1}) \\
        &= \wt{D}^V(I_{\mathcal{H}(M_0)} \oplus R)(I_{\mathcal{H}(M_0)} \oplus U_{M_1}) \\
        &= \wt{D}^V(I_{\mathcal{H}(M_0)} \oplus RU_{M_1})
         = \wt{D}_{RU_{M_1}}^V.
    \end{align*}
\end{remark}

\begin{theorem}\label{thm:exact-main}
    Let $T,M,N\in \N$ with $1\leq N<M$ and $T<M$. There is an efficiently implementable, stateful simulator $\Sim^{(\cdot)}$ that makes at most one query to its oracle per invocation and has the following property.
    For every isometry $V$ sending $\mathcal{H}([N])$ to $\mathcal{H}([M])$ and every oracle algorithm $\A^{(\cdot)}$ making at most $T$ forward queries, we have
    $$\TD\left(\E_{U_{M_1}\gets \Haar(\mathcal{H}(M_1))}[\A^{\Phi^{D_{U_{M_1}}^V}}],\A^{\Sim^{\Phi^V}}\right) \leq O\left(T\sqrt{\frac{N}{M-T}}\right).$$
\end{theorem}

\begin{proof}
    The case $T=0$ is immediate. Fix a sufficiently small absolute
    constant $c>0$. If $T\sqrt{N/M}\geq c$, then
    $T\sqrt{N/(M-T)}\geq c$, so any fixed channel is a valid
    simulator with the stated asymptotic error. We therefore assume that
    \begin{equation}\label{eq:exact-small-error}
        T\sqrt{\frac{N}{M}}<c.
    \end{equation}
    In particular, $T<M/2$ and $T<M-N$.

    Fix a completion $D^V$. 
    Use $PR_{[M],T}$ from
    \Cref{def:exact-path-recording}, in ambient dimension $M$, with
    recording register $\Reg{St}=\bigoplus_{t=0}^T\mathcal R_t$,
    initially in $\ket{\varnothing}$. All operations on the register $\mathcal{H}([M])$ will be implicitly tensored with the identity on $\Reg{St}$.

    Let $\theta\gets [0,2\pi)$ be uniformly random. Define the distribution over  contractions
    \[
    W_{f} \coloneqq PR_{[M],T}\Pi_{M_1}
                +e^{i\theta}\left(
                I-PR_{[M],T}
                    (PR_{[M],T})^\dagger
                \right)D^V\Pi_{M_0}
    \]
    $W_f$ will implicitly define our simulator. In particular, our final simulator $\Sim^{\Phi^V}$ will satisfy
    \begin{equation}\label{eq:efficient-hyb}
        \TD\left(\A^{\Sim^{\Phi^V}}, \E_{\theta\gets [0,2\pi)}\left[\A^{W_{f}}\right]\right) \leq T\frac{N}{M}
    \end{equation}

    \paragraph{Efficient implementation.}
    In particular, $W_f$ is a distribution over contractions of the nice form of~\Cref{cor:randimp}. To see this, define
    \begin{align*}
        \Pi_1&=I-PR_{[M],T}
            (PR_{[M],T})^\dagger,
        &\Pi_2&=\Pi_{M_0},\\
        V_1&=D^V\Pi_{M_0},
        &V_2&=PR_{[M],T}.
    \end{align*}
    and observe that $W_f = (I - \Pi_1)V_2(I-\Pi_2) + e^{i\theta}\Pi_1V_1\Pi_2$.
    It is clear that $M^{\Pi_2}$ is efficiently implementable, and $M^{\Pi_1},\Phi^{\ctrl[V_2]}$ are efficiently implementable by
    \Cref{lem:exact-pr-efficient}. Furthermore, $V_1=D^V\Pi_{M_0}$ is implementable by measuring $\{\Pi_{M_0},\Pi_{M_1}\}$, outputting $\bot$ if the result is $\Pi_{M_1}$, and otherwise applying $V$. Setting all appropriate errors to accumulate to $O\left(T\frac{N}{M}\right)$, by~\Cref{cor:randimp} there exists an efficient simulator $\Sim^{\Phi^V}$ making exactly one oracle query per call satisfying~\Cref{eq:efficient-hyb} running in time $\poly(T,\log M)$.

    \paragraph{The hybrid argument.}
    {\emergencystretch=2em Thus, it remains to show that query access to $W_f$ simulates query access to $D^V_{U_{M_1}}$. We will do this by a series of hybrids $W_0,\dots,W_f$ where $\E[\A^{\Phi^{W_0}}] = \E_{U_{M_1}\gets \Haar(M_1)}[\A^{\Phi^{D^V_{U_{M_1}}}}]$.\par}

    As intermediate tools, we will consider two other path recording oracles. $PR_{M_1,T}$ will be a path recording oracle operating on $\mathcal{H}(M_1)$. We will view this as a partial isometry on $\mathcal{H}([M])\otimes \Reg{St}$ by extending the input and state registers via the natural inclusion $\mathcal{H}(M_1) \hookrightarrow \mathcal{H}([M])$.

    We will also consider a path recording oracle which intuitively will simulate a random unitary $\mathcal{H}(M_1) \to \Ima(D^V \Pi_{M_1})$. We will call this map $\widetilde{PR}$. Formally, for $x,y\in M_1$, define the append map with relabeled outputs by
    \[
        \widetilde{\operatorname{App}}^{(t)}_{x,y}
            \bigl(\ket{\vec x}\otimes\overline{\ket{\vec y}}\bigr)
        =\ket{\vec x,x}\otimes
            \bigl(\overline{\ket{\vec y}}\otimes
                \overline{D^V\ket{y}}\bigr).
    \]
    This is essentially the standard append map but with $D^V$ applied to the appended output. Let $\widetilde\Lambda_t$ and $\widetilde\Lambda_t^+$ be the reweighting
    operators from \Cref{def:exact-path-recording} in dimension $M-N$ (i.e. the reweighting operators used by $PR_{M_1,T}$). 
    We then define the relabeled exact recorder on recording length $t<T$ by
    \[
        \widetilde{PR}
        =\sum_{x,y\in M_1}D^V\ketbra{y}{x}\otimes
            \widetilde\Lambda_{t+1}\Omega_{t+1}
            \widetilde{\operatorname{App}}^{(t)}_{x,y}
            \widetilde\Lambda_t^+\Omega_t,
    \]

    We have now developed the necessary tools to define our hybrid oracles. For all of the following, $U_{M_1}$ will be a random variable representing a sample from the Haar distribution over $\mathcal H(M_1)$, and $\theta$ will be a random variable representing a uniformly random value from $[0,2\pi)$. 
    
    Define the following distributions of contractions:
    \begin{align*}
        W_0&=D_{U_{M_1}}^V,\\
        W_1&=D^VU_{M_1}\Pi_{M_1}+e^{i\theta}D^V\Pi_{M_0},\\
        W_2&=D^VPR_{M_1,T}\Pi_{M_1}
                +e^{i\theta}D^V\Pi_{M_0},\\
        W_3&=D^VPR_{M_1,T}\Pi_{M_1}\\
            &\quad+e^{i\theta}\left(
                    I-D^VPR_{M_1,T}\Pi_{M_1}
                        (PR_{M_1,T})^\dagger(D^V)^\dagger
                \right)D^V\Pi_{M_0},\\
        W_4&=\widetilde{PR}\Pi_{M_1}
                +e^{i\theta}\left(
                    I-\widetilde{PR}\Pi_{M_1}\widetilde{PR}^{\dagger}
                \right)D^V\Pi_{M_0},\\
        W_5&=PR_{[M],T}\Pi_{M_1}
                +e^{i\theta}\left(
                I-PR_{[M],T}
                    (PR_{[M],T})^\dagger
                \right)D^V\Pi_{M_0}.
    \end{align*}
    Note that indeed all of these maps are contractions of the form of~\Cref{cor:randimp} where all corresponding maps are efficient given query access to $U_{M_1},D^V$ and the relevant path recording oracle.

    First, $\E[\A^{\Phi^{W_0}}]=\E[\A^{\Phi^{W_1}}]$. Indeed,
    \[
        D^VU_{M_1}\Pi_{M_1}+e^{i\theta}D^V\Pi_{M_0}
        =e^{i\theta}D^V\bigl(\Pi_{M_0}+e^{-i\theta}U_{M_1}\Pi_{M_1}\bigr).
    \]
    Since $e^{-i\theta}U_{M_1}$ is Haar
    distributed and a global phase does not affect an ordinary oracle
    channel, the two experiments agree for every fixed $\theta$.

    Next, $\E[\A^{\Phi^{W_1}}]=\E[\A^{\Phi^{W_2}}]$. For this comparison,
    regard the fixed completion $D^V$ as part of the circuit and apply
    \Cref{thm:efficiently-implementable-partial-isometry} with
    \[
        \Pi_1=\Pi_{M_1}^{D^V},\qquad \Pi_2=\Pi_{M_1},\qquad
        V_1=D^V\cdot U_{M_1},\qquad V_2=e^{i\theta}D^V.
    \]
    This realizes $W_1$ with one controlled query to $U_{M_1}$ on
    $\mathcal H(M_1)$ per invocation. Both projector measurements
    and all remaining operations are independent of $U_{M_1}$ and leave the
    private recording register untouched.
    Apply~\Cref{thm:exact-prsim} in dimension $M-N$ to
    replace these queries by controlled queries to
    $PR_{M_1,T}$. The resulting contraction is exactly the construction of~\Cref{thm:efficiently-implementable-partial-isometry} applied to $W_2$.
    Consequently, for every fixed $\theta$,
    \begin{equation}\label{eq:exact-dilation-recording}
        \E_{U_{M_1}\gets\Haar(\mathcal H(M_1))}
            \bigl[\A^{\Phi^{W_1}}\bigr]
        =\A^{\Phi^{W_2}}.
    \end{equation}

    Next, $W_2=W_3$. This follows immediately from the fact that $PR_{M_1,T}$ commutes with $\Pi_{M_1}$. And so
    \begin{equation}
        \begin{split}
            W_3&=D^VPR_{M_1,T}\Pi_{M_1}\\
            &\quad+e^{i\theta}\left(
                    I-D^VPR_{M_1,T}\Pi_{M_1}
                        (PR_{M_1,T})^\dagger(D^V)^\dagger
                \right)D^V\Pi_{M_0},\\
               &=W_2 - e^{i\theta}D^V PR_{M_1,T}\Pi_{M_1}
                        (PR_{M_1,T})^\dagger(D^V)^\dagger D^V \Pi_{M_0}\\
               &=W_2 - e^{i\theta}D^V PR_{M_1,T}
                        (PR_{M_1,T})^\dagger \Pi_{M_1} \Pi_{M_0}\\
               &=W_2 - 0=W_2.
        \end{split}
    \end{equation}

    Next, $\A^{\Phi^{W_3}}=\A^{\Phi^{W_4}}$ for every fixed $\theta$.
    To see this, define the basis change unitary acting on $\Reg{St}$ by
    \[
        BC^{D^V}
            =\bigoplus_{t=0}^T
                I^{\otimes t}\otimes\overline{D^V}^{\otimes t}.
    \]
    For $x,y\in M_1$, on recording length $t<T$ we have
    \begin{align*}
        &BC^{D^V}\operatorname{App}^{(t)}_{x,y}
            (BC^{D^V})^\dagger\\
        &\quad=
            \left(I^{\otimes(t+1)}
                \otimes\overline{D^V}^{\otimes(t+1)}\right)
            \operatorname{App}^{(t)}_{x,y}
            \left(I^{\otimes t}
                \otimes(\overline{D^V}^{\dagger})^{\otimes t}\right)\\
        &\quad=I_{\mathcal R_t}\otimes\ket x
            \otimes\overline{D^V\ket y}
            =\widetilde{\operatorname{App}}^{(t)}_{x,y}.
    \end{align*}
    Therefore, expanding out $PR_{M_1,T}$ gives, on recording length $t<T$,
    \begin{align*}
        &(I\otimes BC^{D^V})D^VPR_{M_1,T}
            (I\otimes (BC^{D^V})^\dagger)\\
        &\quad=\sum_{x,y\in M_1}D^V\ketbra{y}{x}\otimes
            \widetilde\Lambda_{t+1}\Omega_{t+1}
            \widetilde{\operatorname{App}}^{(t)}_{x,y}
            \widetilde\Lambda_t^+\Omega_t\\
        &\quad=\widetilde{PR}.
    \end{align*}
    
    Note that $BC^{D^V}$ acts only on the state register, and so
    conjugating the two terms of $W_3$ now gives
    \begin{align*}
        &(I\otimes BC^{D^V})W_3
            (I\otimes (BC^{D^V})^\dagger)\\
        &\quad=\widetilde{PR}\Pi_{M_1}
            +e^{i\theta}\left(
                I-\widetilde{PR}\Pi_{M_1}\widetilde{PR}^{\dagger}
            \right)D^V\Pi_{M_0}\\
        &\quad=W_4.
    \end{align*}
    Applying
    \Cref{lem:oracle-isometric-equivalence} with $J=BC^{D^V}$ then gives
    \[
        \A^{\Phi^{W_3}}=\A^{\Phi^{W_4}}
    \]
    for every fixed $\theta$.

    Next, we will show that 
    $$\TD\left(\E[\A^{\Phi^{W_4}}],\E[\A^{\Phi^{W_5}}]\right) = O\left(T\sqrt{\frac{N}{M-T}}\right)$$
    We first give a characterization of the internal path recording state in the hybrid corresponding to $W_4$. For $t\leq T$, define the projector $\Gamma_t$ on $\Reg{St}$ by
    \[
        \Gamma_t
        =\Omega_t\left(
            \Pi_{M_1}^{\otimes t}
                \otimes\left(\overline{\Pi_{M_1}^{D^V}}\right)^{\otimes t}
        \right).
    \]
    The two factors commute, so $\Gamma_t$ is indeed a projector. Define $\Gamma = \bigoplus_{t=0}^{T-1} \Gamma_t$.
    
    To argue this hybrid, we will rely on the following three lemmas. The first shows that after the first $t<T$ applications of $\wt{PR}$ the internal state will remain in $\Gamma$. The second shows that on this subspace, $\wt{PR}$ and $PR_{[M],T}$ are close in operator norm. The final lemma shows that on this subspace, the right-hand terms of $W_4$ and $W_5$ are both equal to $D^V\Pi_{M_0}$ (without the following projector onto states orthogonal to the image of the path recording oracle). We defer the proofs of all these lemmas to~\Cref{sec:exact-pr-comparison}.

    \begin{lemma}\label{lem:exact-pr-support}
        For $0\leq t <T$,
        $$\widetilde{PR}(I\otimes \Gamma_t) = (I\otimes \Gamma_{t+1})\widetilde{PR}(I\otimes \Gamma_t)$$
    \end{lemma}

    \paragraph{Proof sketch} This follows from the fact that $\widetilde{PR}$ intuitively samples $y$ from $M_1$ and applies $D^V$ to it. And so the new output state stored in the recording registers will always lie in the image of $\ol{\Pi_{M_1}^{D^V}}$. Thus, the non-trivial portion of the proof is to show that the reweighting operators $\widetilde\Lambda_t$ and $\widetilde\Lambda_{t+1}$ do not change this property, which follows from the fact that $\bigl(\Pi_{M_1}^{D^V}\bigr)^{\otimes t}$ is permutation invariant.

    \begin{lemma}\label{lem:exact-pr-comparison}
        For $0\leq t<T$,
        \[
            \norm{\bigl(\widetilde{PR}-PR_{[M],T}\bigr)
                (\Pi_{M_1}\otimes\Gamma_t)}_{op}
                \leq 3\sqrt{\frac{2N}{M-t}}.
        \]
        In particular, since $\widetilde{PR}$ and $PR_{[M],T}$ map $\Gamma_i,\Gamma_j$ to orthogonal subspaces for $i\neq j$, we have that 
        \[
            \norm{\bigl(\widetilde{PR}-PR_{[M],T}\bigr)
                (\Pi_{M_1}\otimes\Gamma)}_{op}
                \leq 3\sqrt{\frac{2N}{M-T+1}}.
        \]
    \end{lemma}

    \paragraph{Proof sketch} The key idea here is that $PR_{[M],T}$ is basis invariant. In particular, we can view it as acting on a basis where the first $N$ basis vectors are the image of $D^V\Pi_{M_0}$ and the remaining $M-N$ basis vectors are the image of $D^V\Pi_{M_1}$. In this basis, the primary difference between $\widetilde{PR}$ and $PR_{[M],T}$ is that $\widetilde{PR}$ samples its outputs from the image of $\Pi_{M_1}^{D^V}$ while $PR_{[M],T}$ samples its outputs from the entire space. But the image of $\Pi_{M_1}^{D^V}$ is a very large proportion of the total space, and so the two maps are very close in operator norm. The reweighting operators $\widetilde\Lambda_t$ and $\Lambda_{t}$ are similarly fairly close to each other, and so the two maps are close in operator norm after reweighting as well.

    \begin{lemma}\label{lem:exact-pr-orthogonality}
        For $0\leq t<T$, 
        $$PR_{[M],T}^\dagger D^V (\Pi_{M_0} \otimes \Gamma_t) = \widetilde{PR}^\dagger D^V (\Pi_{M_0} \otimes \Gamma_t) = 0$$
    \end{lemma}

    \paragraph{Proof sketch} Here, the key idea is that both $(\Pi_{M_0}^{D^V} \otimes I) PR_{[M],T}$ and $(\Pi_{M_0}^{D^V} \otimes I)\widetilde{PR}$ will always have the output register be orthogonal to the image of $\Pi_{M_1}^{D^V}$. After conjugating, the same will be true of the internal register storing this output, and so $\Gamma_t$ will always be orthogonal to the image of $(\Pi_{M_0}^{D^V} \otimes I) PR_{[M],T}$ and $(\Pi_{M_0}^{D^V} \otimes I)\widetilde{PR}$. The lemma then follows by the fact that $\Pi_{M_0}^{D^V}D^V = D^V\Pi_{M_0}$ and taking adjoints.

    Together,~\Cref{lem:exact-pr-support,lem:exact-pr-orthogonality} give that after $t < T$ queries to $W_4$ the register on $\Reg{St}$ is in the image of $\Gamma$. Similarly, \Cref{lem:exact-pr-comparison,lem:exact-pr-orthogonality} give that $\norm{(W_4-W_5)(I\otimes \Gamma)}_{op} \leq 3\sqrt{2N/(M-T+1)}$.
    
    And so, by induction we can replace each query to $W_5$ with a query to $W_4$ one by one. Applying~\Cref{lem:contraction-oracle-distance} gives a total error of at most $2\cdot 3\sqrt{2N/(M-T+1)}$ per query replaced. Since there are at most $T$ queries, this gives a total error of at most $6T\sqrt{2N/(M-T+1)}$. And so for each fixed $\theta$,
    $$\TD\left(\A^{\Phi^{W_4}},\A^{\Phi^{W_5}}\right) \leq 6T\sqrt{2N/(M-T+1)}$$

    Since $W_5=W_f$, combining the hybrid bounds with the
    implementation error gives
    \[
        \TD\left(
            \E_{U_{M_1}}[\A^{\Phi^{D_{U_{M_1}}^V}}],
            \A^{\Sim^{\Phi^V}}\right)
        \leq 6T\sqrt{\frac{2N}{M-T+1}}+T\sqrt{\frac NM}
        =O\left(T\sqrt{\frac{N}{M-T}}\right).
    \]
    This proves the stated bound.
\end{proof}

\subsection{Comparing the Exact Path Recording Oracles}\label{sec:exact-pr-comparison}

We use the notation from the proof of \Cref{thm:exact-main}. In particular, we will assume that $T\sqrt{N/M} < c$ for some sufficiently small constant $c$.

\noindent\textbf{\Cref{lem:exact-pr-support} (restated).}
\textit{For $0\leq t<T$,}
\[
    \widetilde{PR}(I\otimes\Gamma_t)
    =(I\otimes\Gamma_{t+1})\widetilde{PR}(I\otimes\Gamma_t).
\]

\begin{proof}
    Let $Test_t = \Pi_{M_1}^{\otimes t}\otimes(\overline{\Pi_{M_1}^{D^V}})^{\otimes t}$ so $\Gamma_t = \Omega_t Test_t$. We will show that $Test_t$ commutes with $\Omega_t,\widetilde\Lambda_t,\widetilde\Lambda_t^+$. We will also show that $\widetilde{\operatorname{App}}^{(t)}_{x,y} Test_t = Test_{t+1}\widetilde{\operatorname{App}}^{(t)}_{x,y} Test_t$ for all $x,y\in M_1$. Since $\Omega_t$ also commutes with $\widetilde\Lambda_t,\widetilde\Lambda_t^+$, this will imply that $(I\otimes \Gamma_{t+1})\widetilde{PR}(I\otimes\Gamma_t) = \widetilde{PR}(I\otimes\Gamma_t)$, as required.

    Note that it is easy to see that $Test_t$ commutes with $\Omega_t$. By Lemma 7.23 of~\cite{FLMNW26}, to show that $Test_t$ commutes with $\widetilde\Lambda_t,\widetilde\Lambda_t^+$, it suffices to show that $P_{\pi}\otimes I$ commutes with this map for all permutations $\pi$. $P_{\pi}$ commutes with all projectors of the form $\Pi^{\otimes t}$, so this is also immediate.
    
    Finally, it remains to be seen that $Test_{t+1}\widetilde{\operatorname{App}}^{(t)}_{x,y} Test_t= \widetilde{\operatorname{App}}^{(t)}_{x,y} Test_t$ for all $x,y\in M_1$. But $\widetilde{\operatorname{App}}^{(t)}_{x,y}$ simply adds an input register $\ket{x}$ and an output register $\overline{D^V\ket{y}}$. Since $x,y\in M_1$, these registers lie in $\Pi_{M_1}\otimes \overline{\Pi_{M_1}^{D^V}}$. Thus, $Test_{t+1}\widetilde{\operatorname{App}}^{(t)}_{x,y} Test_t = \widetilde{\operatorname{App}}^{(t)}_{x,y} Test_t$ and so we are done.
\end{proof}

\noindent\textbf{\Cref{lem:exact-pr-comparison} (restated).}
\textit{For $0\leq t<T$,}
\[
    \norm{\bigl(\widetilde{PR}-PR_{[M],T}\bigr)
        (\Pi_{M_1}\otimes\Gamma_t)}_{op}
        \leq 3\sqrt{\frac{2N}{M-t}}.
\]

\begin{proof}
    Extend $\widetilde{\operatorname{App}}^{(t)}_{x,y}$ to all
    $x,y\in[M]$ by the same formula as before. The full recorder is
    independent of the choice of output basis, so on recording length $t$,
    \begin{equation}\label{eq:exact-pr-relabeled-full}
        PR_{[M],T}
        =\sum_{x,y\in[M]}D^V\ketbra{y}{x}\otimes
            \Lambda_{t+1}\Omega_{t+1}
            \widetilde{\operatorname{App}}^{(t)}_{x,y}
            \Lambda_t^+\Omega_t.
    \end{equation}

    This follows from the fact that $\Lambda_{t+1},\Omega_{t+1},\Lambda_t^+,\Omega_t$ all commute with operations acting only on the input register as well as the fact that for all unitaries $U$,
    $$\sum_{y} \ket{y}\otimes \ol{\ket{y}} = \sum_y U\ket{y} \otimes \ol{U \ket{y}}$$

    In full,
    \begin{align*}
        &\sum_{x,y\in[M]}\ketbra{y}{x}\otimes
            \operatorname{App}^{(t)}_{x,y}\\
        &\quad=\sum_{x,y\in[M]}\ketbra{y}{x}\otimes
            I_{\mathcal R_t}\otimes\ket x\otimes\overline{\ket y}\\
        &\quad=\sum_{x,y\in[M]}D^V\ketbra{y}{x}\otimes
            I_{\mathcal R_t}\otimes\ket x\otimes\overline{D^V\ket y}\\
        &\quad=\sum_{x,y\in[M]}D^V\ketbra{y}{x}\otimes
            \widetilde{\operatorname{App}}^{(t)}_{x,y}.
    \end{align*}

    Recall that
    \begin{equation*}
        \widetilde{PR} = \sum_{x,y\in M_1}D^V\ketbra{y}{x}\otimes
            \widetilde{\Lambda_{t+1}}\Omega_{t+1}
            \widetilde{\operatorname{App}}^{(t)}_{x,y}
            \widetilde{\Lambda_t^+}\Omega_t.
    \end{equation*}

    We will break $PR_{[M],T}$ into two parts: the terms that are common to both $PR_{[M],T}$ and $\widetilde{PR}$, and the additional terms.
    \begin{equation}
        \begin{split}
            PR_{[M],T} =& \sum_{x\in[M],y\in M_0}D^V\ketbra{y}{x}\otimes
            \Lambda_{t+1}\Omega_{t+1}
            \widetilde{\operatorname{App}}^{(t)}_{x,y}
            \Lambda_t^+\Omega_t\\
            &+\sum_{x\in[M],y\in M_1}D^V\ketbra{y}{x}\otimes
            \Lambda_{t+1}\Omega_{t+1}
            \widetilde{\operatorname{App}}^{(t)}_{x,y}
            \Lambda_t^+\Omega_t\\
            =&\Pi_{M_0}^{D^V} PR_{[M],T} + \Pi_{M_1}^{D^V} PR_{[M],T}.
        \end{split}
    \end{equation}

\begin{claim}\label{clm:exact-pr-common-output}
    For $0\leq t<T$,
    \[
        \norm{\Pi_{M_1}^{D^V}\bigl(PR_{[M],T}
            -\widetilde{PR}\bigr)(\Pi_{M_1}\otimes\Gamma_t)}_{op}
            \leq\frac{3(t+1)N}{M-t}.
    \]
\end{claim}

    \begin{proof}[Proof of the claim]
    The visible output of $\widetilde{PR}$ lies in
    $\Ima(\Pi_{M_1}^{D^V})$, so
    $\Pi_{M_1}^{D^V}\widetilde{PR}=\widetilde{PR}$.
    Projecting \eqref{eq:exact-pr-relabeled-full} onto this subspace
    restricts the output sum to $y\in M_1$. Thus, on
    $\mathcal H(M_1)\otimes\Ima(\Gamma_t)$,
    \[
        \Pi_{M_1}^{D^V}PR_{[M],T}
        =(I\otimes\Lambda_{t+1}\widetilde\Lambda_{t+1}^+)
            \Pi_{M_1}^{D^V}\widetilde{PR}
            (I\otimes\widetilde\Lambda_t\Lambda_t^+).
    \]
    On the same input subspace, adding and subtracting
    $(I\otimes\Lambda_{t+1}\widetilde\Lambda_{t+1}^+)
        \Pi_{M_1}^{D^V}\widetilde{PR}$
    gives
    \begin{align*}
        &\Pi_{M_1}^{D^V}\bigl(PR_{[M],T}
            -\widetilde{PR}\bigr)\\
        &\quad=
            \bigl(I\otimes(\Lambda_{t+1}\widetilde\Lambda_{t+1}^+-I)\bigr)
                \Pi_{M_1}^{D^V}\widetilde{PR}\\
        &\qquad+
            (I\otimes\Lambda_{t+1}\widetilde\Lambda_{t+1}^+)
                \Pi_{M_1}^{D^V}\widetilde{PR}
                \bigl(I\otimes(\widetilde\Lambda_t\Lambda_t^+-I)\bigr).
    \end{align*}
    The triangle inequality and submultiplicativity of the operator
    norm give
    \begin{equation}\label{eq:exact-pr-common-output-error}
        \begin{aligned}
        &\norm{\Pi_{M_1}^{D^V}\bigl(PR_{[M],T}
                -\widetilde{PR}\bigr)
            (\Pi_{M_1}\otimes\Gamma_t)}_{op}\\
        &\quad\leq
            \norm{(\Lambda_{t+1}\widetilde\Lambda_{t+1}^+-I)
                \Gamma_{t+1}}_{op}\\
        &\qquad+
            \norm{\Lambda_{t+1}\widetilde\Lambda_{t+1}^+
                \Gamma_{t+1}}_{op}
            \norm{(\widetilde\Lambda_t\Lambda_t^+-I)\Gamma_t}_{op}.
        \end{aligned}
    \end{equation}

    We first bound
    $\norm{(\widetilde\Lambda_t\Lambda_t^+-I)\Gamma_t}_{op}$
    directly from the coefficients. On
    $\mathcal H(M_1)^{\otimes t}$, the projectors $\Pi_{\lambda,t}$
    in the two reweightings agree, since they are defined by the
    same permutations of the input slots. Thus, on $\Ima(\Gamma_t)$,
    \begin{align*}
        \widetilde\Lambda_t\Lambda_t^+
        &=\sum_\lambda
            \sqrt{\frac{f_\lambda}{d_{\lambda,M-N}}}
            \sqrt{\frac{d_{\lambda,M}}{f_\lambda}}
            (\Pi_{\lambda,t}\otimes I)\\
        &=\sum_\lambda
            \sqrt{\frac{d_{\lambda,M}}{d_{\lambda,M-N}}}
            (\Pi_{\lambda,t}\otimes I),
    \end{align*}
    where $\lambda$ ranges over Young diagrams with $t$ boxes and at
    most $M-N$ rows. These projectors are mutually orthogonal and sum to the
    identity on the input slots, so
    \[
        \norm{(\widetilde\Lambda_t\Lambda_t^+-I)\Gamma_t}_{op}
        \leq\max_\lambda
            \left|\sqrt{\frac{d_{\lambda,M}}{d_{\lambda,M-N}}}-1\right|.
    \]
    Applying the dimension formula \eqref{eq:unitary-dimension}
    from \cite[Lemma~3.27]{FLMNW26} in dimensions $M-N$ and $M$,
    and canceling the hook lengths, gives
    \[
        \frac{d_{\lambda,M-N}}{d_{\lambda,M}}
        =\prod_{(i,j)\in \lambda}\frac{M-N+j-i}{M+j-i}
        =\prod_{(i,j)\in \lambda}
            \left(1-\frac{N}{M+j-i}\right).
    \]
    By the union bound,
    \[
        \frac{d_{\lambda,M-N}}{d_{\lambda,M}} \geq \prod_{(i,j)\in \lambda}
            \left(1-\frac{N}{M-t}\right) \geq 1 - \frac{tN}{M-t}
    \]
    and since each product is $\leq 1$, we have
    \[\frac{d_{\lambda,M-N}}{d_{\lambda,M}} \leq 1\]

    Our assumption $T\sqrt{N/M}<c$ implies
    $tN/(M-t)\leq2T^2N/M<2c^2\leq1/2$.
    Hence every coefficient satisfies
    \[
        0\leq\sqrt{\frac{d_{\lambda,M}}{d_{\lambda,M-N}}}-1
        \leq\left(1-\frac{tN}{M-t}\right)^{-1/2}-1
        \leq\frac{2tN}{M-t},
    \]
    where the last inequality follows from
    \[
        (1-u)^{-1/2}-1
        =\frac{u}{\sqrt{1-u}(1+\sqrt{1-u})}\leq2u
        \qquad(0\leq u\leq1/2).
    \]
    Taking the maximum over $\lambda$ gives
    \begin{equation}
        \norm{(\widetilde\Lambda_t\Lambda_t^+-I)\Gamma_t}_{op}
            \leq\frac{2tN}{M-t}.
        \label{eq:exact-pr-relative-reweighting}
    \end{equation}
    For $t=0$, the products are empty and both reweightings are the
    identity, so the same bound holds.

    Analogously, on $\Ima(\Gamma_{t+1})$,
    \begin{align*}
        \Lambda_{t+1}\widetilde\Lambda_{t+1}^+
        &=\sum_\lambda
            \sqrt{\frac{f_\lambda}{d_{\lambda,M}}}
            \sqrt{\frac{d_{\lambda,M-N}}{f_\lambda}}
            (\Pi_{\lambda,t+1}\otimes I)\\
        &=\sum_\lambda
            \sqrt{\frac{d_{\lambda,M-N}}{d_{\lambda,M}}}
            (\Pi_{\lambda,t+1}\otimes I),
    \end{align*}
    where $\lambda$ ranges over Young diagrams with $t+1$ boxes and
    at most $M-N$ rows. For these diagrams, $M+j-i\geq M-t$, so
    the same product bound gives
    \[
        1-\frac{(t+1)N}{M-t}
        \leq\frac{d_{\lambda,M-N}}{d_{\lambda,M}}\leq1.
    \]
    Since $(t+1)N/(M-t)\leq2T^2N/M<2c^2\leq1/2$, it follows that
    \begin{equation}\label{eq:exact-pr-output-reweighting}
        \begin{aligned}
            &\norm{(\Lambda_{t+1}\widetilde\Lambda_{t+1}^+-I)
                \Gamma_{t+1}}_{op}\\
            &\quad\leq\max_\lambda
                \left(1-\sqrt{\frac{d_{\lambda,M-N}}{d_{\lambda,M}}}\right)\\
            &\quad\leq1-\sqrt{1-\frac{(t+1)N}{M-t}}
                \leq\frac{(t+1)N}{M-t}.
        \end{aligned}
    \end{equation}
    Here we used $1-\sqrt{1-u}\leq u$ for $0\leq u\leq1$.
    This calculation also shows that all coefficients of
    $\Lambda_{t+1}\widetilde\Lambda_{t+1}^+$ on
    $\Ima(\Gamma_{t+1})$ lie in $[0,1]$, so
    \[
        \norm{\Lambda_{t+1}\widetilde\Lambda_{t+1}^+
            \Gamma_{t+1}}_{op}\leq1.
    \]
    Substituting this bound,
    \eqref{eq:exact-pr-relative-reweighting}, and
    \eqref{eq:exact-pr-output-reweighting} into
    \eqref{eq:exact-pr-common-output-error} gives
    \[
        \norm{\Pi_{M_1}^{D^V}\bigl(PR_{[M],T}
                -\widetilde{PR}\bigr)(\Pi_{M_1}\otimes\Gamma_t)}_{op}
        \leq\frac{(t+1)N}{M-t}+\frac{2tN}{M-t}
        \leq\frac{3(t+1)N}{M-t},
    \]
    as claimed.
\end{proof}

\begin{claim}\label{clm:exact-pr-diff-output}
    For $0\leq t<T$,
    \[
        \norm{\Pi_{M_0}^{D^V}\bigl(PR_{[M],T}
            -\widetilde{PR}\bigr)(\Pi_{M_1}\otimes\Gamma_t)}_{op}
            \leq3\sqrt{\frac{N}{M-t}}.
    \]
\end{claim}

\begin{proof}
    Since $\Pi_{M_0}^{D^V} \widetilde{PR} = 0$, it suffices to bound 
    \[
        \norm{\Pi_{M_0}^{D^V}PR_{[M],T}(\Pi_{M_1}\otimes\Gamma_t)}_{op}
    \]

    It remains to bound the additional terms with $y\in M_0$.
    For every unit vector $\ket\psi$ in the input subspace, we will show
    \[
        \norm{(I\otimes\Omega_{t+1})
            \left(\sum_{\substack{x\in M_1\\y\in M_0}}
                D^V\ketbra{y}{x}\otimes
                    \widetilde{\operatorname{App}}^{(t)}_{x,y}
            \right)\ket\psi}^2
        =\frac{N}{t+1}.
    \]
    To see this algebraically, let $x,x'\in M_1$, $y,y'\in M_0$,
    and $\pi\in S_{t+1}$. We will first show
    \begin{equation}\label{eq:slots-delta}
        \Gamma_t
        \bigl(\widetilde{\operatorname{App}}^{(t)}_{x',y'}\bigr)^\dagger
        (P_\pi\otimes\overline{P_\pi})
        \widetilde{\operatorname{App}}^{(t)}_{x,y}\Gamma_t
        =
        \begin{cases}
            \delta_{x,x'}\delta_{y,y'}\Gamma_t,
                &\pi(t+1)=t+1,\\
            0,  &\pi(t+1)\ne t+1.
        \end{cases}
    \end{equation}
    Indeed, if $\pi(t+1)=t+1$, then $P_{\pi} \otimes \overline{P_\pi}$ commutes with $\widetilde{\operatorname{App}}^{(t)}_{x',y'}$. We get
    \[
        (P_\pi\otimes\overline{P_\pi})\widetilde{\operatorname{App}}^{(t)}_{x',y'}\Gamma_t=\widetilde{\operatorname{App}}^{(t)}_{x',y'}(P_\pi\otimes\overline{P_\pi})\Gamma_t=\widetilde{\operatorname{App}}^{(t)}_{x',y'}\Gamma_t
    \]
    Note that this is a slight abuse of notation, $P_{\pi}$ and $\overline{P_\pi}$ in the second term are operators on $\Gamma_t$ applying the permutation $\pi$ restricted to its first $t$ terms.
    Combining this with the fact that
    \[
        \bigl(\widetilde{\operatorname{App}}^{(t)}_{x',y'}\bigr)^\dagger
        \widetilde{\operatorname{App}}^{(t)}_{x,y}
            =\delta_{x,x'}\delta_{y,y'}I
    \]
    we get
    \[
        \Gamma_t
        \bigl(\widetilde{\operatorname{App}}^{(t)}_{x',y'}\bigr)^\dagger
        (P_\pi\otimes\overline{P_\pi})
        \widetilde{\operatorname{App}}^{(t)}_{x,y}\Gamma_t
        =\delta_{x,x'}\delta_{y,y'}\Gamma_t
    \]
    If $\pi(t+1)\neq t+1$, let $i\in[t]$ be such that $\pi(i) = t+1$. Recall that
    $\Gamma_t=\Omega_t(\Pi_{M_1}^{\otimes t}\otimes
        (\overline{\Pi_{M_1}^{D^V}})^{\otimes t}) = (\Pi_{M_1}^{\otimes t}\otimes
        (\overline{\Pi_{M_1}^{D^V}})^{\otimes t})\Omega_t$,
    so every output register of states in $\Ima(\Gamma_t)$ lies in $\Ima(\overline{\Pi_{M_1}^{D^V}})$. Furthermore, after applying $\wt{\operatorname{App}^{(t)}}$ to a state in $\Ima(\Gamma_t)$, the last output register lies in $\Ima(\overline{\Pi_{M_0}^{D^V}})$ and the other output registers remain in $\Ima(\overline{\Pi_{M_1}^{D^V}})$. Write $(\Pi)_{(j)}$ for the projector $\Pi$ applied to the $j$th output register, tensored with the identity on all
    other registers. Formally, we have
    \begin{align*}
        \bigl(\overline{\Pi_{M_0}^{D^V}}\bigr)_{(t+1)}
            \widetilde{\operatorname{App}}^{(t)}_{x,y}\Gamma_t
        &=\widetilde{\operatorname{App}}^{(t)}_{x,y}\Gamma_t,\\
        \bigl(\overline{\Pi_{M_1}^{D^V}}\bigr)_{(i)}
            \widetilde{\operatorname{App}}^{(t)}_{x,y}\Gamma_t
        &=\widetilde{\operatorname{App}}^{(t)}_{x,y}\Gamma_t,
    \end{align*}
    and so
    \[
    \begin{aligned}
        &\Gamma_t
        \bigl(\widetilde{\operatorname{App}}^{(t)}_{x',y'}\bigr)^\dagger
        (P_\pi\otimes\overline{P_\pi})
        \widetilde{\operatorname{App}}^{(t)}_{x,y}\Gamma_t\\
        =&\Gamma_t
        \bigl(\widetilde{\operatorname{App}}^{(t)}_{x',y'}\bigr)^\dagger
        \bigl(\overline{\Pi_{M_0}^{D^V}}\bigr)_{(t+1)}
        (P_\pi\otimes\overline{P_\pi})
        \bigl(\overline{\Pi_{M_1}^{D^V}}\bigr)_{(i)}
        \widetilde{\operatorname{App}}^{(t)}_{x,y}\Gamma_t\\
        =&\Gamma_t
        \bigl(\widetilde{\operatorname{App}}^{(t)}_{x',y'}\bigr)^\dagger
        \bigl(\overline{\Pi_{M_0}^{D^V}}\bigr)_{(t+1)}
        \bigl(\overline{\Pi_{M_1}^{D^V}}\bigr)_{(t+1)}
        (P_\pi\otimes\overline{P_\pi})
        \widetilde{\operatorname{App}}^{(t)}_{x,y}\Gamma_t
        =0
    \end{aligned}
    \]
    Here
    $\overline{\Pi_{M_0}^{D^V}}\,\overline{\Pi_{M_1}^{D^V}}
        =\overline{D^V(\Pi_{M_0}\Pi_{M_1})(D^V)^\dagger}=0$.

    There are $t!$ permutations fixing the new slot, so applying~\Cref{eq:slots-delta} and averaging gives
    \[
        \Gamma_t
        \bigl(\widetilde{\operatorname{App}}^{(t)}_{x',y'}\bigr)^\dagger
        \Omega_{t+1}
        \widetilde{\operatorname{App}}^{(t)}_{x,y}\Gamma_t
        =\frac{\delta_{x,x'}\delta_{y,y'}}{t+1}\Gamma_t.
    \]
    Since $\Omega_{t+1}^2=\Omega_{t+1}$ and
    $\bra{y'}(D^V)^\dagger D^V\ket y=\delta_{y,y'}$, it follows that
    \begin{align*}
        &\norm{(I\otimes\Omega_{t+1})
            \left(\sum_{\substack{x\in M_1\\y\in M_0}}
                D^V\ketbra{y}{x}\otimes
                    \widetilde{\operatorname{App}}^{(t)}_{x,y}
            \right)\ket\psi}^2\\
        &\quad=\sum_{\substack{x,x'\in M_1\\y\in M_0}}
            \bra\psi\left(\ketbra{x'}{x}\otimes
                \Gamma_t
                \bigl(\widetilde{\operatorname{App}}^{(t)}_{x',y}\bigr)^\dagger
                \Omega_{t+1}
                \widetilde{\operatorname{App}}^{(t)}_{x,y}
                \Gamma_t\right)\ket\psi\\
        &\quad=\frac{1}{t+1}
            \sum_{\substack{x\in M_1\\y\in M_0}}
                \bra\psi(\ketbra{x}\otimes\Gamma_t)\ket\psi\\
        &\quad=\frac{N}{t+1}
            \bra\psi(\Pi_{M_1}\otimes\Gamma_t)\ket\psi
            =\frac{N}{t+1}.
    \end{align*}

    By \cite[Lemma~6.5]{FLMNW26}, $\Lambda_s$ differs from
    $\binom Ms^{-1/2}I$ by a relative error $O(s^2/M)$.
    Since $T^2/M\leq T^2N/M<c^2$, choosing $c$ sufficiently small
    gives, for $0\leq s\leq T$,
    \begin{equation}\label{eq:exact-pr-scalar-reweighting}
        \norm{\Lambda_s}_{op}\leq\frac32\binom Ms^{-1/2},
        \qquad
        \norm{\Lambda_s^+}_{op}\leq2\binom Ms^{1/2}.
    \end{equation}
    The initial reweighting $\Lambda_t^+$ preserves
    $\Ima(\Gamma_t)$, so applying the preceding norm identity after
    this reweighting and then bounding the final reweighting gives
    \begin{align*}
        &\norm{\Pi_{M_0}^{D^V}PR_{[M],T}
            (\Pi_{M_1}\otimes\Gamma_t)}_{op}\\
        &\quad\leq\norm{\Lambda_{t+1}}_{op}
            \sqrt{\frac{N}{t+1}}\,
            \norm{\Lambda_t^+}_{op}\\
        &\quad\leq3\sqrt{
            \frac{\binom Mt}{\binom M{t+1}}\frac{N}{t+1}}
            =3\sqrt{\frac{N}{M-t}},
    \end{align*}
    as claimed.
\end{proof}

    Finally, $\Pi_{M_0}^{D^V}$ and $\Pi_{M_1}^{D^V}$ are orthogonal
    and sum to the identity. Therefore
    \Cref{clm:exact-pr-common-output,clm:exact-pr-diff-output} give
    \begin{align*}
        &\norm{\bigl(\widetilde{PR}-PR_{[M],T}\bigr)
            (\Pi_{M_1}\otimes\Gamma_t)}_{op}^2\\
        &\quad\leq
            \norm{\Pi_{M_0}^{D^V}
                \bigl(\widetilde{PR}-PR_{[M],T}\bigr)
                (\Pi_{M_1}\otimes\Gamma_t)}_{op}^2\\
        &\qquad+
            \norm{\Pi_{M_1}^{D^V}
                \bigl(\widetilde{PR}-PR_{[M],T}\bigr)
                (\Pi_{M_1}\otimes\Gamma_t)}_{op}^2\\
        &\quad\leq
            \frac{9(t+1)^2N^2}{(M-t)^2}+\frac{9N}{M-t}
            \leq\frac{18N}{M-t}.
    \end{align*}
    For the last inequality, we used $t+1\leq T$, $T<M/2$, and
    \[
        \frac{(t+1)^2N}{M-t}
            \leq\frac{2T^2N}{M}<2c^2\leq1.
    \]
    Taking square roots proves the lemma.
\end{proof}

\noindent\textbf{\Cref{lem:exact-pr-orthogonality} (restated).}
\textit{For $0\leq t<T$,}
\[
    PR_{[M],T}^\dagger D^V(\Pi_{M_0}\otimes\Gamma_t)
    =\widetilde{PR}^\dagger D^V(\Pi_{M_0}\otimes\Gamma_t)=0.
\]

\begin{proof}
    The case $t=0$ is trivial, since $\widetilde{PR}^\dagger (I\otimes \Gamma_0) = PR_{[M],T}^\dagger (I\otimes \Gamma_0) = 0$.

    We recall $\widetilde{PR}$ and the expression $PR_{[M],T}$ under the basis corresponding to $D^V$ from the proof of \Cref{lem:exact-pr-comparison}.
    \begin{align}
        \widetilde{PR} = \sum_{x,y\in M_1}D^V\ketbra{y}{x}\otimes
            \widetilde{\Lambda_{t+1}}\Omega_{t+1}
            \widetilde{\operatorname{App}}^{(t)}_{x,y}
            \widetilde{\Lambda_t^+}\Omega_t,\\
        PR_{[M],T} = \sum_{x,y\in[M]}D^V\ketbra{y}{x}\otimes
            \Lambda_{t+1}\Omega_{t+1}
            \widetilde{\operatorname{App}}^{(t)}_{x,y}
            \Lambda_t^+\Omega_t.
    \end{align}
    
    Note that since $\Pi^{D^V}_{M_0}\wt{PR}=0$, we also have $\wt{PR}^\dagger D^V \Pi_{M_0} = 0$. We thus focus on the more complicated case of $PR_{[M],T}^\dagger D^V (\Pi_{M_0} \otimes \Gamma_t)$.

    As in the previous lemma, we divide $PR_{[M],T} = \Pi_{M_0}^{D^V}PR_{[M],T} + \Pi_{M_1}^{D^V}PR_{[M],T}$. Since $\Pi_{M_0}^{D^V}\Pi_{M_1}^{D^V}=0$ it suffices to only consider the first term.

    Let $(\overline{\Pi_{M_b}^{D^V}})_{(t)}$ be the projector $\overline{\Pi_{M_b}^{D^V}}$ applied to the
    final register of $\Reg{St}$ containing the $t$th output and tensored with the
    identity on all other registers. Our argument will show that 
    \begin{equation}\label{eq:exact-pr-orthogonality-output}
        \Lambda_t\Gamma_t\Omega_t = \Lambda_t\Gamma_t\Omega_t(\overline{\Pi_{M_1}^{D^V}})_{(t)}
    \end{equation}
    and that for all $x\in[M]$ and $y\in M_0$, we have
    \begin{equation}\label{eq:exact-pr-orthogonality-append}
        \bigl(\overline{\Pi_{M_0}^{D^V}}\bigr)_{(t)}
            \widetilde{\operatorname{App}}^{(t-1)}_{x,y}
        =\widetilde{\operatorname{App}}^{(t-1)}_{x,y}.
    \end{equation}
    Consequently, for all input states supported on $\mathcal{R}_{t-1}$, we have
    \begin{align*}
        (I\otimes\Gamma_t)\Pi_{M_0}^{D^V}PR_{[M],T}\\
        =\sum_{\substack{x\in[M]\\y\in M_0}}
            D^V\ketbra{y}{x}\otimes
            \Gamma_t\Lambda_t\Omega_t
            \widetilde{\operatorname{App}}^{(t-1)}_{x,y}
            \Lambda_{t-1}^+\Omega_{t-1}\\
        =\sum_{\substack{x\in[M]\\y\in M_0}}
            D^V\ketbra{y}{x}\otimes
            \Gamma_t\Lambda_t\Omega_t
            \bigl(\overline{\Pi_{M_1}^{D^V}}\bigr)_{(t)}
            \bigl(\overline{\Pi_{M_0}^{D^V}}\bigr)_{(t)}
            \widetilde{\operatorname{App}}^{(t-1)}_{x,y}
            \Lambda_{t-1}^+\Omega_{t-1}\\
        =0.
    \end{align*}
    and so by taking the adjoint we get
    $$PR_{[M],T}^\dagger D^V(\Pi_{M_0}\otimes\Gamma_t)
        =PR_{[M],T}^\dagger\Pi_{M_0}^{D^V}
            (I\otimes\Gamma_t)D^V\Pi_{M_0}=0,$$
    proving the lemma.

    To see~\Cref{eq:exact-pr-orthogonality-output} formally, define $Proj_t = I^{\otimes t}\otimes(\overline{\Pi_{M_1}^{D^V}})^{\otimes t}$, the projector onto the output registers of the path recording internal state containing only states in $\Ima(\overline{\Pi_{M_1}^{D^V}})$. This commutes with $\Omega_t$ since it is invariant under permutations on either the input or output registers of the path recording internal state. It also commutes with $\Lambda_t$ since they act on disjoint registers. Consequently, we have
    \begin{equation*}
        \Lambda_t\Gamma_t\Omega_t\\
        =\Lambda_tProj_t\Gamma_t\Omega_t\\
        =\Lambda_t\Gamma_t\Omega_tProj_t\\
        =\Lambda_t\Gamma_t\Omega_tProj_t(\overline{\Pi_{M_1}^{D^V}})_{(t)}\\
        =\Lambda_t\Gamma_t\Omega_t(\overline{\Pi_{M_1}^{D^V}})_{(t)}.
    \end{equation*}
    
    To see~\Cref{eq:exact-pr-orthogonality-append} formally, we observe that for $y\in M_0$
    \begin{align*}
        &\bigl(\overline{\Pi_{M_0}^{D^V}}\bigr)_{(t)}
            \widetilde{\operatorname{App}}^{(t-1)}_{x,y}\\
        &\quad=
            \bigl(I^{\otimes(t-1)}\otimes\ket x\bigr)\otimes
            \bigl(I^{\otimes(t-1)}\otimes
                \overline{\Pi_{M_0}^{D^V}}\,
                \overline{D^V\ket y}\bigr)\\
        &\quad=
            \bigl(I^{\otimes(t-1)}\otimes\ket x\bigr)\otimes
            \bigl(I^{\otimes(t-1)}\otimes
                \overline{D^V\Pi_{M_0}(D^V)^\dagger D^V\ket y}\bigr)\\
        &\quad=
            \bigl(I^{\otimes(t-1)}\otimes\ket x\bigr)\otimes
            \bigl(I^{\otimes(t-1)}\otimes
                \overline{D^V\Pi_{M_0}\ket y}\bigr)\\
        &\quad=
            \bigl(I^{\otimes(t-1)}\otimes\ket x\bigr)\otimes
            \bigl(I^{\otimes(t-1)}\otimes
                \overline{D^V\ket y}\bigr)\\
        &\quad=\widetilde{\operatorname{App}}^{(t-1)}_{x,y}.
    \end{align*}
\end{proof}

\section{Applications of purifying random isometries}\label{sec:applications}

\subsection{Learning isometry channels}

Our simulator transfers lower bounds for forward queries to an isometry to lower bounds for forward queries to its unitary completion. The bounds hold already for Haar-random completions, and hence for algorithms required to succeed for every compatible completion. Queries are uncontrolled and forward only.

More precisely, let $V:\mathcal{H}([N])\to\mathcal{H}([M])$, where $N<M$. A $q$-query algorithm given a unitary completion of $V$ can be simulated using at most $q$ queries to $V$, with output trace-distance error
\[
    O\left(q\sqrt{\frac{N}{M-q}}\right)
\]
for $q<M$, by~\Cref{thm:exact-main}. This gives the following lower-bound transformation.

\begin{corollary}[Lifting isometry query lower bounds]\label{thm:isometry-lifting-lb}
    Let $N<M$, and consider a task involving an unknown isometry $V:\mathcal{H}([N])\to\mathcal{H}([M])$. Suppose solving the task requires $\Omega(L)$ forward queries to $V$, and this lower bound remains valid under a sufficiently small constant loss in the success guarantee. Then solving the same task using uncontrolled forward queries to a Haar-random unitary completion of $V$ requires
    \[
        \Omega\left(\min\left\{L,\sqrt{M/N}\right\}\right)
    \]
    queries. The same lower bound holds for algorithms required to succeed for every compatible completion.
\end{corollary}

\begin{proof}
Let $q$ be the number of completion queries. If $q$ exceeds a sufficiently small constant times $\sqrt{M/N}$, the dimension-dependent lower bound already holds. Otherwise,~\Cref{thm:exact-main} simulates the algorithm using at most $q$ isometry queries with a sufficiently small constant loss in its success guarantee. The assumed isometry-query lower bound then gives $q=\Omega(L)$.
\end{proof}

To see some examples of this corollary, we observe that lower bounds for isometry tomography and isometry inversion extend to the setting where the adversary can query an arbitrary completion.

\begin{theorem}[Isometry tomography, Theorem~1.1 of~\cite{CZY26tomography}]\label{thm:isometry-tomography-lb}
    Let $M\geq2N$ and $0<\epsilon\leq\epsilon_0$ for a sufficiently small universal constant $\epsilon_0>0$. Given forward query access to an unknown isometry $V:\mathcal{H}([N])\to\mathcal{H}([M])$, outputting a classical description of a channel $\widehat\Phi$ satisfying
    \[
        \|\widehat\Phi-\Phi^V\|_\diamond\leq\epsilon,
        \qquad \Phi^V(\rho)=V\rho V^\dagger,
    \]
    with probability at least $2/3$ requires $\Omega(NM/\epsilon^2)$ queries, even allowing adaptive coherent queries.
\end{theorem}

\begin{corollary}[Tomography from a unitary completion]\label{cor:isometry-tomography-completion-lb}
    Let $M\geq2N$ and $0<\epsilon\leq\epsilon_0$ for a sufficiently small universal constant $\epsilon_0>0$. Tomography of an unknown isometry $V:\mathcal{H}([N])\to\mathcal{H}([M])$ to diamond norm error $\epsilon$, with success probability at least $2/3$, requires
    \[
        \Omega\left(\min\left\{\frac{NM}{\epsilon^2},\sqrt{M/N}\right\}\right) = \Omega(\sqrt{M/N})
    \]
    forward queries to a Haar-random unitary completion of $V$. The same lower bound holds for algorithms required to succeed for every compatible completion.
\end{corollary}

\begin{theorem}[Isometry inversion, Corollary~11 of~\cite{YSM25adjointation} and Theorem~3.7 of~\cite{CYZ26inversion}]\label{thm:isometry-inversion-lb}
    Let $2\leq N<M$. Given forward query access to an unknown isometry $V:\mathcal{H}([N])\to\mathcal{H}([M])$, implementing a decoder $\mathcal R^V$ satisfying
    \[
        \|\mathcal R^V\circ\Phi^V-\mathrm{id}_{\mathcal{H}([N])}\|_\diamond\leq1/10
    \]
    requires $\Omega(N^2)$ queries.
\end{theorem}

\begin{corollary}[Inversion from a unitary completion]\label{cor:isometry-inversion-completion-lb}
    Let $2\leq N<M$. Implementing a decoder for an unknown isometry $V:\mathcal{H}([N])\to\mathcal{H}([M])$ to diamond norm error at most $1/10$ requires
    \[
        \Omega\left(\min\left\{N^2,\sqrt{M/N}\right\}\right)
    \]
    forward queries to a Haar-random unitary completion of $V$, where the error is measured for the decoder averaged over the completion. The same lower bound holds for algorithms required to decode for every compatible completion. In contrast, one inverse query to the completion decodes exactly.
\end{corollary}

\begin{proof}
The lower bound in~\Cref{thm:isometry-inversion-lb} also holds
with error threshold $1/5$, by Theorem~3.7
of~\cite{CYZ26inversion}.
For $q\leq c\sqrt{M/N}$, \Cref{thm:exact-main},
applied uniformly over inputs entangled with a reference,
gives diamond-norm simulation error $O(c)$.
Choosing $c>0$ sufficiently small increases the decoding
error from $1/10$ to at most $1/5$, so
$q=\Omega(N^2)$.
\end{proof}

Since isometry inversion can be solved exactly with $O(N^2)$ queries~\cite{CMLZW24inversion,YSM25adjointation}, the bound for isometry inversion is tight in the case when $M = \Omega(N^5)$.

\subsection{State preparation problems}

\begin{definition}
    A state preparation unitary for an unknown mixed quantum state $\rho$ is a unitary $U^\rho$ such that $U^\rho\ket{0} = \ket{\psi_\rho}$ for some state satisfying $\rho = \Tr_{\Reg{B}}(\ketbra{\psi_\rho}_{\Reg{A}\Reg{B}})$. We say that an algorithm has purified forward quantum query access to $\rho$ if it has uncontrolled forward oracle access to $U^\rho$.
\end{definition}

In particular, let $\ket{\psi}$ be any pure state, and let $V^{\ket{\psi}}$ be the state preparation isometry $\ket{0}\mapsto \ket{\psi}$. Then, we can implement any adversary making queries to random completions $D^{V^{\ket{\psi}}}$ of $V^{\ket{\psi}}$ by replacing these queries with our simulator applied to $V^{\ket{\psi}}$. But note that each completion $D^{V^{\ket{\psi}}}$ is indeed a state preparation unitary for $\ketbra{\psi}$. Since $V^{\ket{\psi}}$ can be implemented efficiently using copies of $\ket{\psi}$ (by simply swapping in a fresh copy of $\ket{\psi}$ whenever a query is made), we immediately get the following corollary:

\begin{corollary}\label{cor:purestateprep}
    Let $t<M$, $0<\epsilon<1$, and let $\A^{(\cdot)}$ be an oracle circuit which makes at most $t$ uncontrolled forward queries to a state preparation unitary acting on $\mathcal{H}([M])$. There exists a circuit $\B$ of size $\poly(|\A|,\log(M),\log(1/\epsilon))$ taking in $t$ copies of a pure state such that for every pure state $\ket{\psi}\in\mathcal{H}([M])$, there exists a distribution $\D$ over state preparation unitaries for $\ketbra{\psi}$ satisfying
    $$\TD\left(\E_{U^{\ket{\psi}} \gets \D}\left[\A^{U^{\ket{\psi}}}\right], \B(\ketbra{\psi}^{\otimes t})\right) \leq \epsilon + O\left(\frac{t}{\sqrt{M - t}}\right)$$
\end{corollary}

Combining the pure-state simulation of~\Cref{cor:purestateprep}, in total dimension $N\cdot D$, with the random purification channel~\cite{TWZ26,GML26} allows us to extend this argument to state preparation unitaries for arbitrary mixed states.

The random purification channel satisfies the following property. For an $N$-dimensional mixed state and a purification register of dimension $D\geq N$, this channel transforms $t$ copies of $\rho$ into the average of $t$ copies of the same Haar-random purification. Its efficient implementation gives the following simulation principle.

\begin{lemma}[Replacing mixed states with pure states, \cite{TWZ26}]\label{lem:randompurificationlb}
    Let $D\geq N$, $0<\epsilon<1$, and let $\A$ be a circuit taking in at most $t$ copies of a pure state on $\mathcal{H}([N])\otimes\mathcal{H}([D])$. Then there exists a circuit $\B$ of size $\poly(|\A|,\log(N),\log(D),\log(1/\epsilon))$ such that for every mixed state $\rho$ on $\mathcal{H}([N])$, there exists a distribution $\D^\rho$ over purifications $\ket{\psi_\rho}$ on $\mathcal{H}([N])\otimes\mathcal{H}([D])$ satisfying
    $$\TD\left(\E_{\ket{\psi_\rho} \gets \D^\rho}\left[\A(\ketbra{\psi_\rho}^{\otimes t})\right], \B(\rho^{\otimes t})\right) \leq \epsilon.$$
    The distribution $\D^\rho$ is obtained by applying a Haar-random unitary to the purifying register of any fixed purification of $\rho$.
\end{lemma}

By applying~\Cref{cor:purestateprep} to a random purification of an arbitrary mixed state, this gives the following:

\begin{theorem}\label{thm:perfectlb}
    Let $D\geq N$, $t<N\cdot D$, and $\epsilon>0$. Let $\A^{(\cdot)}$ be an oracle circuit which makes at most $t$ uncontrolled forward queries to a unitary acting on $\mathcal{H}([N])\otimes \mathcal{H}([D])$. There exists a circuit $\B$ of size $\poly(|\A|,\log(N),\log(D),\log(1/\epsilon))$, independent of $\rho$ and taking exactly $t$ copies of $\rho$, such that for every $N$-dimensional mixed state $\rho$, there exists a distribution $\D^\rho$ over state preparation unitaries satisfying
    $$\TD\left(\E_{U^\rho\gets \D^\rho}[\A^{U^\rho}], \B(\rho^{\otimes t})\right) \leq \epsilon + O\left(\frac{t}{\sqrt{N\cdot D - t}}\right)$$

    The distribution $\D^\rho$ first chooses a Haar-random purification $\ket{\psi_\rho}$ on the specified registers, and then chooses a Haar-random unitary completion of $\ket{0}\mapsto\ket{\psi_\rho}$.
\end{theorem}

\begin{proof}
Apply the random purification channel to $\rho^{\otimes t}$, to trace-distance accuracy $\epsilon/2$. Conditional on its random purification, run the simulator of~\Cref{thm:exact-main} for the one-dimensional-input isometry preparing that purification, replacing each isometry query with a fresh copy. The same purification is used throughout all $t$ calls. The triangle inequality and contractivity of trace distance give the stated error, allocating the remaining $\epsilon/2$ to implementation accuracy. Both circuits are efficient and independent of $\rho$.
\end{proof}

Now, just as our simulation of random completions allows us to lift isometry query bounds to completion query bounds, a similar technique allows us to lift copy complexity bounds to state preparation query bounds.

\begin{corollary}[Lifting copy complexity lower bounds]\label{cor:state-lifting-lb}
    Let $D\geq N$, and consider a decision task involving an unknown $N$-dimensional state $\rho$. Suppose solving the task with success probability at least $2/3$ requires $\Omega(L)$ copies of $\rho$. Then solving the same task with success probability at least $2/3$ using uncontrolled forward queries to a state preparation unitary for $\rho$, with a purification register of dimension $D$, requires
    \[
        \Omega\left(\min\left\{L,\sqrt{N\cdot D}\right\}\right)
    \]
    queries, for algorithms required to succeed for every compatible preparation unitary and every purification. For pure-state problems, the same bound holds for every $D\geq1$. In particular, an empty purification register has dimension $D=1$. Conversely, any algorithm using $t$ copies of $\rho$ can be implemented using $t$ forward preparation queries.
\end{corollary}

\begin{proof}
To prove the lower bound, amplify a query algorithm's success to $0.99$ with a constant number of repetitions. If its query count is at most a sufficiently small constant times $\sqrt{N\cdot D}$, the simulation of~\Cref{thm:perfectlb} loses at most $0.01$ in success probability and uses only a constant factor more copies. Otherwise, the query count already satisfies the dimension-dependent term. For pure-state problems, apply~\Cref{cor:purestateprep} to $\ket{\psi}\otimes\ket{0}$ in total dimension $N\cdot D$. For the upper bound, query on fresh zero registers and discard the purifying registers.
\end{proof}

The same reduction applies to estimation and learning tasks whose copy-complexity lower bounds remain valid under a sufficiently small constant loss in success probability. Since the simulation is efficient, it also preserves computational restrictions, up to polynomial overhead. Compared with the forward-and-inverse lifting theorem of~\cite{TWZ26} and its property-testing applications~\cite{CWZ25}, this avoids a quadratic loss in copies, with the dimension-dependent term shown explicitly in~\Cref{cor:state-lifting-lb}.

We observe that for many applications, one wants an algorithm that works regardless of the dimension of the ambient space. In such cases, $D$ can be set appropriately large in which case the copy complexity lower bound lifts directly to the purified forward quantum query setting.

\paragraph{Comparison to~\cite{IYM26PAC} and PAC learning}

The recent work of~\cite{IYM26PAC} demonstrated a similar technique for transforming copy complexity lower bounds to state preparation query lower bounds. In particular, they proved a version of~\Cref{cor:purestateprep} for pure states restricted to a small dimensional subspace. We restate their theorem below for completeness.

\begin{theorem}[Copy simulation of forward preparation queries, Theorem~3.3 of~\cite{IYM26PAC}]\label{thm:iym-state-preparation-simulation}
    Let $\mathcal S\subseteq\mathcal H([M])$ be a known $N$-dimensional subspace orthogonal to $\ket0$, and suppose $q(N-1)<M-1$. For every oracle algorithm $\A^{(\cdot)}$ making $q$ uncontrolled forward queries to an $M$-dimensional state preparation unitary, there exists a quantum channel $\B$, independent of $\ket\psi$, such that for every unit vector $\ket\psi\in\mathcal S$,
    \[
        \TD\left(
            \E_{U\gets\D^\psi}\left[\A^U\right],
            \B(\ketbra\psi^{\otimes q})
        \right)
        \leq\frac{q(N-1)}{M-1}.
    \]
    Here $\D^\psi$ is the Haar distribution over unitary completions of $\ket0\mapsto\ket\psi$, and the same sampled unitary is used in every query.
\end{theorem}

Note that this restriction can be removed by instead appending the prepared states with some high dimensional state $\ket{0^n}$. Doing so can recover a version of~\Cref{cor:state-lifting-lb}. Our result has two advantages. First, our simulator is efficient and so the lifting of lower bounds also applies to the computationally bounded setting with polynomial blowup. Second, our simulator maintains good error even for pure states with no purification register, as long as the dimensions of the states themselves are large. In particular, extending the theorem of~\cite{IYM26PAC} by the padding argument above will still not recover the $D=1$ case when restricting to pure states. However, the concrete bounds achieved by~\cite{IYM26PAC} do not have a square root loss and so are better than ours for some parameter regimes.

The techniques of~\cite{IYM26PAC} were developed specifically for the context of quantum PAC learning. We show that~\Cref{cor:state-lifting-lb} can be used to recover their results.

For a Boolean concept class $\mathcal C$ of VC dimension $v$, a quantum example in the realizable model is
\[
    \ket{\psi_{c,p}}=\sum_x\sqrt{p(x)}\ket{x}\ket{c(x)},
\]
where $c\in\mathcal C$ and $p$ is an unknown distribution. A learner must output a hypothesis of error at most $\epsilon$ with probability at least $1-\delta$. In the agnostic model, examples coherently encode an arbitrary distribution on labeled pairs, and the goal is error at most $\epsilon$ above the best concept in $\mathcal C$.

Lower bounds are known for quantum PAC learning in the copy complexity model.

\begin{theorem}[Quantum PAC learning, \cite{AdW18PAC}]\label{thm:pac-copy-lb}
    Let $\mathcal C$ have VC dimension $v\geq2$, and let $0<\epsilon\leq1/32$ and $0<\delta\leq1/4$. Realizable and agnostic quantum PAC learning of $\mathcal C$ require, respectively,
    \[
        \Omega\left(\frac{v+\log(1/\delta)}{\epsilon}\right)
        \qquad\text{and}\qquad
        \Omega\left(\frac{v+\log(1/\delta)}{\epsilon^2}\right)
    \]
    copies of the quantum example state. The agnostic lower bound also holds for $v=1$.
\end{theorem}

\begin{corollary}[PAC learning from forward preparation queries]\label{cor:pac-preparation-lb}
    Under the conditions of~\Cref{thm:pac-copy-lb}, let $M$ denote the total dimension on which the example-preparation unitary acts. Realizable and agnostic PAC learning using uncontrolled forward preparation queries require, respectively,
    \begin{align*}
        q_{\mathrm{real}}&=\Omega\left(\min\left\{\frac{v+\log(1/\delta)}{\epsilon},\delta\sqrt M\right\}\right),\\
        q_{\mathrm{agn}}&=\Omega\left(\min\left\{\frac{v+\log(1/\delta)}{\epsilon^2},\delta\sqrt M\right\}\right).
    \end{align*}
    The learner must succeed for every compatible preparation unitary.
\end{corollary}

\begin{proof}
The copy lower bounds in~\Cref{thm:pac-copy-lb} hold uniformly for failure probabilities at most $3/8$ (Lemma~2.7 of~\cite{IYM26PAC}).
Choose the simulation accuracy to be at most $\delta/2$. A learner of failure probability $\delta$ is then converted into a copy learner of failure probability at most $3\delta/2\leq3/8$. Applying these copy lower bounds gives the bounds, with the dimension-dependent term accounting for the simulation error.
\end{proof}

Taking arbitrarily large $M$ recovers the same parameters as the forward query lower bounds in Theorems 1.1 and 1.2 of~\cite{IYM26PAC}, which are optimal and match the classical sample bounds.

\paragraph{State distinguishing}

Another recent work directly proved bounds in the forward query model for the problem of state distinguishing~\cite{WCC26distinguishing}. In particular, they considered the following task: given a unitary $U$ with the promise that either $U\ket{0} = \ket{0}$ or $U\ket{0} = \sqrt{1-\epsilon^2}\ket{0} + \epsilon\ket{1}$, determine which holds. They showed that as long as the dimension of $U$ is $\Omega(\epsilon^{-8})$, this task requires $\Theta(\epsilon^{-2})$ queries.

Our technique immediately recovers this theorem from copy-complexity bounds on distinguishing two known quantum states.

\begin{theorem}[Pure-state distinguishing, consequence of Theorem~3.4 of~\cite{Watrous18TQI}]\label{thm:state-distinguishing-copy-lb}
    Distinguishing two known pure states $\ket{\psi_0},\ket{\psi_1}$ of trace distance $\epsilon$ with success probability at least $2/3$ requires $\Theta(1/\epsilon^2)$ copies.
\end{theorem}

Applying~\Cref{cor:state-lifting-lb} with $D=1$ gives the following:

\begin{corollary}[Distinguishing from forward preparation queries]\label{cor:state-distinguishing-preparation-lb}
    Let a preparation unitary acting in total dimension $M$ be promised to prepare one of two known pure states of trace distance $\epsilon$. Distinguishing the two cases with success probability at least $2/3$, for every compatible preparation unitary, requires
    \[
        q=\Omega\left(\min\left\{\epsilon^{-2},\sqrt M\right\}\right)
    \]
    uncontrolled forward queries.
\end{corollary}

In particular, our bound applies as well for the case where the two pure states are $\ket{0}$ and $\sqrt{1-\epsilon^2}\ket{0} + \epsilon\ket{1}$. It says that as long as the dimension of the unitary is $\Omega(\epsilon^{-4})$, solving this task with success at least $2/3$ requires $\Omega(\epsilon^{-2})$ queries. This improves on the requirement of the dimension being $\Omega(\epsilon^{-8})$.

Their separate continuous-time result, allowing arbitrary short forward Hamiltonian evolutions, is outside of the model we consider.

\paragraph{State tomography}

One of the most general learning questions in the quantum setting is the problem of state tomography. The question of state tomography is the following: ``how many copies of a state does it take to learn its classical description?'' To make this question feasible, some error is allowed in the described state.

Combining the random purification channel with pure state tomography gives an efficient, sample-optimal algorithm for mixed state tomography~\cite{PSTW25}. Its guarantee implies the following trace-distance upper bound.

\begin{theorem}[Efficient state tomography, Theorem~1.3 of~\cite{PSTW25}]\label{thm:state-tomography-copy-ub}
    Given copies of an unknown rank-at-most-$r$ state $\rho$ in dimension $N$, there is an algorithm using $O(Nr/\epsilon^2)$ copies that outputs a classical description of $\widehat\rho$ satisfying $\TD(\rho,\widehat\rho)\leq\epsilon$ with probability at least $2/3$. The algorithm runs in time $\poly(N,r,1/\epsilon)$.
\end{theorem}

We will show that this algorithm is also essentially optimal even when given forward query access to state preparation unitaries. First, recall the known lower bound on mixed state tomography:

\begin{theorem}[State tomography lower bound, Theorem~1.1 of~\cite{SSW25tomography} and Corollary~1.3 of~\cite{CZY26tomography}]\label{thm:state-tomography-copy-lb}
    Let $N\geq2$, $1\leq r\leq N$, and $0<\epsilon\leq\epsilon_0$ for a sufficiently small universal constant $\epsilon_0>0$. Given copies of an unknown rank-at-most-$r$ state $\rho$ in dimension $N$, outputting a classical description of $\widehat\rho$ satisfying $\TD(\rho,\widehat\rho)\leq\epsilon$ with probability at least $2/3$ requires $\Omega(Nr/\epsilon^2)$ copies.
\end{theorem}

Applying~\Cref{cor:state-lifting-lb} then gives:

\begin{corollary}[Tomography from forward preparation queries]\label{cor:state-tomography-preparation-lb}
    Let $N\geq2$, $1\leq r\leq N$, and $0<\epsilon\leq\epsilon_0$ for a sufficiently small universal constant $\epsilon_0>0$. Tomography of rank-at-most-$r$ states in dimension $N$ to trace-distance error $\epsilon$, with success probability at least $2/3$ for every compatible preparation unitary and every purification, requires
    \[
        q=\Omega\left(\min\left\{\frac{Nr}{\epsilon^2},\sqrt{N\cdot D}\right\}\right)
    \]
    uncontrolled forward preparation queries with purification dimension $D\geq N$. For arbitrary mixed states take $r=N$. For pure states take $r=1$, and the bound holds for every $D\geq1$, giving $\Omega(\min\{N/\epsilon^2,\sqrt N\})$ when $D=1$.
\end{corollary}

\begin{proof}
The copy lower bound in~\Cref{thm:state-tomography-copy-lb}
remains valid with success probability $2/3-1/100$, as follows
from the success-probability bound in the proof
of~\cite{CZY26tomography}.
For $q\leq c\sqrt{ND}$, with $c>0$ sufficiently small,
\Cref{thm:perfectlb} simulates the query algorithm using
$q$ copies with trace-distance error at most $1/100$.
Its tomography success probability is therefore at least
$2/3-1/100$, giving $q=\Omega(Nr/\epsilon^2)$.
\end{proof}

Preparing one copy per forward query and applying~\Cref{thm:state-tomography-copy-ub} uses $O(Nr/\epsilon^2)$ forward preparation queries, which is optimal whenever the purification dimension $D$ of the state preparation unitary satisfies $D = \Omega\left(\frac{Nr^2}{\epsilon^4}\right)$. That is, it is optimal for algorithms which are agnostic to the ambient dimension of the unitary.

\paragraph{Mixedness testing}

In the setting of mixedness testing, given a mixed state one tests whether it is close to the maximally mixed state. In this setting, one would expect a state preparation unitary preparing a state of dimension $N$ to have purification register also of dimension $N$, since this will be the case for the maximally mixed state. We will see that our techniques give optimal lower bounds for this setting in the constant error regime, even without blowing up the space of the ambient dimension.

\begin{theorem}[Mixedness testing, Theorem~1.10 of~\cite{OW21spectrum}]\label{thm:mixedness-copy-lb}
    Let $N\geq2$ and $0<\epsilon\leq\epsilon_0$ for a sufficiently small universal constant $\epsilon_0>0$. Distinguishing $\rho=I/N$ from $\TD(\rho,I/N)\geq\epsilon$ with success probability at least $2/3$ requires $\Theta(N/\epsilon^2)$ copies of the unknown state $\rho$.
\end{theorem}

Applying~\Cref{cor:state-lifting-lb} gives:

\begin{corollary}[Mixedness testing from forward preparation queries]\label{cor:mixedness-preparation-lb}
    Let $N\geq2$ and $0<\epsilon\leq\epsilon_0$ for a sufficiently small universal constant $\epsilon_0>0$. For purification dimension $D\geq N$, mixedness testing with success probability at least $2/3$ for every compatible preparation unitary and every purification requires
    \[
        q=\Omega\left(\min\left\{\frac{N}{\epsilon^2},\sqrt{N\cdot D}\right\}\right)
    \]
    uncontrolled forward preparation queries.
\end{corollary}

In particular, when $\epsilon$ is a constant and $D=N$, this gives a lower bound of $\Omega(N)$, which is optimal by~\Cref{thm:mixedness-copy-lb}.

\section{Impossibility of Simulating Isometry Dilations of Channels}
\label{sec:negative-result}
We demonstrate a pair of channels $\SAME,\DIFF$ which can be distinguished perfectly by query access to any isometry dilation, but cannot be efficiently distinguished by channel query access. In particular, any algorithm which distinguishes the channels with constant advantage must make an exponential (in the number of qubits) number of channel queries.

In particular, the isometry access distinguisher will work by applying a superoperator known as the quantum switch. Thus, our argument will also give a lower bound on implementations of the quantum switch (resolving Conjecture~1 of~\cite{BKM+25} up to a cubic loss in parameters).

\begin{theorem}\label{thm:channel-dilation-separation}
    For every $D,q\in\N$ with $D\geq3$ and $q\geq1$, there are
    two ensembles of quantum channels $\SAME$ and $\DIFF$, with input
    and output dimension $D$, such that the following hold.
    \begin{enumerate}
        \item One application of the quantum switch to two copies
        of the same unknown channel distinguishes any pair of channels from the supports of $\SAME$ and
        $\DIFF$ perfectly. That is, there exists a state $\rho$ such that for all $\Phi_{\SAME} \in \SAME,\Phi_{\DIFF}\in \DIFF$,
        $$\TD\left(\Sw[\Phi_{\SAME},\Phi_{\SAME}](\rho),\Sw[\Phi_{\DIFF},\Phi_{\DIFF}](\rho)\right) = 1$$

        \item For every oracle algorithm $\A^{(\cdot)}$ independent
        of the hidden pairs and making at most $q$ forward channel
        queries,
        \[
            \TD\left(\E[\A^{\SAME}],\E[\A^{\DIFF}]\right)
            =O\left(\frac{q^{3/2}}{D^{1/2}}\right),
        \]
        where the expectations are over the hidden pairs, sampled
        once and fixed throughout each experiment, and the implicit
        constant is absolute. Consequently, constant distinguishing
        advantage requires $\Omega(D^{1/3})$ channel queries in the
        worst case.
    \end{enumerate}
\end{theorem}

This theorem immediately gives lower bounds on simulating the quantum switch, even with any fixed error below
$1/2$.

\begin{corollary}\label{cor:quantum-switch-query-lower-bound}
    Fix $0\leq\epsilon<1/2$. Any universal simulator that uses only
    ordinary adaptive forward channel queries to implement one
    quantum switch on $D$-dimensional channels, with trace-distance
    error at most $\epsilon$ on every input, requires
    $\Omega(D^{1/3})$ total channel queries in the worst case.
    In particular, for $n$-qubit channels this requires
    $\Omega(2^{n/3})$ queries. This holds even when the two channels
    supplied to the switch are promised to be identical, and with
    arbitrary computation and reference registers between queries.
\end{corollary}

\begin{remark}
    Since classical control of causal order is equivalent to query access when only one channel is present, this also gives a lower bound on implementing the quantum switch with classical control of causal order.
\end{remark}

As the quantum switch can be implemented with query access to any isometry dilation of the unknown channel, we similarly get the following bound on simulating random dilations of an arbitrary quantum channel.

\begin{corollary}\label{cor:no-channel-dilation-simulator}
    Fix $0\leq\epsilon<1/2$. Any universal simulator that, using only
    forward queries to an unknown channel with input and output
    dimension $D$, simulates two adaptive forward queries to a fixed
    random isometry dilation with total trace-distance error at most
    $\epsilon$ requires $\Omega(D^{1/3})$ channel queries in the worst
    case. This holds even for stateful simulators with arbitrary
    computation between queries, and with the dilation environment
    fixed to dimension $D^2$.
\end{corollary}

We can also turn the decision problem into a search problem by
encoding a hidden string $z\in\{0,1\}^m$ in independent channel
instances. One switch application, or two dilation queries,
recovers each bit with certainty. A hybrid argument shows that channel
queries have exponentially small recovery probability, even with
exponentially many queries. This yields a simulation lower bound
even when the allowed error approaches one.

\begin{corollary}\label{cor:channel-dilation-near-unit-error}
    There is an absolute constant $c>0$ such that, for all sufficiently
    large $n$, there is an ensemble of channels $\mathcal C_z$, indexed
    by $z\leftarrow\{0,1\}^n$, with input, output, and a common
    dilation environment each containing $O(n)$ qubits, with the
    following properties.
    \begin{enumerate}
        \item An efficient algorithm makes $2n$ adaptive forward
        queries to any fixed isometry dilation of $\mathcal C_z$
        and recovers $z$ with probability one. Alternatively,
        $n$ applications of the quantum switch suffice.

        \item Every algorithm making at most $2^{cn}$ forward queries
        to $\mathcal C_z$ recovers $z$ with probability at most
        $2^{-2cn}$, averaged over $z$, the hidden pairs, and the
        algorithm's randomness, even with arbitrary computation
        between queries.
    \end{enumerate}
    Consequently, any universal simulator of $2n$ adaptive forward
    queries to a fixed random isometry dilation with total
    trace-distance error at most $1-2^{-cn}$ requires
    $\Omega(2^{cn})$ channel queries in the worst case. The same
    lower bound holds for simulating $n$ adaptive uses of the
    quantum switch with this total error, even when the same
    unknown channel is supplied to both slots on every use.
\end{corollary}

\paragraph{The hard to distinguish channels}
We now define the distributions over channels. Let $L\geq1$ and $d\geq2$. The randomness of both channels will consist entirely of the following set of quantum states. In detail,
for each $x\in[L]$, sample an ordered orthonormal pair
\[
    \ket{\phi_{x,0}},\ket{\phi_{x,1}}\in\C^d
\]
as the first two columns of a Haar-random unitary.

We write
$P_{x,b}=\ketbra{\phi_{x,b}}$.
Here, $\ket{x}\ket{\phi_{x,b}}$ is a state in $\mathcal{H}([L])\otimes \mathcal{H}([d])$. We will also add in a special state $\ket{s}$ orthogonal to the above. That is, both channels will act on the space $$\Reg{In}=\Span(\ket{s})\oplus \mathcal{H}([L])\otimes \mathcal{H}([d])$$
of dimension $Ld+1$.

We will define the two channels $\SAME$ and $\DIFF$ by defining corresponding isometry dilations. In particular, define internal registers $\Reg{X} = \mathcal{H}([L])$, $\Reg{B} = \mathcal{H}(\{0,1\})$, and $\Reg{Extra}$ a copy of $\Reg{In}$.

Define $V^{\SAME}$ and $V^{\DIFF}$ sending $\Reg{In}$ to $\Reg{In}\otimes\Reg{X}\otimes\Reg{B} \otimes \Reg{Extra}$ as follows.

    Define $V^{\SAME}$ by
    \[
        \begin{aligned}
            V^{\SAME}\ket{s}_{\Reg{In}}
                &=\frac1{\sqrt{2L}}\sum_{x\in[L],\,b\in\{0,1\}}
                    \ket{x,\phi_{x,b}}_{\Reg{In}}\ket{x,b}_{\Reg{X}\Reg{B}}\ket{s}_{\Reg{Extra}},\\
            V^{\SAME}\ket{x,\phi_{x,b}}_{\Reg{In}}
                &=\ket{s}_{\Reg{In}}\ket{x,b\xor 1}_{\Reg{X}\Reg{B}}\ket{s}_{\Reg{Extra}},\\
            V^{\SAME}\ket{x,\psi}_{\Reg{In}}
                &=\ket{s}_{\Reg{In}}\ket{0}_{\Reg{X}\Reg{B}}\ket{x,\psi}_{\Reg{Extra}},
        \end{aligned}
    \]
    where the last line applies when $\ket{\psi}$ is orthogonal
    to $\ket{\phi_{x,0}},\ket{\phi_{x,1}}$.
    Define $V^{\DIFF}$ by
    \[
        \begin{aligned}
            V^{\DIFF}\ket{s}_{\Reg{In}}
                &=\frac1{\sqrt{2L}}\sum_{x\in[L],\,b\in\{0,1\}}
                    \ket{x,\phi_{x,b}}_{\Reg{In}}\ket{x,b}_{\Reg{X}\Reg{B}}\ket{s}_{\Reg{Extra}},\\
            V^{\DIFF}\ket{x,\phi_{x,b}}_{\Reg{In}}
                &=(-1)^b\ket{s}_{\Reg{In}}\ket{x,b\xor 1}_{\Reg{X}\Reg{B}}\ket{s}_{\Reg{Extra}},\\
            V^{\DIFF}\ket{x,\psi}_{\Reg{In}}
                &=\ket{s}_{\Reg{In}}\ket{0}_{\Reg{X}\Reg{B}}\ket{x,\psi}_{\Reg{Extra}},
        \end{aligned}
    \]
    with the same orthogonality condition on $\ket{\psi}$.
    By construction, on orthogonal inputs these maps produce orthogonal outputs. Thus, they are both isometries. We then define the distributions over channels $\SAME$ and $\DIFF$ to be the action of $V^{\SAME}$ and $V^{\DIFF}$ with the internal registers traced out.

\begin{definition}[The channels $\SAME$ and $\DIFF$]
    The channels act as follows.
    \[
        \begin{aligned}
            \SAME(\rho)&=\Tr_{\Reg{X},\Reg{B},\Reg{Extra}}
                \bigl(V^{\SAME}\rho(V^{\SAME})^\dagger\bigr),\\
            \DIFF(\rho)&=\Tr_{\Reg{X},\Reg{B},\Reg{Extra}}
                \bigl(V^{\DIFF}\rho(V^{\DIFF})^\dagger\bigr).
        \end{aligned}
    \]
    All measurement outcomes are private to the channel.
\end{definition}

\subsection{Distinguishing with the Quantum Switch}
\label{sec:quantum-switch-lower-bound}

It turns out that applying the self-switch to the state $\ket{+}\ket{s}$ (where the $\ket{+}$ is on the control register) perfectly distinguishes $\SAME$ and $\DIFF$.

\begin{proposition}\label{prop:self-switch-distinguisher}
    For every choice of $\Phi_{\SAME}$ in the support of $ \SAME$ and $\Phi_{\DIFF}$ in the support of $\DIFF$, 
    $$\Sw[\Phi_{\SAME},\Phi_{\SAME}](\ketbra{+}\otimes \ketbra{s}) = \ketbra{+}\otimes \ketbra{s}$$
    and
    $$\Sw[\Phi_{\DIFF},\Phi_{\DIFF}](\ketbra{+}\otimes \ketbra{s}) = \ketbra{-}\otimes \ketbra{s}$$
    Thus one application of the quantum switch
    distinguishes the two channels perfectly.
\end{proposition}

\begin{proof}
    Fix the hidden pairs $\ket{\phi_{x,b}}$ for all $x,b$. 

    By \Cref{lem:self-switch-from-dilation}, we can evaluate each
    self-switch using two queries to its dilation followed by a
    controlled swap of the environments. In particular, since this fact is agnostic to the choice of dilation, we will work with $V^{\SAME}$ and $V^{\DIFF}$ for $\SAME$ and $\DIFF$ respectively. For $\SAME$, starting from $\ket{s}$, we get
    \[
        \begin{aligned}
            \ket{s}
                &\longmapsto\frac1{\sqrt{2L}}
                    \sum_{x,b}\ket{x,\phi_{x,b}}
                        \ket{x,b}_{\Reg {X}_1\Reg{B}_1}\ket{s}_{\Reg{Extra}_1}\\
                &\longmapsto\frac{\ket{s}}{\sqrt{2L}}
                    \sum_{x,b}
                        \ket{x,b}_{\Reg{X}_1\Reg{B}_1}\ket{s}_{\Reg{Extra}_1}
                        \ket{x,b\xor 1}_{\Reg{X}_2\Reg{B}_2}\ket{s}_{\Reg{Extra}_2}
                    =\ket{s}\ket{\Omega_{\SAME}},
        \end{aligned}
    \]
    where
    \[
        \ket{\Omega_{\SAME}}=\frac1{\sqrt{2L}}\sum_{x\in[L]}
            \left(\ket{x,0}\ket{s}\ket{x,1}\ket{s}
                +\ket{x,1}\ket{s}\ket{x,0}\ket{s}\right).
    \]
    For $\DIFF$, the two queries instead give
    \[
        \begin{aligned}
            \ket{s}
                &\longmapsto\frac1{\sqrt{2L}}
                    \sum_{x,b}\ket{x,\phi_{x,b}}_{\Reg{In}}
                        \ket{x,b}_{\Reg{X}_1\Reg{B}_1}\ket{s}_{\Reg{Extra}_1}\\
                &\longmapsto\frac{\ket{s}}{\sqrt{2L}}
                    \sum_{x,b}(-1)^b
                        \ket{x,b}_{\Reg{X}_1\Reg{B}_1}\ket{s}_{\Reg{Extra}_1}
                        \ket{x,b\xor 1}_{\Reg{X}_2\Reg{B}_2}\ket{s}_{\Reg{Extra}_2}
                    =\ket{s}\ket{\Omega_{\DIFF}},
        \end{aligned}
    \]
    where
    \[
        \ket{\Omega_{\DIFF}}=\frac1{\sqrt{2L}}\sum_{x\in[L]}
            \left(\ket{x,0}\ket{s}\ket{x,1}\ket{s}
                -\ket{x,1}\ket{s}\ket{x,0}\ket{s}\right).
    \]
    The first environment state is symmetric under swapping
    $\Reg{X}_1\Reg{B}_1\Reg{Extra}_1$ and $\Reg{X}_2\Reg{B}_2\Reg{Extra}_2$, while the second is antisymmetric:
    \[
        \operatorname{SWAP}\ket{\Omega_{\SAME}}
            =\ket{\Omega_{\SAME}},
        \qquad
        \operatorname{SWAP}\ket{\Omega_{\DIFF}}
            =-\ket{\Omega_{\DIFF}}.
    \]
    Thus the controlled swap, with switch control $\Reg C$
    initialized to $\ket{+}$, gives
    \[
        \begin{aligned}
            \ket{+}_{\Reg C}\ket{s}\ket{\Omega_{\SAME}}
                &\longmapsto
                \ket{+}_{\Reg C}\ket{s}\ket{\Omega_{\SAME}},\\
            \ket{+}_{\Reg C}\ket{s}\ket{\Omega_{\DIFF}}
                &\longmapsto
                \ket{-}_{\Reg C}\ket{s}\ket{\Omega_{\DIFF}}.
        \end{aligned}
    \]
    Discarding the environments leaves $\ket{+}\ket{s}$ for
    $\SAME$ and $\ket{-}\ket{s}$ for $\DIFF$, proving both
    identities in the proposition. The dilation lemma identifies
    this procedure with the self-switch for any fixed dilation,
    so the conclusion is independent of the chosen dilation.
\end{proof}

\begin{corollary}\label{cor:channel-dilation-distinguisher}
    There is an efficient algorithm that makes two adaptive forward
    queries to any fixed isometry dilation $V^\Phi$ of a channel
    $\Phi$ in the support of $\SAME$ or $\DIFF$, and identifies
    which ensemble it belongs to with probability one. This holds
    for every choice of the hidden pairs and every fixed dilation
    into a common environment space.
\end{corollary}

\begin{proof}
    Implement $\Sw[\Phi,\Phi]$ on $\ket{+}\ket{s}$ using
    \Cref{lem:self-switch-from-dilation}, then measure the control
    in the $X$ basis. By \Cref{prop:self-switch-distinguisher}, the
    outcome is $+$ for $\SAME$ and $-$ for $\DIFF$ with certainty.
\end{proof}

\subsection{Lower Bound for Channel Access}
\label{sec:impossibility-channel-access}

\Cref{thm:channel-dilation-separation} will follow from~\Cref{prop:self-switch-distinguisher} along with the following theorem by setting $d=2$, $L=\floor{(D-1)/2}$ (and padding the input dimension to $D$ if necessary).

\begin{theorem}\label{thm:channel-query-lower-bound}
    Let $\A$ make at most $q\geq1$ forward queries to the unknown
    channel. Let $\rho_{\SAME}$ and $\rho_{\DIFF}$ be its final states
    in the two experiments, averaged over the choice of the hidden
    pairs. Then
    \[
        \TD(\rho_{\SAME},\rho_{\DIFF})
        \leq\frac{8q^{3/2}}{\sqrt L}.
    \]
    Consequently, achieving a constant distinguishing advantage
    requires $\Omega(L^{1/3})$ queries.
\end{theorem}

\begin{proof}
    We may pad the algorithm with queries whose outputs it discards
    so that it makes exactly $q$ queries. If $q>L$, the claimed
    bound is larger than one and hence trivial, so assume $q\leq L$.
    Write
    \[
        P_s=\ketbra{s},
        \qquad Q_x=\ketbra{x}\otimes (\ketbra{\phi_{x,0}}+\ketbra{\phi_{x,1}}),
    \]

    We will give a sequence of stateful hybrid oracles $\Hyb_{1,0},\dots,\Hyb_{6,0}$ and $\Hyb_{1,1},\dots,\Hyb_{6,1}$. We will then bound for each $i$,
    $$\TD(\A^{\Hyb_{i,0}},\A^{\Hyb_{i+1,0}})$$
    and
    $$\TD(\A^{\Hyb_{i,1}},\A^{\Hyb_{i+1,1}})$$
    We will also give the following equalities
    $$\E[\A^{\SAME}] = \A^{\Hyb_{1,0}},\quad \E[\A^{\DIFF}] = \A^{\Hyb_{1,1}},\quad \A^{\Hyb_{6,0}} = \A^{\Hyb_{6,1}}$$
    together, these will prove the theorem.

    Throughout a query, $S$ denotes a private copy of the set at the start of that query. All membership tests and sampling exclusions use this pre-query copy, even if $\Reg{St}$ has already been updated. Every measured index is recorded, including those from invalid inputs. All work registers are discarded after the response; only $\Reg{St}$ and the hidden-pair descriptions specified below are reused between queries. In a purification, discarded registers and classical measurement records are retained in the environment.

    \begin{enumerate}
        \item $\Hyb_{1,b_{\DIFF}}$ will first choose $\{\ket{\phi_{x,b}}\}_{x\in [L],b\in \{0,1\}}$, then if $b_\DIFF=0$ implement $\SAME$ and if $b_\DIFF=1$ implement $\DIFF$. In addition, it will keep track of a list $S$ of queried points $x$.

        {\emergencystretch=5em In particular, at the beginning of the game it will sample Haar random states $\{\ket{\phi_{x,b}}\}_{x\in [L],b\in \{0,1\}}$ subject to $\ket{\phi_{x,0}},\ket{\phi_{x,1}}$ being orthogonal. The set $S$ will be stored in a register $\Reg{St} = \mathcal{H}(2^{[L]})$ where $2^{[L]}$ refers to the set of all subsets of $[L]$. $\Reg{St}$ will begin initialized to $\ket{\emptyset}$. When $\A$ makes a query, $\Hyb_{1,b_{\DIFF}}$ will implement $\SAME$ or $\DIFF$ according to $b_{\DIFF}$ for the choice of $\{\ket{\phi_{x,b}}\}_{x\in [L],b\in \{0,1\}}$. Afterwards, it will add the measured input $x$ to $\Reg{St}$ if it is not already in there. In particular, we will use the following operational description of this map.\par}
        \begin{enumerate}
            \item Initialize private register $\Reg{C} = \mathcal{H}(\{0,1\})$, $\Reg{X} = \mathcal{H}([L])$, $\Reg{B}=\mathcal{H}(\{0,1\})$ all to $\ket{0}$. Define an extra register $\Reg{Extra}$ as a copy of $\Reg{In}$ initialized to $\ket{s}$.
            \item Apply $M^{P_s}$ to registers $\Reg{In},\Reg{C}$. Recall that this is the map which measures $\{P_s,I-P_s\}$, coherently XORing the result with the value in $\Reg{C}$.
            \item Controlled on $\Reg{C} = 1$, apply the map
            $$\ket{0}_{\Reg{X}}\ket{0}_{\Reg{B}} \mapsto \frac{1}{\sqrt{2L}}\sum_{x\in [L],b\in \{0,1\}}\ket{x}_{\Reg{X}}\ket{b}_{\Reg{B}}$$
            \item Controlled on $\Reg{C} = 0$, copy the input into register $\Reg{X}$. That is, apply the map
            $$\ket{x,\psi}_{\Reg{In}}\ket{0}_{\Reg{X}} \mapsto \ket{x,\psi}_{\Reg{In}}\ket{x}_{\Reg{X}}$$
            \item Controlled on $\Reg{C}=1$, apply the map
            $$\ket{s}_{\Reg{In}}\ket{x}_{\Reg{X}}\ket{b}_{\Reg{B}}\mapsto \ket{x,\phi_{x,b}}_{\Reg{In}}\ket{x}_{\Reg{X}}\ket{b}_{\Reg{B}}$$
            Controlled on $\Reg{C} = 0$, apply the map which for all $x\in [L],b\in \{0,1\}$ acts as
            $$\ket{x,\phi_{x,b}}_{\Reg{In}}\ket{x}_{\Reg{X}}\ket{0}_{\Reg{B}}\ket{s}_{\Reg{Extra}} \mapsto (-1)^{b\cdot b_{\DIFF}}\ket{s}_{\Reg{In}}\ket{x}_{\Reg{X}}\ket{b\xor 1}_{\Reg{B}}\ket{s}_{\Reg{Extra}}$$
            and for all $x\in [L]$ and $\ket{\psi}$ orthogonal to $\ket{\phi_{x,0}},\ket{\phi_{x,1}}$ acts as
            $$\ket{x,\psi}_{\Reg{In}}\ket{x}_{\Reg{X}}\ket{0}_{\Reg{B}}\ket{s}_{\Reg{Extra}}\mapsto \ket{s}_{\Reg{In}}\ket{x}_{\Reg{X}}\ket{0}_{\Reg{B}}\ket{x,\psi}_{\Reg{Extra}}$$
            \item Apply $M^{I-P_s}$ to registers $\Reg{In},\Reg{C}$. This will set the value in $\Reg{C}$ to be $\ket{0}$ regardless of all other operations.
            \item Measure $\Reg{X}$ in the standard basis, producing result $x$. Set $S = S \cup \{x\}$ in $\Reg{St}$.
            \item Output $\Reg{In}$.
        \end{enumerate}
        Note that for all partial isometries defined above, by construction it will always be the case that at the point the channel applies each partial isometry, the input state will be in its corresponding support.
        \item $\Hyb_{2,b_{\DIFF}}$ will be almost the same as $\Hyb_{1,b_{\DIFF}}$, except that we require that there are no collisions in sampled points. That is, we replace the map in (c) with
        $$\ket{0}_{\Reg{X}}\ket{0}_{\Reg{B}}\ket{S}_{\Reg{St}}\mapsto \frac{1}{\sqrt{2(L - |S|)}} \sum_{x\in [L]\setminus S,b\in \{0,1\}}\ket{x}_{\Reg{X}}\ket{b}_{\Reg{B}}\ket{S}_{\Reg{St}}$$
        \item $\Hyb_{3,b_{\DIFF}}$ will act the same as $\Hyb_{2,b_{\DIFF}}$ except that it will measure $\Reg{X}$ and add it to $\Reg{St}$ immediately after step (d). That is, on a query it will act as
        \begin{enumerate}
            \item to (d): same as $\Hyb_{2,b_{\DIFF}}$ (a) to (d)
            \item[(e)]: same as $\Hyb_{2,b_{\DIFF}}$ (g)
            \item[(f)] to (g): same as $\Hyb_{2,b_{\DIFF}}$ (e) to (f)
            \item[(h)]: Output $\Reg{In}$
        \end{enumerate}
        \item $\Hyb_{4,b_{\DIFF}}$ will, immediately after measuring $\Reg{X}$, check if the received value $x$ is in the pre-query set $S$. If it is, then it will immediately terminate and output $\ket{s}$. In a purified, operational form, this looks like
        \begin{enumerate}
            \item to (d): same as $\Hyb_{2,b_{\DIFF}}$.
            \item[(e)]: Measure $\Reg{X}$ in the standard basis, producing result $x$, and update $\Reg{St}$ to $S\cup\{x\}$.
            \item[(f)]: If $x\in S$, swap registers $\Reg{In}$ and $\Reg{Extra}$.
            \item[(g)]: Otherwise, apply steps (f) to (g) of $\Hyb_{3,b_{\DIFF}}$.
            \item[(h)]: Output $\Reg{In}$.
        \end{enumerate}
        \item $\Hyb_{5,b_{\DIFF}}$ will be the same as $\Hyb_{4,b_{\DIFF}}$, except in the case that $x\notin S$, $\ket{\phi_{x,b}}$ will be sampled freshly on the spot rather than at initialization. Note that the case $x\in S$ does not depend on $\ket{\phi_{x,b}}$ at all.
        \item $\Hyb_{6,b_{\DIFF}}$ will be the same as $\Hyb_{5,b_{\DIFF}}$, except that we sample a random phase $(-1)^r$ and apply it to $\ket{\phi_{x,1}}$ in the case that $x\notin S$. In full, this can operationally be described by
        \begin{enumerate}
            \item Initialize private register $\Reg{C} = \mathcal{H}(\{0,1\})$, $\Reg{X} = \mathcal{H}([L])$, $\Reg{B}=\mathcal{H}(\{0,1\})$ all to $\ket{0}$. Define an extra register $\Reg{Extra}$ as a copy of $\Reg{In}$ initialized to $\ket{s}$.
            \item Apply $M^{P_s}$ to registers $\Reg{In},\Reg{C}$. Recall that this is the map which measures $\{P_s,I-P_s\}$, coherently XORing the result with the value in $\Reg{C}$.
            \item Controlled on $\Reg{C} = 1$, apply the map
            $$\ket{0}_{\Reg{X}}\ket{0}_{\Reg{B}} \mapsto \frac{1}{\sqrt{2(L-|S|)}}\sum_{x\in [L]\setminus S,b\in \{0,1\}}\ket{x}_{\Reg{X}}\ket{b}_{\Reg{B}}$$
            \item Controlled on $\Reg{C} = 0$, copy the input into register $\Reg{X}$. That is, apply the map
            $$\ket{x,\psi}_{\Reg{In}}\ket{0}_{\Reg{X}} \mapsto \ket{x,\psi}_{\Reg{In}}\ket{x}_{\Reg{X}}$$
            \item Measure $\Reg{X}$ in the standard basis, producing result $x$, and update $\Reg{St}$ to $S\cup\{x\}$.
            \item If $x\in S$, swap registers $\Reg{In}$ and $\Reg{Extra}$.
            \item If $x\notin S$, do the following
            \begin{enumerate}
                \item Sample $r \gets \{0,1\}$.
                \item Sample orthogonal Haar random states $\ket{\phi_{x,0}},\ket{\phi_{x,1}}\in \mathcal{H}([d])$.
                \item Controlled on $\Reg{C}=1$, apply the map
                $$\ket{s}_{\Reg{In}}\ket{x}_{\Reg{X}}\ket{b}_{\Reg{B}}\mapsto (-1)^{r\cdot b}\ket{x,\phi_{x,b}}_{\Reg{In}}\ket{x}_{\Reg{X}}\ket{b}_{\Reg{B}}$$
                \item Controlled on $\Reg{C} = 0$, apply the map which for all $x\in [L],b\in \{0,1\}$ acts as
                $$\ket{x,\phi_{x,b}}_{\Reg{In}}\ket{x}_{\Reg{X}}\ket{0}_{\Reg{B}}\ket{s}_{\Reg{Extra}} \mapsto (-1)^{r\cdot b}\cdot (-1)^{b\cdot b_{\DIFF}}\ket{s}_{\Reg{In}}\ket{x}_{\Reg{X}}\ket{b\xor 1}_{\Reg{B}}\ket{s}_{\Reg{Extra}}$$
                and for all $x\in [L]$ and $\ket{\psi}$ orthogonal to $\ket{\phi_{x,0}},\ket{\phi_{x,1}}$ acts as
                $$\ket{x,\psi}_{\Reg{In}}\ket{x}_{\Reg{X}}\ket{0}_{\Reg{B}}\ket{s}_{\Reg{Extra}}\mapsto \ket{s}_{\Reg{In}}\ket{x}_{\Reg{X}}\ket{0}_{\Reg{B}}\ket{x,\psi}_{\Reg{Extra}}$$
            \end{enumerate}
            \item Apply $M^{I-P_s}$ to registers $\Reg{In},\Reg{C}$, setting $\Reg C$ to $\ket{0}$.
            \item Output $\Reg{In}$.
        \end{enumerate}
    \end{enumerate}

    Write $\rho_{i,b_{\DIFF}}$ for the algorithm's final state
    with oracle $\Hyb_{i,b_{\DIFF}}$, averaged over the
    challenger's randomness.

    \begin{claim}
        $\rho_{1,0}=\rho_{\SAME}$ and
        $\rho_{1,1}=\rho_{\DIFF}$.
    \end{claim}
    \begin{proof}
        To see this, fix any choice $\{\ket{\phi_{x,b}}\}_{x\in [L],b\in \{0,1\}}$. Consider the purified action of $\Hyb_{1,b_{\DIFF}}$ up to step (h). In particular, on input $\ket{s}$ it acts as follows
        \begin{equation}
            \begin{split}
                \ket{s}_{\Reg{In}}&\xrightarrow{(a)} \ket{s}_{\Reg{In}}\ket{0}_{\Reg{X}}\ket{0}_{\Reg{B}}\ket{0}_{\Reg{C}}\ket{s}_{\Reg{Extra}}\\
                &\xrightarrow{(b)} \ket{s}_{\Reg{In}}\ket{0}_{\Reg{X}}\ket{0}_{\Reg{B}}\ket{1}_{\Reg{C}}\ket{s}_{\Reg{Extra}}\\
                &\xrightarrow{(c)} \frac{1}{\sqrt{2L}}\sum_{x\in [L],b\in \{0,1\}}\ket{s}_{\Reg{In}}\ket{x}_{\Reg{X}}\ket{b}_{\Reg{B}}\ket{1}_{\Reg{C}}\ket{s}_{\Reg{Extra}}\\
                &\xrightarrow{(e)}\frac{1}{\sqrt{2L}}\sum_{x\in [L],b\in \{0,1\}}\ket{x,\phi_{x,b}}_{\Reg{In}}\ket{x}_{\Reg{X}}\ket{b}_{\Reg{B}}\ket{1}_{\Reg{C}}\ket{s}_{\Reg{Extra}}\\
                &\xrightarrow{(f)}\frac{1}{\sqrt{2L}}\sum_{x\in [L],b\in \{0,1\}}\ket{x,\phi_{x,b}}_{\Reg{In}}\ket{x}_{\Reg{X}}\ket{b}_{\Reg{B}}\ket{0}_{\Reg{C}}\ket{s}_{\Reg{Extra}}\\
            \end{split}
        \end{equation}
        on input $\ket{x,\phi_{x,b}}$ for $x\in[L],b\in\{0,1\}$, it acts as
        \begin{equation}
            \begin{split}
                \ket{x,\phi_{x,b}}_{\Reg{In}}&\xrightarrow{(a)} \ket{x,\phi_{x,b}}_{\Reg{In}}\ket{0}_{\Reg{X}}\ket{0}_{\Reg{B}}\ket{0}_{\Reg{C}}\ket{s}_{\Reg{Extra}}\\
                &\xrightarrow{(b)}\ket{x,\phi_{x,b}}_{\Reg{In}}\ket{0}_{\Reg{X}}\ket{0}_{\Reg{B}}\ket{0}_{\Reg{C}}\ket{s}_{\Reg{Extra}}\\
                &\xrightarrow{(d)}\ket{x,\phi_{x,b}}_{\Reg{In}}\ket{x}_{\Reg{X}}\ket{0}_{\Reg{B}}\ket{0}_{\Reg{C}}\ket{s}_{\Reg{Extra}}\\
                &\xrightarrow{(e)}(-1)^{b\cdot b_{\DIFF}}\ket{s}_{\Reg{In}}\ket{x}_{\Reg{X}}\ket{b\xor 1}_{\Reg{B}}\ket{0}_{\Reg{C}}\ket{s}_{\Reg{Extra}}\\
                &\xrightarrow{(f)}(-1)^{b\cdot b_{\DIFF}}\ket{s}_{\Reg{In}}\ket{x}_{\Reg{X}}\ket{b\xor 1}_{\Reg{B}}\ket{0}_{\Reg{C}}\ket{s}_{\Reg{Extra}}\\
            \end{split}
        \end{equation}
        and finally on input $\ket{x}\ket{\psi}$ for $\ket{\psi}$ orthogonal to $\ket{\phi_{x,0}},\ket{\phi_{x,1}}$, it outputs
        $$\ket{x,\psi}_{\Reg{In}} \mapsto \ket{s}_{\Reg{In}}\ket{x}_{\Reg{X}}\ket{0}_{\Reg{B}}\ket{0}_{\Reg{C}}\ket{x,\psi}_{\Reg{Extra}}$$
        This map is exactly the same as the purified action of $\SAME,\DIFF$ up to the isometry defined by, for all $x,y\in [L]$ and $\ket{\psi}\in \mathcal{H}([d])$,
        \begin{equation}\label{eq:hyb1-intertwiner}
            \begin{split}
        \ket{y}_{\Reg{X}}\ket{x,\psi}_{\Reg{Extra}} \mapsto \ket{y+ x\mod L}_{\Reg{X}}\ket{0}_{\Reg{C}}\ket{x,\psi}_{\Reg{Extra}}\\
        \ket{x}_{\Reg{X}}\ket{s}_{\Reg{Extra}} \mapsto \ket{x}_{\Reg{X}}\ket{0}_{\Reg{C}}\ket{s}_{\Reg{Extra}}
            \end{split}
        \end{equation}

        Thus, $\Hyb_{1,0}$ is identical to the following process. First, apply a purification of $\SAME$. Then, apply the isometry from~\Cref{eq:hyb1-intertwiner}, then measure $\Reg{X}$ in the standard basis and add it to $S$. Output $\Reg{In}$. Since the isometry and measuring applications act only on the internal registers, this is exactly the same action as $\SAME$. The same argument applies for $\Hyb_{1,1}$ and $\DIFF$.
    \end{proof}

    \begin{claim}
        For each $b_{\DIFF}\in\{0,1\}$,
        \[
            \TD(\rho_{1,b_{\DIFF}},\rho_{2,b_{\DIFF}})
            \leq\frac{4q^{3/2}}{\sqrt L}.
        \]
    \end{claim}
    \begin{proof}
        For $S$ a subset of $[L]$ with $|S|<L$, let $V_S$ be the partial isometry with support $\Span(\ket{s})$ defined by
        $$V_S\ket{s}=\frac{1}{\sqrt{L-|S|}}\ket{s}\sum_{x\in [L]\setminus S}\ket{x}$$
        and write $\ket{\gamma_S} = V_S\ket{s}$.
        It suffices to show that for $|S|\leq q-1$, $V_S$ and $V_{\emptyset}$ are close in operator norm.

        \begin{equation}
            \begin{split}
                \norm{V_S - V_{\emptyset}}_{op}^2
                &=\norm{\ket{\gamma_S}-\ket{\gamma_{\emptyset}}}^2\\
                &=2-2\bra{\gamma_S}\ket{\gamma_{\emptyset}}\\
                &=2-2\sqrt{1-\frac{|S|}{L}}
                \leq\frac{2|S|}{L} \leq \frac{2q}{L}.
            \end{split}
        \end{equation}
        
        Then, applying \Cref{lem:contraction-oracle-distance} and inducting over the number of queries, we get that 
        \[
            \TD(\rho_{1,b_{\DIFF}},\rho_{2,b_{\DIFF}})
            \leq 2q\cdot \sqrt{\frac{2q}{L}} \leq \frac{4q^{3/2}}{\sqrt{L}}.
        \]
    \end{proof}

    \begin{claim}
        $\rho_{2,b_{\DIFF}}=\rho_{3,b_{\DIFF}}$.
    \end{claim}
    \begin{proof}
        Note that after (d) in $\Hyb_{2,b_{\DIFF}}$, the register $\Reg{X}$ is only used as a control for other operations. Since measurement commutes with controlling, $\Hyb_{2,b_{\DIFF}}$ and $\Hyb_{3,b_{\DIFF}}$ apply equivalent channels.
    \end{proof}

    \begin{claim}
        $\rho_{3,b_{\DIFF}}=\rho_{4,b_{\DIFF}}$.
    \end{claim}
    \begin{proof}
        Fix any family $\{\ket{\phi_{x,b}}\}$. We will look at the purified actions of $\Hyb_{3,b_{\DIFF}}$ and $\Hyb_{4,b_{\DIFF}}$ after $\Reg{X}$ is measured in the case where $x\in S$. In particular, we will show that these are equivalent up to an isometry applied to the traced out internal registers.

        In particular, the residual state after $\Reg{X}$ is measured, conditioned on $x\in S$, will be a superposition of states of the following form
        $$\ket{\beta_{x,\psi}} = \ket{x,\psi}_{\Reg{In}}\ket{x}_{\Reg{X}}\ket{0}_{\Reg{B}}\ket{0}_{\Reg{C}}\ket{s}_{\Reg{Extra}}$$
        with $\ket{\psi}\in \mathcal{H}([d])$. Note that the case where $\Reg{In}$ is in state $\ket{s}$ is impossible since in that case $\Reg{X}$ will be supported on $x\notin S$.

        Let $U_{3,b_{\DIFF}},U_{4,b_{\DIFF}}$ be the unitary corresponding to the purified action of $\Hyb_{3,b_{\DIFF}}$ and $\Hyb_{4,b_{\DIFF}}$ after $\Reg{X}$ is measured in the case where $x\in S$. 
        
        We then have for $\ket{\psi}=\ket{\phi_{x,b}}$,
        $$U_{3,b_{\DIFF}}\ket{\beta_{x,\phi_{x,b}}}= (-1)^{b\cdot b_{\DIFF}}\ket{s}_{\Reg{In}}\ket{x}_{\Reg{X}}\ket{b\xor 1}_{\Reg{B}}\ket{0}_{\Reg{C}}\ket{s}_{\Reg{Extra}}$$
        $$U_{4,b_{\DIFF}}\ket{\beta_{x,\phi_{x,b}}}= \ket{s}_{\Reg{In}}\ket{x}_{\Reg{X}}\ket{0}_{\Reg{B}}\ket{0}_{\Reg{C}}\ket{x,\phi_{x,b}}_{\Reg{Extra}}$$
        and for all orthogonal $\ket{\psi}$,
        $$U_{3,b_{\DIFF}}\ket{\beta_{x,\psi}} = \ket{s}_{\Reg{In}}\ket{x}_{\Reg{X}}\ket{0}_{\Reg{B}}\ket{0}_{\Reg{C}}\ket{x,\psi}_{\Reg{Extra}}$$
        $$U_{4,b_{\DIFF}}\ket{\beta_{x,\psi}} = \ket{s}_{\Reg{In}}\ket{x}_{\Reg{X}}\ket{0}_{\Reg{B}}\ket{0}_{\Reg{C}}\ket{x,\psi}_{\Reg{Extra}}$$

        Now, define the unitary map $J$ acting on registers $\Reg{X},\Reg{B},\Reg{C},\Reg{Extra}$ defined such that for all $x\in [L],b\in \{0,1\}$, $J$ swaps the following two states
        $$\ket{x}_{\Reg{X}}\ket{0}_{\Reg{B}}\ket{0}_{\Reg{C}}\ket{x,\phi_{x,b}}_{\Reg{Extra}}\leftrightarrow \ket{x}_{\Reg{X}}(-1)^{b_{\DIFF}\cdot b}\ket{b\xor 1}_{\Reg{B}}\ket{0}_{\Reg{C}}\ket{s}_{\Reg{Extra}}$$
        and acts as the identity on all orthogonal states.

        We have that 
        \begin{equation}
            \begin{aligned}
            JU_{4,b_{\DIFF}}\ket{\beta_{x,\phi_{x,b}}} &= (-1)^{b\cdot b_{\DIFF}} \ket{s}_{\Reg{In}}\ket{x}_{\Reg{X}}\ket{b\xor 1}_{\Reg{B}}\ket{0}_{\Reg{C}}\ket{s}_{\Reg{Extra}}\\
            &= U_{3,b_{\DIFF}}\ket{\beta_{x,\phi_{x,b}}}\\
            JU_{4,b_{\DIFF}}\ket{\beta_{x,\psi}} &= U_{4,b_{\DIFF}}\ket{\beta_{x,\psi}} = U_{3,b_{\DIFF}}\ket{\beta_{x,\psi}},\quad \ket{\psi} \bot \ket{\phi_{x,0}},\ket{\phi_{x,1}}
            \end{aligned}
        \end{equation}
    \end{proof}

    \begin{claim}
        $\rho_{4,b_{\DIFF}}=\rho_{5,b_{\DIFF}}$.
    \end{claim}
    \begin{proof}
        Each $\ket{\phi_{x,b}}$ will only be used in a round where $\Reg{X}$ measures to $x$ and $x\notin S$. In particular, since $x$ is new at this point, $\ket{\phi_{x,b}}$ can't have been used in an earlier query. Since $x$ is added to $S$ at the end of the query, $\ket{\phi_{x,b}}$ also can't have been used in a later query. Since all the pairs are independent, sampling the pair at its first (and only) use gives the same distribution as sampling all the pairs at the beginning. Since the pairs are only used once, discarding them after this point also does not affect the resulting states.
    \end{proof}

    \begin{claim}
        $\rho_{5,b_{\DIFF}}=\rho_{6,b_{\DIFF}}$.
    \end{claim}
    \begin{proof}
        For fixed $r,x$, sampling the orthonormal pair $(\ket{\phi_{x,0}},\ket{\phi_{x,1}})$ in step (g)(ii) and applying the phase factors $(-1)^{r\cdot b}$ in steps (g)(iii)--(iv) is equivalent to sampling a Haar random unitary $U$ and using the pair
        $$\ket{\phi_{x,b}} = U Z_{r} \ket{b},\qquad b\in\{0,1\},$$
        where 
        $$Z_{0} = I$$
        $$Z_{1} = I - 2\ketbra{1}$$
        is the map which flips the phase of $\ket{1}$ if $r=1$.

        By Haar invariance, this is identical to sampling a Haar random $U$ and setting 
        $$\ket{\phi_{x,b}} = U \ket{b}$$
        which is the same distribution as $\Hyb_{5,b_{\DIFF}}$.
    \end{proof}

    \begin{claim}
        $\rho_{6,0}=\rho_{6,1}$.
    \end{claim}
    \begin{proof}
        Since the two oracles only differ in the $x\notin S$ case, it suffices to show that the reduced actions of steps iii. and iv. of step (g) agree after averaging over $r$ and tracing out the register $\Reg{B}$ (which is unused after this point).

        We will purify the maps corresponding to steps iii. and iv. by adding an extra register $\Reg{R}=\mathcal{H}(\{0,1\})$ containing $r$, initialized to $\ket{+}$. Let the corresponding maps be $U_{6,0}$ and $U_{6,1}$. The internal state after the register $\Reg{R}$ is created will be a superposition of states of the following forms
        $$\ket{\alpha_b} \coloneqq \ket{s}_{\Reg{In}} \ket{x}_{\Reg{X}}\ket{b}_{\Reg{B}}\ket{1}_{\Reg{C}}\ket{s}_{\Reg{Extra}}\ket{+}_{\Reg{R}}$$
        $$\ket{\beta_{x,\psi}} = \ket{x,\psi}_{\Reg{In}}\ket{x}_{\Reg{X}}\ket{0}_{\Reg{B}}\ket{0}_{\Reg{C}}\ket{s}_{\Reg{Extra}}\ket{+}_{\Reg{R}}$$
        We have
        \[
            \begin{aligned}
                U_{6,0}\ket{\alpha_b}
                    &=U_{6,1}\ket{\alpha_b}\\
                    &=\ket{x,\phi_{x,b}}_{\Reg{In}}\ket{x}_{\Reg X}
                        \ket{b}_{\Reg B}\ket{1}_{\Reg C}
                        \ket{s}_{\Reg{Extra}}
                        \bigl(Z_{\Reg R}^b\ket{+}_{\Reg R}\bigr).
            \end{aligned}
        \]

        We will now define a unitary $J$ acting only on $\Reg{B}\Reg{R}$ as follows
        $$J\ket{b}_{\Reg{B}}\ket{r}_{\Reg{R}} = \ket{b}_{\Reg{B}}\ket{r\xor(b\xor 1)}_{\Reg{R}}$$
        
        We will see now that $JU_{6,0} = U_{6,1}$ on the
        input subspace described above. On $\ket{\alpha_0}$,
        the output registers $\Reg B,\Reg R$ contain
        $\ket{0}\ket{+}$, which $J$ fixes. On
        $\ket{\alpha_1}$ the output has $\Reg B=1$, so $J$
        acts as the identity. Thus
        $$JU_{6,0}\ket{\alpha_b}=U_{6,1}\ket{\alpha_b}.$$

        For valid challenge inputs, we have
        \[
            \begin{aligned}
                U_{6,0}\ket{\beta_{x,\phi_{x,b}}}
                    &=\ket{s}_{\Reg{In}}\ket{x}_{\Reg X}
                        \ket{b\xor 1}_{\Reg B}\ket{0}_{\Reg C}
                        \ket{s}_{\Reg{Extra}}
                        \bigl(Z_{\Reg R}^b\ket{+}_{\Reg R}\bigr),\\
                U_{6,1}\ket{\beta_{x,\phi_{x,b}}}
                    &=(-1)^b U_{6,0}\ket{\beta_{x,\phi_{x,b}}}.
            \end{aligned}
        \]
        For $b=0$, the output has $\Reg B=1$ and $J$ does
        nothing. For $b=1$, the output has $\Reg B=0$ and
        $\Reg R=\ket{-}$, so $J$ applies $X_{\Reg R}$ and
        supplies the required minus sign. Therefore
        $$JU_{6,0}\ket{\beta_{x,\phi_{x,b}}}
            =U_{6,1}\ket{\beta_{x,\phi_{x,b}}}.$$

        Finally, for $\ket{\psi}$ orthogonal to both hidden
        states,
        \[
            \begin{aligned}
                U_{6,0}\ket{\beta_{x,\psi}}
                    &=U_{6,1}\ket{\beta_{x,\psi}}\\
                    &=\ket{s}_{\Reg{In}}\ket{x}_{\Reg X}
                        \ket{0}_{\Reg B}\ket{0}_{\Reg C}
                        \ket{x,\psi}_{\Reg{Extra}}\ket{+}_{\Reg R}.
            \end{aligned}
        \]
        Here $J$ fixes $\ket{0}_{\Reg B}\ket{+}_{\Reg R}$,
        so the same equality holds. By linearity,
        $JU_{6,0}=U_{6,1}$ on the full input subspace,
        including with arbitrary reference registers.

        The register $\Reg R$ purifies the uniform choice of
        $r$, so tracing it out implements the required average.
        Since $J$ acts only on the discarded registers
        $\Reg B,\Reg R$ and commutes with the final control
        erasure, the two hybrids give identical reduced
        responses. Both registers are unused after this query,
        and the updates to $S$ agree. Induction over the
        queries gives $\rho_{6,0}=\rho_{6,1}$.
    \end{proof}

    Only the transition from the first to the second hybrid
    incurs an error. The triangle inequality gives
    \[
        \begin{aligned}
            \TD(\rho_{\SAME},\rho_{\DIFF})
                &\leq
                    \TD(\rho_{1,0},\rho_{2,0})
                    +\TD(\rho_{2,1},\rho_{1,1})\\
                &\leq\frac{8q^{3/2}}{\sqrt L}.
        \end{aligned}
    \]
    This proves the theorem.
\end{proof}
\subsection{A Search Lower Bound}
\label{sec:channel-search-lower-bound}

We turn the decision problem into a search problem by encoding a
hidden string in independent instances of the channels. Let $m\geq1$,
$L\geq1$, and $d\geq2$. Sample $z\leftarrow\{0,1\}^m$, and
independently sample the hidden pairs for $m$ instances
$(\SAME_i,\DIFF_i)$, indexed by $i\in[m]$.
Define a channel $\mathcal C_z$ that measures an input index register
in the computational basis. On outcome $i$, it applies $\SAME_i$
to the remaining input if $z_i=0$, and $\DIFF_i$ if $z_i=1$.
It returns $i$ together with the channel output.
The search problem is to output $z$.

\begin{theorem}\label{thm:channel-search-lower-bound}
    An efficient algorithm recovers $z$ with probability one using
    $2m$ forward queries to any fixed isometry dilation of
    $\mathcal C_z$, or $m$ applications of
    $\Sw[\mathcal C_z,\mathcal C_z]$.
    In contrast, any algorithm $\A$ making at most $q\geq0$ forward
    channel queries satisfies
    \[
        \Pr[\A^{\mathcal C_z}=z]
            \leq2^{-m}+\frac{4m q^{3/2}}{\sqrt L},
    \]
    where the probability is averaged over $z$, the hidden pairs,
    and the algorithm's randomness. The input and output dimensions
    are $m(Ld+1)$, and a common dilation environment can be fixed to
    the square of this dimension.
\end{theorem}

\Cref{cor:channel-dilation-near-unit-error} follows from the above
theorem by setting $m=n$, $L=2^{3n}$, $d=2$, and $c=1/4$.
For $q\leq2^{cn}$, the recovery probability is at most
\[
    2^{-n}+4n2^{-9n/8}\leq2^{-n/2}=2^{-2cn}
\]
for all sufficiently large $n$, while the channel and its dilation
environment each contain $O(n)$ qubits.
A simulator with total trace-distance error at most $1-2^{-cn}$
would turn the perfect recovery algorithm into a channel-query
algorithm with recovery probability at least $2^{-cn}$,
contradicting this bound.

\begin{proof}
    Let $W$ be any fixed dilation of $\mathcal C_z$. For a fixed
    input index $i$, the output index is always $i$, so there is
    an isometry $W_i$ such that
    \[
        W\bigl(\ket{i}\ket{\psi}\bigr)
            =\ket{i}W_i\ket{\psi}.
    \]
    Tracing out the environment shows that $W_i$ is a fixed
    dilation of $\SAME_i$ if $z_i=0$, and of $\DIFF_i$ if $z_i=1$.
    Run the two-query test from
    \Cref{cor:channel-dilation-distinguisher} once at each index.
    Each test identifies its bit with certainty, so $2m$ dilation
    queries recover $z$ with probability one. Alternatively,
    applying $\Sw[\mathcal C_z,\mathcal C_z]$ at a
    fixed input index $i$ applies the self-switch of the
    corresponding channel and preserves the index. Thus
    \Cref{prop:self-switch-distinguisher} recovers $z$ with $m$
    switch applications.

    For the channel-query lower bound, the case $q=0$ is immediate:
    the algorithm's output is independent of the uniform string
    $z$, so its recovery probability is at most $2^{-m}$.
    For $q\geq1$, compare the following hybrid games. In each game,
    the challenger samples $z$ and all hidden pairs independently,
    and measures the index on every query.
    \begin{enumerate}
        \item $\Hyb_0$. At every index $i$, the challenger applies
        $\SAME_i$, independently of $z_i$, and returns the index
        together with the output.

        \item $\Hyb_j$, for $j=1,\ldots,m$. At indices $i\leq j$,
        the challenger applies $\SAME_i$ if $z_i=0$ and $\DIFF_i$
        if $z_i=1$. At indices $i>j$, it applies $\SAME_i$.
        It returns the index together with the output.
        Thus $\Hyb_m$ gives access to $\mathcal C_z$.
    \end{enumerate}
    Let $p_j$ be the probability that the algorithm outputs $z$
    in $\Hyb_j$. In $\Hyb_0$, all responses are independent of
    $z$, so $p_0\leq2^{-m}$.

    The games $\Hyb_{j-1}$ and $\Hyb_j$ agree when $z_j=0$.
    Conditioned on $z_j=1$, they differ only in whether index
    $j$ applies $\SAME_j$ or $\DIFF_j$. A distinguisher can
    sample $z$ with this conditioning and all other instances,
    embed its unknown channel at index $j$, and simulate the
    remaining indices using the sampled descriptions. It measures
    each query index and calls the unknown channel only on outcome
    $j$. This exactly simulates the relevant game using at most
    $q$ queries to the unknown channel. It outputs $1$ exactly when
    the algorithm's answer equals the sampled $z$.
    Since $\Pr[z_j=1]=1/2$, \Cref{thm:channel-query-lower-bound} gives
    \[
        \abs{p_j-p_{j-1}}
            \leq\frac12\cdot\frac{8q^{3/2}}{\sqrt L}
            =\frac{4q^{3/2}}{\sqrt L}.
    \]
    Summing over the hybrids proves the claimed recovery bound.
    The assertions about dimensions follow from the indexed
    construction and the standard environment dimension bound.
\end{proof}

\clearpage
\section*{Acknowledgments}

The author would like to thank Marco Fanizza for his talk on the "Random Stinespring Superchannel" at the Random Purification Workshop at STOC 2026. Key ideas for the positive simulation result were developed because of and during the course of this talk. The author would also like to thank John Wright and Ewin Tang for helpful discussions.
\section*{AI Disclosure}

This document was prepared with the help of ChatGPT 5.6 Sol, 6 Astra, and 6.1 Sol. The theorems involve a combination of human ideas and model assisted insights. AI assisted with drafting technical details for all proofs, with the final results being both independently verified and substantially rewritten.

For the isometry simulation result, the description of the simulator as well as a proof sketch (including the order of hybrids) were developed without AI assistance using the approximate path recording framework of~\cite{STOC:MaHua25}. After this point, ChatGPT assisted heavily with drafting the proofs and pointing out issues. Its most substantive use for this portion of the paper was to point out that the parameters of the construction could be improved by considering exact path recording~\cite{FLMNW26,GY25}. In particular, it developed the original versions of the proofs in~\Cref{sec:exact-pr-comparison}. It also pointed out the necessity for an isometry version of decontrolling unitaries, although this proof was primarily human-developed. ChatGPT also heavily assisted with finding applications of the main theorem and performing literature search.

For the lower bound result, we began with a human proposed sketch of an initial lower bound. This lower bound gave a distinguishing problem which could be solved with advantage $\frac{1}{4}$ by an algorithm making dilation queries, while distinguishing with advantage better than $\negl(n)$ required exponential queries. Afterwards, the model was asked to improve the bound to perfect distinguishability with two queries in the dilation query case, which it succeeded at. The model was then asked if the bound could be extended to apply to the quantum switch, which was nearly immediate.

\bibliographystyle{alpha}
\bibliography{refs/abbrev0,refs/crypto,refs/refs,impossibility/references}

\clearpage
\appendix
\section{Approximating the quantum switch by classical control}
\label{app:switch-half-error}

\begin{theorem}\label{thm:switch-half-error}
    For any channels $\Phi$ and $\Psi$ on a common space, the quantum switch $\Sw[\Phi,\Psi]$ can be simulated using one forward query to each channel and classically controlled causal order, with trace-distance error at most $1/2$ on every input, including inputs entangled with an arbitrary reference register.
\end{theorem}

\begin{proof}
Let $\Reg C$ be the switch control qubit, and define its dephasing channel by
\[
    \Delta_{\Reg C}(\sigma)
        =\sum_{c\in\{0,1\}}\ketbra{c}_{\Reg C}\sigma\ketbra{c}_{\Reg C}
        =\frac12\left(\sigma+Z_{\Reg C}\sigma Z_{\Reg C}\right),
\]
with identity operators on all other registers implicit. Measuring $\Reg C$ in the computational basis and using the outcome to choose the order of $\Phi$ and $\Psi$ implements
\[
    \widetilde{\Sw}[\Phi,\Psi]
        =\Sw[\Phi,\Psi]\circ\Delta_{\Reg C}
        =\Delta_{\Reg C}\circ\Sw[\Phi,\Psi].
\]
The last equality follows as measuring $\Reg C$ commutes with controlled operations. This procedure uses one query to each channel and classically controlled causal order; it is the fully dephased classical switch considered in~\cite{Branciard16switch}. Note that the distance from the non-dephased switch is not directly considered in~\cite{Branciard16switch}.

For any input, including one entangled with an arbitrary reference register, let $\sigma$ be the ideal switch output. The simulated output is $\Delta_{\Reg C}(\sigma)$, so
\[
    \TD\left(\sigma,\Delta_{\Reg C}(\sigma)\right)
        =\frac12\TD\left(\sigma,Z_{\Reg C}\sigma Z_{\Reg C}\right)
        \leq\frac12.
\]
Thus the simulation has trace-distance error at most $1/2$ on every input.
\end{proof}

\end{document}